\documentclass[article]{biometrika}

\usepackage{amsmath}
\usepackage{amsfonts}
\usepackage{xcolor}
\usepackage[T1]{fontenc}
\usepackage{graphicx}
\usepackage{subcaption}

\usepackage{times}
\usepackage{bm}
\usepackage{natbib} 

\usepackage[plain,noend]{algorithm2e}

\usepackage[colorlinks=true, linkcolor=blue, citecolor=blue, urlcolor=blue]{hyperref}
  
\makeatletter
\renewcommand{\algocf@captiontext}[2]{#1\algocf@typo. \AlCapFnt{}#2} 
\def\@algocf@capt@plain{top}
\renewcommand{\algocf@makecaption}[2]{%
 \addtolength{\hsize}{\algomargin}%
 \sbox\@tempboxa{\algocf@captiontext{#1}{#2}}%
 \ifdim\wd\@tempboxa >\hsize
  \hskip .5\algomargin%
  \parbox[t]{\hsize}{\algocf@captiontext{#1}{#2}}
 \else%
  \global\@minipagefalse%
  \hbox to\hsize{\box\@tempboxa}
 \fi%
 \addtolength{\hsize}{-\algomargin}%
}
\makeatother

\def\T{{ \mathrm{\scriptscriptstyle T} }}

\usepackage{threeparttable}

\begin{document}





\title{High-dimensional semi-supervised learning: in search of optimal inference of the mean}
\author{Yuqian Zhang}
\affil{Institute of Statistics and Big Data, Renmin University of China,\\ 59 Zhongguancun Avenue, Haidian District, Beijing 100872, P. R. China}
\author{\and Jelena Bradic}\thanks{This version corresponds to the accepted manuscript. The final published version is available at https://doi.org/10.1093/biomet/asab042.}
\affil{Department of Mathematics, University of California San Diego,\\ 9500 Gilman Drive, La Jolla, California 92093 0112, U.S.A.
\email{jbradic@ucsd.edu}}
\maketitle

\begin{abstract}
A fundamental challenge in semi-supervised learning lies in the observed data's disproportional size when compared with the size of the data collected with missing outcomes. An implicit understanding is that the dataset with missing outcomes, being significantly larger, ought to improve estimation and inference. However, it is unclear to what extent this is correct. We illustrate one clear benefit: root-$n$ inference of the outcome’s mean is possible while only requiring a consistent estimation of the outcome, possibly at a rate slower than root $n$. This is achieved by a novel $k$-fold, cross-fitted, double robust estimator. We discuss both linear and nonlinear outcomes. Such an estimator is particularly suited for models that naturally do not admit root-$n$ consistency, such as high-dimensional, nonparametric or semiparametric models. We apply our methods to estimating heterogeneous treatment effects.

\end{abstract}

\begin{keywords}
Coefficient of determination; Double-robustness; Missing data; Model-lean inference.
\end{keywords}

\section{Introduction}

We consider a semi-supervised setting with $n$ independent and identically distributed pairs $(X_i, Y_i)_{i=1}^n \sim P_{(X,Y)}$ of observations, with covariates $X_i \in \mathbb{R}^{p-1}$ and the outcome $Y_i \in \mathbb{R}$. We presuppose the existence of an additional set of $m$ observations, $(X_i)_{i=n+1}^{n+m}$. With $\tau=\lim_{m,n\to\infty}n/(m+n)\in[0,1]$ denoting the ratio of the fully observed data and data with the missing outcomes, we are particularly focused on the case of $\tau=0$, i.e., $m\gg n$. The semi-supervised learning setting can be viewed as a particular missing data setting, where the outcome is missing completely-at-random. Although the missing data literature, in general, addresses a more general setting of the outcomes missing-at-random \citep{scharfstein1999adjusting}, semi-supervised learning has a particular caveat that the missing data's size is enormous, $m \gg n$. With $m \gg n$, typical missing-at-random approaches \citep{bang2005doubly} no longer apply. The positivity/overlap condition (see, e.g., \citet{rotnitzky2012improved}), is no longer satisfied; with $\tau=0$, the probability of observing the outcome converges to zero, therefore implying that the semi-supervised setting is not a simple subset of the missing-at-random setting.
Instead, we treat the missingness size, an impediment from the missing-at-random perspective, as a semi-supervised strength. In the case of infinite missingness of the response, we are left with infinite additional information regarding the covariates' distribution, $P_X$. Mimicking the known $P_X$ setting, we remove the bias in estimating the outcome model and show that  semi-supervised-double-robust inference is achievable.

Our main contribution is in constructing new semi-supervised estimates of $\theta=E(Y)$ and in providing root-$n$ inferential guarantees while allowing for misspecification of the distribution of $Y\mid X$. An impediment to providing optimal inferences about $\theta$ lies in the inability to estimate $E(Y\mid X)$ with root-$n$ guarantees. Sparse regularizers, random forests, nonparametric (smoothing) estimators, or neural networks do not admit root-$n$ consistency. While there is vast literature on semi-supervised learning, comparatively little is known about making inferences about $\theta$; see \cite{zhu2005semi}. Recent results of \cite{wasserman2008statistical}, \cite{el2016asymptotic} and \cite{mai2018random} consider the class of low-dimensional graph-oriented semi-supervised algorithms. Semi-supervised learning in the context of classification has had a long tradition; see \cite{chapelle2009semi} and \cite{grandvalet2005semi}. 
A small but growing literature has considered the development of semi-supervised inferential procedures. The recent work of \cite{zhang2019semi} is a special case of our construction. Authors utilize the least-squares approach in linear models whenever $p = o( n^{1/2})$. Our results are based on $n^{-1} \log(p)=o(1)$ together with many possible estimators, e.g., random forests and neural networks. \cite{chakrabortty2018efficient} develop the semi-supervised regression method with improved efficiency when the linear model is misspecified. \cite{gronsbell2018semi} consider semi-supervised prediction, while \cite{cai2018semi} propose semi-supervised explained variance estimates. We, therefore, view our contribution as complementary to this growing literature. 

We believe that our new estimating tools will be useful beyond the specific class of environments studied here. We illustrate this point by applying our findings to heterogeneous treatment effects. The existing approaches of \cite{chernozhukov2017double,chernozhukov2018double} and \cite{kunzel2019metalearners} build learners that can conform to many machine learning methods \citep{athey2018approximate,wager2017estimation}. However, they do not consider the semi-supervised setting with the outcome and the treatment missing. We discover that the asymptotic variance size is reduced regardless of whether additional information on the treatment is available. Moreover, treatment assignment can potentially depend on all covariates with no explicit sparsity requirement. The method also shares the low-dimensional asymptotic efficiency of \cite{cheng2018efficient}.
  
 \section{Efficient estimation of the mean}\label{s2}

\subsection{From de-biasing to double-robustness}\label{s21}

 Let $\beta^* \in \mathbb{R}^p$,  the population slope, be an $l_2$ projection defined as 
$
\beta^*=\arg\min_{\beta\in\mathbb R^p}E \left(Y -\beta_1-X ^\T\beta_{-1} \right)^2.
$
Here, $\beta_{-j}$ denotes
 $\beta$ with the $j$th coordinate removed. For $\varepsilon=Y-\beta_1^*-X^\T\beta_{-1}^*$ and $\sigma^2_{\varepsilon} = \mbox{var}(\varepsilon)$ with $E(\varepsilon\mid X)\neq0$
we do not necessarily assume  that the regression model is linear. 
With $\mu$ and $C$, denoting the mean and the covariance of $X_i$, respectively, we use $V_i=X_i-\mu$, and $Z_i = C^{-{1/2}}(X_i -\mu)$. 
With  $\tilde X_i=(1,X_i^\T)^\T$ and  $\tilde V_i=(1,V_i^\T)^\T$, let $\tilde\mu=(1,\mu^\T)^{\T}$ and $\tilde C=\mathrm{cov}(\tilde X)$ denote the mean and covariance of $\tilde X =(1,X^\T)^\T$.
The mean of the response, $\theta = E(Y)$, 
 can be seen as a linear contrast of $\beta^*$:
$
 \theta = {\tilde \mu }^\T {\beta^*}.
$

 When $p \gg n$, a good candidate estimate of $\beta^*$, is a regularized estimator, $\hat \beta$, e.g., Lasso \citep{tibshirani1997lasso} or square-root Lasso \citep{belloni2011square}. However, such estimators suffer from slower than root-$n$ consistency: when the outcome model is linear, $\| \hat \beta - \beta^*\| _2^2 = o_P\{ {s \log (p) }/ { n}\}$ with $s=|\{j:\beta_j^*\neq0\}|$. Hence, a plug-in estimate will not achieve root-$n$ inference regarding $\theta$, even if the outcome model is correct, unless $s$ is a constant.  Existing literature provides easy solutions with many possible ways to remove the bias of regularization.  Each of these could potentially achieve root-$n$ inference of $\theta$ but would, however, require strong assumptions on the models: the outcome must be well specified as well as sparse enough.
For example, let
 $
\hat\beta_\mathrm{db}=\hat\beta+ n^{-1} \sum_{i=1}^n \hat\Theta\tilde{ X_i} (Y_i-\tilde{ X_i}^\T \hat\beta),
$
denote the de-biased Lasso \citep{van2014asymptotically}. Here, $\hat \Theta$, is a candidate estimate of $\tilde\Sigma^{-1}$, $
\tilde\Sigma= E\tilde X \tilde X ^\T \in \mathbb{R}^{p \times p}$. Root-$n$ inference of $\theta$ would then  require outcome sparsity $s=o\{n^{1/2}/\log(p)\}$  
as well as $ |\{k\neq j:(\tilde\Sigma^{-1})_{j,k}\neq0\}| =o\{ n /\log(p)\}$ \citep{van2014asymptotically}.

However, $\hat\beta_\mathrm{db}$ does not directly use the additional covariate information available in the semi-supervised setting. Let us consider a particular case where $P_X$, and with it, $\tilde \Sigma^{-1}$ and $\tilde \mu$ are known. In this case, we could use an improved de-biased semi-supervised estimator
 $
\tilde \beta=\hat\beta+ n^{-1} \sum_{i=1}^n \tilde\Sigma^{-1} \tilde{ X_i} (Y_i-\tilde{ X_i}^\T \hat\beta),
$
which then leads to 
$
\tilde\mu^\T \tilde \beta=\tilde\mu^\T\hat\beta+n^{-1}\sum_{i=1}^ne_1^\T\tilde X_i^\T(Y_i-\tilde X_i\hat\beta) 
=\tilde\mu^\T\hat\beta+ n^{-1}\sum_{i=1}^n ({Y_i-\tilde X_i^\T\hat\beta}),
$
where $e_1 =(1,0,0,\ldots,0)^\T$. Interestingly, by algebraic  manipulation, it is not difficult to see that the right-hand side above becomes  
$
\bar Y + (\tilde\mu - \bar X) ^\T\hat\beta 
$, where $\bar Y=n^{-1}\sum_{i=1}^nY_i$ and $\bar X=n^{-1}\sum_{i=1}^n\tilde X_i$,
therefore matching with the low-dimensional estimator of \cite{zhang2019semi}. There seems to be an intricate connection between the above estimator and the double-robust, missing-at-random estimators of \cite{bang2005doubly}. However, there is an important difference. If $T_j=1$ for $j=1,\dots,n$ and zero otherwise, i.e., $T$ is the indicator of the observed data. Missing-at-random treats $T$ as a random variable whereas semi-supervised learning treats $T$ as fixed,  non-random. 
Semi-supervised learning can be viewed as missing-at-random conditional on $(T_i)_{i=1}^{m+n}$ being fixed. Then, the missing-at-random average treatment effect of the treated matches the above estimator 
\[
 \tilde \mu^\T \hat \beta + {(n+m)}^{-1} \sum_{i=1}^{n+m} T_i (Y_i - X_i \hat \beta)/\mathrm{pr}(T_i=1\mid X_i),
\]
where $\mathrm{pr}(T_i=1\mid X_i) = \mathrm{pr}(T_i=1)=n/ (n+m)$. 
 However, missing-at-random double-robust estimates require $\mathrm{pr}(T_i=1\mid X_i) > 0 $ whereas in the semi-supervised setting we have $\mathrm{pr}(T_i=1\mid X_i) \to 0 $ with $m \gg n$.

 In the semi-supervised setting we aim to show that the above estimator's sample equivalent will suffice for root-$n$ inference on $\theta$. Let $
\tilde \theta=\hat \mu^\T\hat\beta + n^{-1} \sum_{i=1}^n{(Y_i-\tilde X_i^\T\hat\beta)}$ and $\hat \mu = {(n+m)}^{-1} \sum_{i=1}^n \tilde X_i.
$
 Our estimator will use cross-fitting, which plays a crucial role in establishing the double-robust property of the proposed estimator, i.e., in controlling the term $t_2$ in the decomposition
$
 \tilde \theta - \theta = 
 t_1 + t_2 + t_3,
$
where 
$ t_1 = \theta - n^{-1} \sum_{i=1}^n{Y_i}, 
\quad t_2 =  (n^{-1} \sum_{i=1}^n\tilde X_i -\hat \mu)^\T (\hat\beta - \beta^*)
,
\quad t_3=  (n^{-1} \sum_{i=1}^n\tilde X_i -\hat \mu)^\T \beta^*.
$
The cross-fitting technique helps in removing the bias arising from  $t_2$.
 With the use of cross-fitting,  $\hat \beta$'s and $X_i$'s influences in  $t_2$ are separated and tight control of  $t_2$ is achieved under minimal conditions. Without cross-fitting, 
$
 |\tilde \theta - \theta| \leq \|n^{-1} \sum_{i=1}^n\tilde X_i -\hat \mu \|_\infty \| \hat \beta - \beta^*\|_1
$
 where the right-hand side is $O_P(n^{-1/2})$ as long as $s \leq n^{1/2}/\log(p)$. Instead, with the use of cross-fitting, we can guarantee root-$n$
consistency as long as $s \leq n/\log(p)$.  
Cross-fitting can be traced back to the natural ideas of cross-validation. Historical background is provided by \cite{stone1974cross} and \cite{geisser1975predictive} for example. More recently, \cite{rinaldo2016bootstrapping} showed that sample splitting increases the accuracy and robustness of inference. \cite{chernozhukov2017double} used cross-fitting to define double-robust missing-at-random estimates.

We start by splitting the labeled observations into $K$ sets, $I_{k}$, each of size $N$, and split the unlabeled observations into sets $I_k'$. Let $J_k=I_k\cup I_k'$ with $|J_k| =M$. Let $\hat\beta^{(-k)}$ denote an estimate of $\beta^*$ computed on all but the $k$th labeled observations, 
$
\hat\beta^{(-k)}=\hat\beta(\{(\tilde X_i,Y_i):i\in\{1,2,\ldots,n\}\setminus I_k\}) \in\mathbb R^p.$
Then, we propose
\begin{equation}\label{eEY}
\hat\theta^{(k)}={\ \hat\mu^{(k)}}^\T\hat\beta^{(-k)} + N^{-1}\sum_{i\in I_k} \left(Y_i - \tilde X_i^\T\hat\beta^{(-k)}\right), \qquad \hat\mu^{(k)}=M^{-1}\sum_{i\in J_k}\tilde X_i.
\end{equation} 
Finally, we propose the following semi-supervised estimator, which aggregates the above estimates:
$$
\hat\theta=K^{-1}\sum_{k=1}^K\hat\theta^{(k)}.
$$ We will show  that this estimator becomes an unbiased estimator of $\theta$, even in finite samples.

 \subsection{From the mean to the coefficient of determination}\label{s22} 

A crucial statistical problem 
 is the estimation of the Proportion of Variance Explained (PVE), 
 $\mathrm{PVE}=\mathrm{var}(\tilde X^\T\beta^*)/\sigma_Y^2.
 $
 Estimation of PVE with $p \gg n$ is difficult due to the numerous overfitting issues. 
 In this section, we propose a semi-supervised coefficient of determination, $R^2$, an estimator of $\mathrm{PVE}$.
 The estimation of the explained variance, $b^2=\mathrm{var}(\tilde X^\T\beta^*)$ \citep{cai2018semi}, can be performed with the  cross-fitted residuals
\begin{equation}\label{ev}
\hat b^{2^{(k)}}=\hat\beta^{(-k)^\T} \hat C^{(k)}\hat\beta^{(-k) }  +2N^{-1}\sum_{i\in I_k}\hat\beta^{(-k)^\T} \hat V_i \hat \varepsilon_i, \quad \hat \varepsilon_i =Y_i-\hat\theta-\hat\beta^{(-k)^\T} \hat V_i,
\end{equation}
and $\hat b=K^{-1}\sum_{k=1}^K\hat b^{2^{(k)}}$, where the estimates of $\tilde V_i$ are $\hat V_i=\tilde X_i-\hat\mu^{(k)}$ and their covariance $ \hat C^{(k)}=M^{-1}\sum_{i\in J_k}\hat V_i\hat V_i^\T$. 
The motivation behind this careful construction is governed by bias-propagation in the high-dimensional setting;
 as we will show, the residuals as defined above are, however, root-$n$ consistent.  This, in turn, provides a more stable estimate and enables theoretically weak conditions. To see that the naive estimate $Y_i-\hat\beta^{\T} \tilde X_i $ may not guarantee root-$n$ consistency, we only need to observe that in such a case, 
$Y_i-\hat\beta^{\T}\tilde X_i= \varepsilon_i+ (\theta-\hat\beta^\T\tilde\mu)-(\hat\beta-\beta^*)^\T(\tilde X_i-\tilde\mu)$,
 while the term $\theta-\hat\beta^\T\tilde\mu$ is not necessarily root-$n$ consistent whenever $p \gg n$.
Our cross-fitted construction can be seen as a bias-corrected estimate of the residuals. 
We propose a new estimator of the variance of the response, $\sigma_Y^2=\mathrm{var}(Y)$,
\begin{equation}\label{eq:var}
\hat\sigma_Y^{2^{(k)}} =N^{-1}\sum_{i\in I_k} (Y_i-\hat\theta)^2  + N^{-1}\sum_{i\in I_k} { \ { \hat\beta^{(-k)}
 } }^\T \left(\hat C^{(k)}  -  \hat V_i {\hat V_i}^\T \right)\hat\beta^{(-k)},
\end{equation}
and with it 
$
\hat\sigma_Y^2=K^{-1}\sum_{k=1}^K\hat\sigma_Y^{2^{(k)}}.
$
Our results  also hold for the truncated version $\hat\sigma_{Y,\mathrm{trunc}}^2=\max(\hat\sigma_Y^2,0)$.
 A classical estimate, the simple sample variance, 
$
 S_Y^2 = n^{-1} \sum_{i=1}^n (Y_i - \bar Y)^2,
$ does not utilize any additional knowledge of the covariates. 
 Alternatively, one may consider 
$
 n^{-1} \sum_{i=1}^n (Y_i - \hat \theta)^2.
$
However, both of these estimates can be improved.
Our theoretical results demonstrate a persistent variance magnification, $
 n^{-1}\sum_{i=1}^n(Y_i-\hat\theta)^2=\sigma_Y^2+n^{-1}\sum_{i=1}^n \{{ \beta^*}^\T ( \tilde V_i\tilde V_i^\T -\tilde C  ) \beta^* \}+ T + O_P(n^{-1}),
$
 where $ E(T)=0 $ and $T = n^{-1}\sum_{i=1}^n({ \ 2{\beta^*} }^\T\tilde V_i\varepsilon_i+\varepsilon_i^2-\sigma_\varepsilon^2)$; details are presented in the Supplement. Hence, our estimator  adds a correction term  so that the contribution of the middle term disappears. 
 Therefore, $R^2$ can be obtained by plugging in the estimators of $b^2 $, \eqref{ev}, and the variance of the response $\sigma_Y^2$,\eqref{eq:var},
\begin{equation}\label{R2}
R^2= K^{-1}\sum_{k=1}^K \hat b^{2^{(k)}}/\hat\sigma_Y^{2^{(k)}}.
\end{equation}

 \subsection{ Root-$n $ consistency} \label{sec:consistency}
 
 We establish the root-$n$ consistency of the proposed semi-supervised estimators. Constants in what follows, possibly changing from line to line, are independent of the sample size.

\begin{condition}\label{a2}
 Let the covariance matrix $C$ be such that $\lambda_{\min}(C)>0$ and $\lambda_{\max}(C) \leq c_1$ and $\sup_{\|a\|_2=1} E |a^\T Z|^{2+c}<c_1$ as well as $ E |Y|^{2+c}<c_1$, for positive constants $c,c_1>0$. 
\end{condition}

\begin{condition}\label{a6}
The responses are such that $ E |Y| ^{4+c}<c_1$ whereas the covariance  matrix $C$ satisfies, $\lambda_{\min}(C)>0$ and $\lambda_{\max}(C) \leq c_1$ and $\sup_{\|a\|_2=1} E |a^\T Z |^{4+c}<c_1$ for positive constants $c,c_1>0$.
\end{condition}

\begin{condition}\label{a3}
$\hat\beta $ is an estimator for $\beta^*$ that satisfies 
$
\|\hat\beta-\beta^*\|_2=O_P(1),
$
 as $n,p\rightarrow\infty$.
\end{condition}

Condition \ref{a2} or \ref{a6}, used one at a time,  provide 
a well-defined linear approximation model $\beta^*$. 
 A bounded variance of $Y$ simplifies exposition; all of the results still hold even if this condition is removed. However, the results would be less interpretable.  
 Condition \ref{a3} allows for
  a wide variety of estimates of $\beta^*$: Lasso, Dantzig, Square-root Lasso, Elastic-net \citep{zou2005regularization} or Slope \citep{bogdan2015slope} are plausible. Similarly, different structural forms of $\beta^*$ are permissible; a considerably weaker form of sparsity, $l_r$ sparsity with $r\in(0,1)$, would be effective as long as  $\|\beta^*\|_r^r=o[\{n/\log(p)\}^{1-r/2}]$ \citep{ye2010rate}, for example. As per  Conditions \ref{a2} and \ref{a6}, bounded $2+c$ and $4+c$ moments allow heavy-tailed distributions for the covariates as well as the noise; see, e.g., the Huber estimate of 
\cite{sun2018adaptive}.

\begin{theorem}\label{c22}
Let Conditions \ref{a2} and \ref{a3} hold. Then, as $m,n,p\rightarrow\infty$,
$
\hat\theta-\theta=O_P(n^{-1/2}).
$
Moreover, if Condition \ref{a6} hold as well, 
$
 \hat\sigma_Y^2-\sigma_Y^2 =O_P(n^{-1/2}).
$
\end{theorem}

Regarding $\hat \theta$, Condition \ref{a2} can be relaxed to bounded $1+c$ moments.  Importantly,
 we do not rely on a strong signal-to-noise ratio to achieve root-$n$ consistency. If $s=p$, one can show that the Lasso estimate equals zero with high-probability, in which case the proposed estimate will be the same as the naive $\bar Y$. Hence, there is no loss in efficiency, and it seems that the semi-supervised mean estimate is advantageous in almost all cases. We discuss some aspects of the variance in the Supplement.

 \subsection{Asymptotic normality} \label{sec:normality}
 
In this section, we proceed to prove that semi-supervised estimates are asymptotically normal and that they improve the efficiency of estimation by borrowing strength from the additional dataset.

\begin{condition}\label{a4}
$\hat\beta$ is an estimator of $\beta^*$ that satisfies
$
\|\hat\beta-\beta^*\|_2=o_P(1),\text{ as }n,p\rightarrow\infty.
$
\end{condition}

\begin{theorem}\label{t23}
Let Conditions \ref{a2} and \ref{a4} hold. Then,  as $m,n,p\rightarrow\infty$,
\begin{equation}\label{asymtheta}
n^{1/2}(\hat\theta-\theta)\rightarrow N \left(0,\sigma_\varepsilon^2+\tau b^2\right)
\end{equation}
in distribution, provided that $\sigma_\varepsilon^2+\tau b^2>c$ for some constant $c>0$.
\end{theorem}

 Compared with requirements for inference in high-dimensional linear models, Conditions \ref{a2} and \ref{a4} are milder. Where we require only moderately sparse regimes $s=o(n/\log{p})$, high-dimensional and even doubly-robust methods require more strict settings; see, e.g., \cite{bradic2019sparsity}, \cite{smucler2019unifying} and \cite{tan2018model,tan2020regularized}.
In particular, we do not require any sparsity structure on $\Sigma^{-1}$, a condition that has been typically assumed throughout the literature, if the variance is unknown. Lastly, we do not require homogeneity of the errors, $\varepsilon$.

Regarding efficiency, observe that
$
\mathrm{var}(n^{1/2}\bar Y)=\sigma_Y^2=\sigma_\varepsilon^2+b^2\geq\sigma_\varepsilon^2+\tau b^2,
$
where $\sigma_\varepsilon^2+\tau b^2$ is the asymptotic variance of $\hat\theta$ as in \eqref{asymtheta}. Hence,
the semi-supervised estimator $\hat\theta$ is asymptotically at least as accurate as $\bar Y$ and is often more accurate. Namely, the additional unlabeled data reduce the asymptotic variance by $(1-\tau)b^2$.
The more unlabeled data we observe, the more accurate the proposed estimator $\hat\theta$ becomes. When $\tau=0$, the asymptotic variance is equivalent to the case of known  $P_X$.
 
Throughout the paper, we mainly focus on the case of the signal-to-noise ratio, $\mathrm{SNR}=b^2/\sigma_\varepsilon^2$,  being bounded away from $0$ and $\infty$. However, observe that the two extremes are not particularly informative. Namely, the case of $\mathrm{SNR}=0$ illustrates that no estimator can improve the naive $\bar Y$. Conversely, the case of $\mathrm{SNR}=\infty$ and $\tau =0$, illustrates that semi-supervised estimator can potentially lead to a better than $ n^{1/2}$ convergence rate.
Set $\rho_j=\mathrm{Corr}(Z_{j},Y)$ for each $j\in\{1,2,\ldots,p-1\}$. Then,
$
b^2={\beta_{-1}^*}^\T C\beta_{-1}^*=\{C^{-1} E (V Y )\}^\T CC^{-1} E (V Y )=\sigma_Y^2\sum_{j=1}^{p-1}\rho_j^2.
$
If $\tau<1$ and $\sigma_Y^2\sum_{j=1}^{p-1}\rho_j^2>c$ for some $c>0$, i.e., when at least one of the covariates has positive marginal correlation with the response, $\hat\theta$ is asymptotically more accurate than $\bar Y$.  

Our estimator is also optimal in the following sense. The asymptotic variance in Theorem \ref{t23} is the same as that of \cite{zhang2019semi}, proved under a low-dimensional setting; see their Theorem 2.4. Moreover,  it also achieves the oracle lower bound presented in their Proposition 3.1.
 The following result presents theoretically valid root-$n$ confidence intervals of $\theta$, while only requiring consistency of $\hat \beta$ at an arbitrarily slow rate. 

\begin{theorem}\label{t31}
Let Conditions \ref{a2} and \ref{a4} hold. 
With $
\hat \varepsilon_i$ defined in \eqref{ev}, we define $\hat\sigma_\varepsilon^2=n^{-1}\sum_{i=1}^n\hat \varepsilon_i ^2.$
Then, whenever $m,n,p\rightarrow\infty$,  $\hat\sigma_\varepsilon^2 = \sigma_\varepsilon^2 + o_P(1)$, 
$
\hat b^2=b^2 + o_P(1),
$
and a  valid 
confidence intervals about $\theta$, at significance level $\alpha$, is defined as
$$
\mathrm{CI}(\theta)= \left[\hat\theta-z_{1-\alpha/2}\{\hat\sigma_\varepsilon^2/n+\hat b^2/(m+n)\}^{1/2}, \ \hat\theta+z_{1-\alpha/2}\{\hat\sigma_\varepsilon^2/n+\hat b^2/(m+n)\}^{1/2}\ \right],
$$
with $z_{1-\alpha/2}$ being $(1-\alpha/2)$-quantile of a standard normal distribution. 
\end{theorem} 

A few comments are in order. 
If we are willing to assume 
 Condition \ref{a6}, we show that 
$\hat b^2-b^2=O_P(\|\hat\beta-\beta^*\|_2^2+n^{-{1/2}})$. In contrast, a naive 
 plug-in estimate of $b^2$, ${ \ { \hat\beta^{(-k)}} }^\T \hat C \hat\beta^{(-k)}$ would only guarantee $O_P(\|\hat\beta-\beta^*\|_2)$. Therefore, our result on $\hat b^2$ can be seen as complementary to \cite{cai2018semi}. We provide the same convergence rate, whenever $b^2>c$, $c>0$, however, with weaker assumptions: we allow heavy-tailed $X$ and  $\varepsilon$ and misspecified  linear model. An asymptotically normal result holds once $\|\hat\beta-\beta^*\|_2=o_P(n^{-{1}/{4}})$; the details of the asymptotic theory regarding $\hat b$ are contained in Theorem S2 of the Supplement under a more general setting.

Next, we discuss the high-dimensional $R^2$ semi-supervised estimate. We begin by highlighting the asymptotic results on the variance estimate, followed by a simple corollary regarding the asymptotics of $R^2$.

 \begin{theorem}\label{c33}
Let Conditions  \ref{a6} and \ref{a4} hold.
Then, as $m,n,p\rightarrow\infty$,
\begin{equation}\label{7}
n^{1/2}(\hat\sigma_Y^2-\sigma_Y^2)\rightarrow N \left\{0,\mathrm{var}(\varepsilon^2+2{\beta^*}^\T\tilde V \varepsilon )+\tau\mathrm{var}({\beta^*}^\T\tilde V )^2 \right\}
\end{equation}
in distribution, provided that $\mathrm{var}(\varepsilon^2+2{\beta^*}^\T\tilde V \varepsilon )+\tau\mathrm{var}({\beta^*}^\T\tilde V )^2>c $ for some constant $c>0$. Moreover,
 for $\hat\sigma_\nu^2$ and $\hat\sigma_\xi^2$ defined in the Supplement, we have
$$
\hat\sigma_\nu^2+n(m+n)^{-1}\hat\sigma_\xi^2 = \mathrm{var}(\varepsilon^2+2{\beta^*}^\T\tilde V \varepsilon )+\tau\mathrm{var}({\beta^*}^\T\tilde V )^2 + o_P(1).
$$
\end{theorem}

A sufficient condition regarding Theorem \ref{c33} includes $\mathrm{var}(\varepsilon^2+2{\beta^*}^\T\tilde V \varepsilon )>0$: whenever $\sigma_{\varepsilon}^2 >c_1$ and $\mathrm{corr}(\varepsilon^2,{\beta^*}^\T\tilde V \varepsilon)>-1+c_2$ for some $c_1,c_2>0$, the asymptotic variance in \eqref{7} is positive. Now we are ready to state the asymptotic normality of $R^2$ as a simple corollary of a more general result; see Theorem S2 in the Supplement.

 \begin{corollary}\label{cor}
Let Conditions 1 and 4 hold. Then, for $R^2$ defined in \eqref{R2}, we have $R^2=PVE+o_P(1)$, whenever $m,n,p\to\infty$. Moreover, if Condition 2 holds with $\|\hat\beta-\beta^*\|_2=o_P(n^{-1/4})$, then, as $m,n,p\to\infty$,
$
n^{1/2}V^{-1/2}(R^2)(R^2-\mathrm{PVE})\to N(0,1)
$
in distribution, provided $V(R^2)>0$, where
$
V(R^2)=\mathrm{var}[\sigma_Y^{-4}b^2\varepsilon^2+\sigma_Y^{-4}\sigma_\varepsilon^2\{2\varepsilon{\beta^*}^\T\tilde V+\tau({\beta^*}^\T\tilde V)^2\}]+\tau\sigma_Y^{-8}\sigma_\varepsilon^4\mathrm{var}\{({\beta^*}^\T\tilde V)^2\}.
$
\end{corollary}

\section{Beyond linear outcome models}\label{sec:nonlinear}
 
Recall that our estimation towards the mean depends on the linear projection of  $g^0(x)=E(Y\mid X=x)$. A question arises naturally: can we use general machine learning algorithms to estimate $g^0(x)$ and design nonlinear projection for optimal estimation of $\theta$?
Are we able to construct confidence intervals, and will the asymptotic variances of the estimators be improved? 
We provide positive answers to both questions.

A natural extension of $\hat \theta$ can be defined as
\begin{equation}\label{thetagen}
\hat\theta_{\mathrm{gen}}=K^{-1}\sum_{k=1}^K\hat\theta_{\mathrm{gen}}^{(k)},\;\;\text{where}\;\hat\theta_{\mathrm{gen}}^{(k)}=M^{-1}\sum_{i\in J_k}\hat g^{(-k)}(X_i)+N^{-1}\sum_{i\in I_k}\left\{Y_i - \hat g^{(-k)}(X_i)\right\}
\end{equation}
and $\hat g^{(-k)}$ is the estimate of $g^0$ computed on all but the $k$th labeled observations. 
We suppose the existence of some $g^*=g_d^*:\mathbb R^p\to\mathbb R$ such that 
$\mu_{2,X}\{\hat g^{(-k)}(x)-g^*(X)\}=o_P(1)$
as $n\to\infty$ and possibly $p,q\to\infty$, and 
where $\mu_r(f)=E\{f-E(f)\}^r$ is the $r$th central moment and $\mu_{r,X}(f)=E_X\{f-E_X(f)\}^r$, with $E_X$ denoting the conditional expectation on the marginal distribution $P_X$. 
Here, $d$ denotes the degree of freedom of the working model. 
 Note that $g^*(x)=g^0(x)$ is unnecessary. Here, $g^*=g_d^*$ can be chosen as the projection of the underlying curve $g^0(x)$ to a functional class $\mathcal G_d$, i.e., 
\begin{equation}\label{eq:gg}
g^*=\arg\min_{g\in\mathcal G_d}E\{g^*(X)-g^0(X)\}^2.
\end{equation} With a small abuse of notation, let $\varepsilon=Y-g^*(X)$ denote the unexplained error of the model. To better interpret our results, we assume that $E(\varepsilon)=0$ and $E\{\varepsilon g^*(X)\}=0$, which is satisfied once $b+ag\in\mathcal G_d$ for all $a,b\in\mathbb R$ and $g\in\mathcal G_d$. We demonstrate in Theorem \ref{gen} that $\hat\theta_{\mathrm{gen}}$ of \eqref{thetagen} is asymptotically normal with asymptotic variance 
$
V_\mathrm{gen}(\theta)=\sigma_{\varepsilon,\mathrm{gen}}^2+\tau b_\mathrm{gen}^2,
$
where $b_\mathrm{gen}^2=\mathrm{var}\{g^*(X)\}$ denotes the explained variance of the model $g$, and $\sigma_{\varepsilon,\mathrm{gen}}^2=E\{Y-g^*(X)\}^2=\mathrm{var}(Y)-b_\mathrm{gen}^2$ denotes the unexplained variance.
When $g^*$ is defined as in \eqref{eq:gg}, $b_\mathrm{gen}^2$ and $\sigma_{\varepsilon,\mathrm{gen}}^2$ are the largest explained variance and smallest unexplained variance among the functional class $\mathcal G_d$, respectively. 
 The unexplained variance can be estimated using a cross-fitting scheme,
\begin{equation}\label{genuev}
\hat\sigma_{\varepsilon,\mathrm{gen}}^2=n^{-1}\sum_{k=1}^K\sum_{i\in I_k}\left\{Y_i-\hat\theta_\mathrm{gen}-\hat h^{(-k)}(X_i)\right\}^2,
\end{equation}
with $\hat h^{(-k)}(X_i)=\hat g^{(-k)}(X_i)-M^{-1}\sum_{i\in J_k}\hat g^{(-k)}(X_i)$.
 As for the explained variance, \eqref{ev} can be generalized through a bias-corrected cross-fitting estimator
\begin{equation}\label{genev}
\hat b_\mathrm{gen}^2=(m+n)^{-1}\sum_{k=1}^K\sum_{i\in J_k}\{\hat h^{(-k)}(X_i)\}^2+2n^{-1}\sum_{k=1}^K\sum_{i\in I_k}\hat h^{(-k)}(X_i)\{Y_i-\hat\theta_\mathrm{gen}-\hat h^{(-k)}(X_i)\}.
\end{equation}
Now, 
$\hat V_\mathrm{gen}(\theta)=\hat\sigma_{\varepsilon,\mathrm{gen}}^2+n\hat b_\mathrm{gen}^2/(m+n)$, and an $\alpha$-level confidence interval can be constructed as
\begin{equation}\label{CIthetagen}
\mathrm{CI}_\mathrm{gen}(\theta)=\left[\hat\theta_\mathrm{gen}-z_{1-\alpha/2}\left\{\hat V_\mathrm{gen}(\theta)/n\right\}^{1/2},\hat\theta_\mathrm{gen}+z_{1-\alpha/2}\left\{\hat V_\mathrm{gen}(\theta)/n\right\}^{1/2}\right].
\end{equation}
Due to space constraints, we relegate the asymptotic normality of nonlinear $R^2$  to the Supplement.

\begin{theorem}\label{gen}
 Suppose that $E|Y|^{2+c}<C$ and $E|g^*(X)|^{2+c}<C$ for some $C<\infty$. Then, as long as  $\mu_{2,X}\{\hat g^{(-k)}(x)-g^*(X)\}=o_P(1)$ for each $k$, as $n,p\to\infty$ or $n,p,d\to\infty$, $\hat\theta_\mathrm{gen}$ satisfies
$
{n^{1/2} V_\mathrm{gen}^{-1/2}(\theta) (\hat\theta_\mathrm{gen}-\theta)}
\to N (0,1)$, $\hat V_\mathrm{gen}(\theta)= V_\mathrm{gen}(\theta)+o_P(1)
$
provided that $V_\mathrm{gen}(\theta)>0$.
\end{theorem}

 The asymptotic variance above depends on the explained variance $b^2_\mathrm{gen}$: the larger the explained variance is, the more efficient estimation of $\theta$ is. In particular, a worst case of $b^2_\mathrm{gen}=0$ corresponds to  the sample mean estimator.  When $g^*(x)=g^0(x)$, the asymptotic variance is optimal; it matches the oracle lower bound of Proposition 3.1 in \cite{zhang2019semi},
 and one can see a clear efficiency gain through
 $b_\mathrm{gen}^2(g^*) \leq b_\mathrm{gen}^2(g^0)$.

  \section{Heterogeneous treatment effects}\label{s4}
 
Suppose that in addition to previous settings we have access to a treatment indicator 
$D_i \in \{ 0,1\}$, $i=1,\cdots, m+n$. Following the potential outcomes framework \citep{rubin1974estimating,holland1988causal,splawa1990application} we then hypothesize the presence of potential outcomes $Y_i{(0)}$ and $Y_i{(1)}$ corresponding to, respectively, the response the $i$th subject would have experienced with and without the treatment. We then observe that the 
average treatment effect
$
\delta =  E\{E(Y\mid X,D=1)-E(Y\mid X,D=0)\}=\tau_1 - \tau_0.
$

Similarly as in \S \ref{s2}, we hypothesize the existence of the $l_2$ slopes 
$
\beta^*_w=\min_{\beta\in\mathbb R^p} E\{(Y-\tilde X^\T\beta)^2\mid D=w\},
$
defined at the population level for $w \in \{ 0,1\}$.
 A standard way of constructing  the average treatment effects estimates is to posit a model on the treatment assignment and then adjust for possible confounding. Treatments are assigned to subjects according to an
underlying scheme that depends on the subjects' features. Their dependence can be captured by $
D_i=e(X_i)+\zeta_i,
$
where $e(X_i)$ is an unknown  propensity score function \citep{rosenbaum1983central}. In the following, we assume two primitive conditions: a widely regarded  overlap condition regarding the treatment missingness and an identifiability condition. 
 
 \begin{condition}\label{cond:4}
 Let $\mathrm{pr}\{c\leq e(X )\leq1-c\}=1$ and $\mathrm{pr}\{c\leq \hat e(X )\leq1-c\}=1$ for some constant $c\in(0,1)$.
 For $\varepsilon_i=Y_i(D_i)-\{D_i\beta_1^*+(1-D_i)\beta_0^*\}^\T\tilde X_i$,
let $ E (\zeta \mid X )=0$, as well as $\mathrm{pr}\{E (\varepsilon^2\mid X)<C\}=1$ with some constant $C>0$. \end{condition}

Let $\hat\beta_1$, $\hat\beta_0$ and $\hat e$ denote estimators for $\beta_1^*$, $\beta_0^*$ and $e$, respectively,  satisfying
$
 E _{P_X} \{(\hat\beta_w^{(-k)}-\beta_w^*)^\T\tilde X \}^2 = O_P(a_{n,p}^2)$, $E _{P_X} \{\hat e^{(-k)}(X)-e(X)\}^2 = O_P(b_{m+n,p}^2)$ and 
$ E [\{E(Y\mid X)-{\ \beta_w^*}^\T\tilde X \}^2 \mid D=w ]=O_P(c_p^2).$
Here, $a_{n,p}$, $b_{m+n,p}$ and $c_p$ are non-negative sequences of numbers, with $c_p$ describing how close the linear model is to the true underlying curve.
 The semi-supervised estimator \eqref{eEY} needs to be adjusted for the confounding effects. To that end, we introduce 
\begin{equation*}
\hat\tau^{(k)}_\omega={\ \hat\mu^{(k)}}^\T\hat\beta^{(-k)}_\omega+ N^{-1}\sum_{i\in I_k} w_i^{(-k)}(\omega)\left( Y_i-\tilde X_i^\T\hat\beta^{(-k)}_\omega\right), \qquad \hat\mu^{(k)}=M^{-1}\sum_{i\in J_k}\tilde X_i.
\end{equation*}
In the above, the weights $w_i^{(-k)}(\omega)$ correspond to the ratio of the observed treatment proportion; then, the framework from \S \ref{s21} will still lead to root-$n$ consistent estimates. We denote these weights as
$w_i^{(-k)}(\omega)=\omega{D_i}/{ \hat e^{(-k)}(X_i)}+(1-\omega){(1-D_i)}/{ \{1 -\hat e^{(-k)}(X_i)\}}.$
Then, the estimate of the average treatment effect can be defined as a difference of 
$
\hat\delta^{(k)}= \hat \tau_1^{(k)} 
- \hat \tau_0^{(k)}
$
and 
$
\hat\delta = K^{-1} \sum_{k=1}^K \hat\delta^{(k)}.
$

An asymptotic $(1-\alpha)$-level confidence interval for the average treatment effect could then be defined as
\begin{equation}\label{CIdelta}
\left(\hat\delta-z_{1-\alpha/2}{\hat V_\delta^{1/2}}n^{-1/2},\ \hat\delta+z_{1-\alpha/2}{\hat V_\delta^{1/2}}n^{-1/2} \right).
\end{equation}
 The estimator of $V_\delta= V_1 +\tau V_2$, \eqref{eq:vdelta}, is defined as 
$
\hat V_\delta=K^{-1}\sum_{k=1}^K \{\hat V_1^{{(k)}}+ {n}{(m+n)^{-1}} \hat V_2^{{(k)}} \}.
$
Observe that $
V_1=\mathrm{var} \{r(Y-{\beta_1^*}^\T\tilde X)-\rho(Y-{\beta_0^*}^\T\tilde X) \}.
$ Then, a natural plug-in estimator can be defined as 
$\hat V_1^{{(k)}}= {N}^{-1} \sum_{i\in I_k}\nu_{\delta,i}^2 $, where
$
\nu_{\delta,i}= r_i^{(-k)}(Y_i-\hat\beta_1^{(-k)^\T}\tilde X_i) - \rho_i^{(-k)} (Y_i-\hat\beta_0^{(-k)^\T}\tilde X_i) -\{\hat\delta -(\hat\beta_1^{(-k)}-\hat\beta_0^{(-k)})^\T\hat\mu^{(k)}\},
$
recall $\hat\mu^{(k)}$ is defined in \eqref{eEY}.
The second component, $V_2=E \{(\beta_1^*-\beta_0^*)^\T (\tilde X-\tilde\mu)\}^2$,  is estimated as $\hat V_2^{{(k)}}= {N}^{-1}\sum_{i\in I_k}\xi_{\delta,i} ^2$
for
$
\xi_{\delta,i}=(\hat\beta_1^{(-k)}-\hat\beta_0^{(-k)} )^\T(\tilde X_i-\hat\mu).
$
The next theorem is the main result of this section.

\begin{theorem}\label{t42}
Let Conditions \ref{a2} and \ref{cond:4} hold. 
Then, under the setting of this section,
$
\hat\delta-\delta=O_P (n^{-{1/2}}+a_{n,p}b_{m+n,p}+b_{m+n,p}c_p)
$
whenever $a_{n,p}=O(1)$. Therefore, whenever $a_{n,p}b_{m+n,p}=o(1)$ and $b_{m+n,p}c_p=o(1)$, $\hat\delta$ is consistent. If, however, $a_{n,p}b_{m+n,p}=O(n^{-{1/2}})$ and $b_{m+n,p}c_p=O(n^{-{1/2}})$, $\hat \delta$ is a $n^{1/2}$-consistent estimate of $\delta$.
 Additionally, 
 the asymptotic normality follows
\begin{equation}\label{deltanormal}
n^{1/2}(\hat\delta-\delta)\rightarrow N(0,V_\delta)
\end{equation}
in distribution, whenever, $
a_{n,p}=o(1),$ $b_{m+n,p}=o(1),$ $a_{n,p}b_{m+n,p}=o(n^{-{1/2}})$ and $b_{m+n,p}c_p=o(n^{-{1/2}}),$
with an asymptotic variance
\begin{equation}\label{eq:vdelta}
V_\delta=\mathrm{var} \left(\varepsilon \zeta/[e(X)\{1-e(X)\}]\right)+\tau(\beta_1^*-\beta_0^*)^\T\tilde C(\beta_1^*-\beta_0^*),
\end{equation}
 provided that $V_\delta>c$ for some $c>0$, and $\tau=\lim_{m,n\to\infty}n/(m+n)$. Moreover, $
\hat V_\delta=V_\delta+o_P(1).
$
\end{theorem}

 Suppose the sparsity of the outcome and the treatment model are $s_Y$ and $s_D$, respectively. For illustration purposes suppose that both models are parametric and linear. Then, $c_p=0$ and the rates $a_{n,p}$ and $b_{m+n,p}$  for a Lasso estimate become $a_{n,p}=O_P[\{s_Y\log(p)/n\}^{1/2}]$, $b_{m+n,p}=O_P[\{s_D\log(p)/(m+n)\}^{1/2}]$. Therefore, $s_Y=o\{n/\log(p)\}$, $s_D=o\{(m+n)/\log(p)\}$ and $s_Ys_D=o[(m+n)/\{\log (p)\}^2]$ are required to achieve asymptotic normality. Then, when $m$ is large enough, in that $s_Dn\log p/m\rightarrow0$, we require $s_Y=o\{n/\log(p)\}$, which is extremely mild, i.e., consistency in estimation of the propensity model at any arbitrary rate. 
 If both   $D$ and $Y$ were unavailable in the unlabeled data, the estimation error on the propensity score would depend on $n$ rather than $m+n$ with the same sparsity assumptions as in \cite{chernozhukov2017double}, \cite{smucler2019unifying} and others. At the same time, we achieve a more efficient estimator, regardless of whether $D$ is available in the unlabeled data or not, i.e., reducing the size of the asymptotic variance. 
When the outcome model  is misspecified, even if $c_p=O(1)$, such that the linear model does not reach the underlying curve as $p$ grows, we can still obtain the asymptotic normality \eqref{deltanormal} provided $m$ is large enough so that $b_{m+n,p}=o(n^{-1/2})$. 
Supervised settings have more stringent conditions; see, e.g., \cite{smucler2019unifying} and \cite{tan2020regularized}. 
If one is only interested in obtaining a root-$n$ consistency, the outcome model can be completely misspecified, including completely dense high-dimensional models. They can be estimated using machine learning methods, such as random forests, Bayesian classification, regression tree, and deep neural networks; one just needs to replace the linear projection $\tilde X_i^\T\hat\beta^{(-k)}$ by any $\hat g^{(-k)}(w,X_i)$, 
\begin{align*}
\hat\delta_\mathrm{gen}=&(m+n)^{-1}\sum_{k=1}^K\sum_{i\in J_k}\left\{\hat g^{(-k)}(1,X_i)-\hat g^{(-k)}(0,X_i)\right\}\nonumber\\
&\qquad+n^{-1}\sum_{k=1}^K\sum_{i\in I_k}\left[\frac{D_i\{Y_i-\hat g^{(-k)}(1,X_i)\}}{\hat e^{(-k)}(X_i)}-\frac{(1-D_i)\{Y_i-\hat g^{(-k)}(0,X_i)\}}{1-\hat e^{(-k)}(X_i)}\right],
\end{align*}
where $\hat g^{(-k)}(w,X_i)$ is an estimate of $E(Y\mid X,D=w)$ trained on $(D_i,Y_i,X_i)_{i\in\{1,2,\dots,n\}\setminus I_k}$ for $w\in\{0,1\}$. Moreover, an asymptotic confidence interval can be extended from \eqref{CIdelta} by replacing the linear outcome model with a general nonlinear model.

 \section{Finite-sample experiments}\label{s5}

 \subsection{Numerical experiments}

 In this section we illustrate the finite-sample properties of $\hat \theta$. The estimation of the variance can be found in the Supplement. We consider semi-supervised estimators based on ordinary least squares (SSL-OLS), the 10-fold cross-validated lasso (SSL-Lasso), the additive model (SSL-Additive), XGBoost (SSL-XGBoost), multilayer perceptron (SSL-MLP), and random forest (SSL-RF) for which vanilla, preset tuning parameters are used. 
We compare with the sample mean $\bar Y=n^{-1}\sum_{i=1}^nY_i$ and with the semi-supervised least squares estimator (SSLS) proposed in \cite{zhang2019semi} whenever $p<n$. 
We consider confidence intervals \eqref{CIthetagen}, where the significance level is $\alpha=0.05$ throughout. Each set of results is based on 200 repetitions with $K=5$. The black solid line in all the plots denotes the optimal  ratio $\{\sigma_Y^2-mb^2_\mathrm{gen}(g^0)/(m+n)\}/\sigma_Y^2$. We will see that, as long as the sample size $n$ is large enough, our proposed semi-supervised estimators $\hat\theta$ is better than the sample mean $\bar Y$ in the sense of mean squared error.
 
\begin{figure*}
  \centering
  \begin{subfigure}[t]{0.485\textwidth}
    \centering
    \includegraphics[height=0.65\linewidth,width=0.9\linewidth]{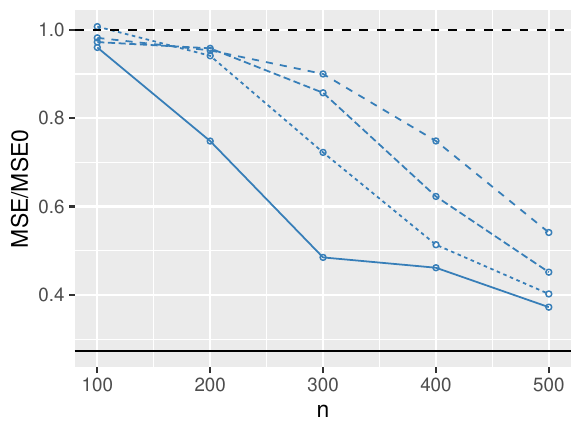}
    \caption{ The ratio of mean squared errors. }
    \label{msem11}
  \end{subfigure}
    ~ 
    \begin{subfigure}[t]{0.485\textwidth}
    \centering
    \includegraphics[height=0.65\linewidth,width=0.9\linewidth]{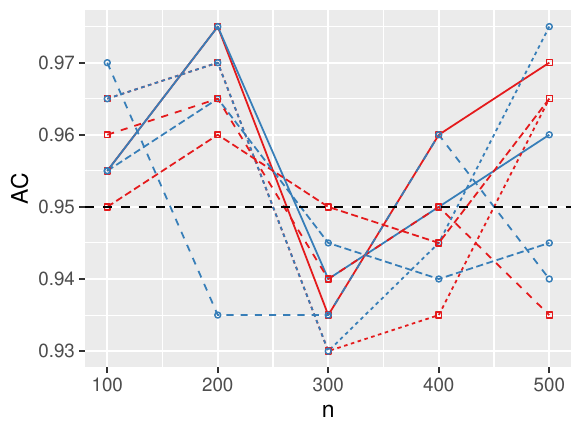}
    \caption{The average coverage of $\bar Y$ and $\hat\theta$. }\label{acm11}
  \end{subfigure}
  \caption{Model 1: Comparison of SSL-Lasso and the sample mean. The mean squared error of the sample mean is denoted as $\mathrm{MSE0}$. The plot includes sample mean (red squares)  and SSL-Lasso (blue circles) estimates.  The sparsity level of the linear coefficients, $s$, is denoted with long dashed, dashed, dotted and solid lines for $s=90$, $s=70$, $s=50$ and $s=30$, respectively.} \label{fig:1}
 \end{figure*}

{Model 1.}  
Let
$
X_i\overset{\mathrm{iid}}{\sim}N_{p-1}(0,I_{p-1})
$
with $p=500$ and $m=10n$, and
$
Y_i=s^{-1/2}\sum_{j=1}^sX_{ij}+\delta_i, \ s\in\{30,50,70,90\},
$
 $\delta_i\overset{\mathrm{iid}}{\sim}N(0,0.25)$.
The results are presented in Fig. \ref{fig:1}, where we observe that our SSL-Lasso estimator  is more efficient than the sample mean, Fig. \ref{msem11}, regardless of the level of sparsity. Fig. \ref{acm11} illustrates robustness in terms of the average coverage probability of the SSL-Lasso estimate.

\begin{figure*}
  \centering
  \begin{subfigure}[t]{0.485\textwidth}
    \centering
    \includegraphics[height=0.65\linewidth,width=0.9\linewidth]{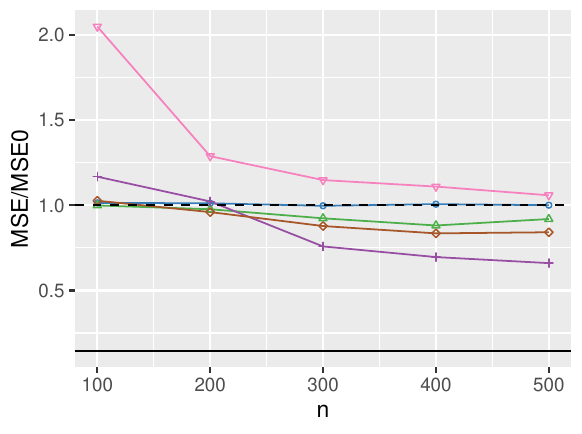}
    \caption{ The ratio of mean squared errors. }\label{msem2}
  \end{subfigure}%
  ~ 
    \begin{subfigure}[t]{0.485\textwidth}
    \centering
    \includegraphics[height=0.65\linewidth,width=0.9\linewidth]{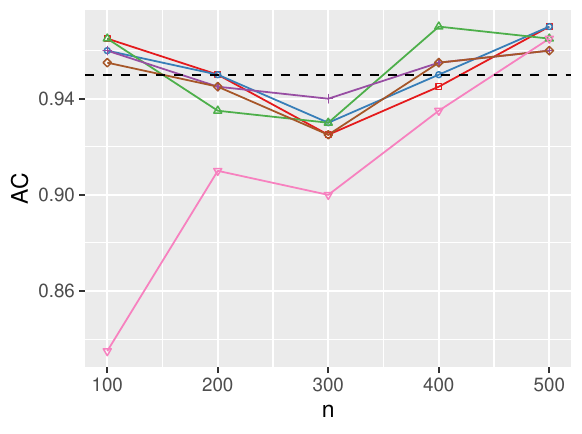}
    \caption{The average coverage.}\label{acm2}
  \end{subfigure}
\caption{Model 2: Comparison between \cite{zhang2019semi} (SSLS) and our SSL estimators. The plot includes sample mean (red squares), SSL-Lasso (blue circles), SSL-Additive (green up triangles), SSL-XGBoost (purple pluses), SSL-RF (brown diamonds) and SSLS (pink down triangles) estimates.}
\end{figure*}

{Model 2.} Let 
$X_i$ and $\delta_i$ be as in {Model 1.} 
  and consider a nonlinear model
 $Y_i=3\cos(X_{i1}+X_{i2}+X_{i3})+\delta_i$,
with $p=51$, $m=10n$. 
We compare our SSL estimator with a variety of baseline procedures and  the semi-supervised least squares estimator $\hat\theta_\mathrm{SSLS}$ of \cite{zhang2019semi}. Fig. \ref{msem2} illustrates that SSLS is less efficient than the sample-mean estimator, that our SSL-Lasso is equivalent to the sample-mean, and that all other SSL-methods are more efficient with SSL-XGBoost outperforming the rest. Fig. \ref{acm2} demonstrates extremely poor finite-sample coverage of SSLS and nominal coverage of our proposal. 

\begin{figure*}
  \centering
  \begin{subfigure}[t]{0.485\textwidth}
    \centering
    \includegraphics[height=0.65\linewidth,width=0.9\linewidth]{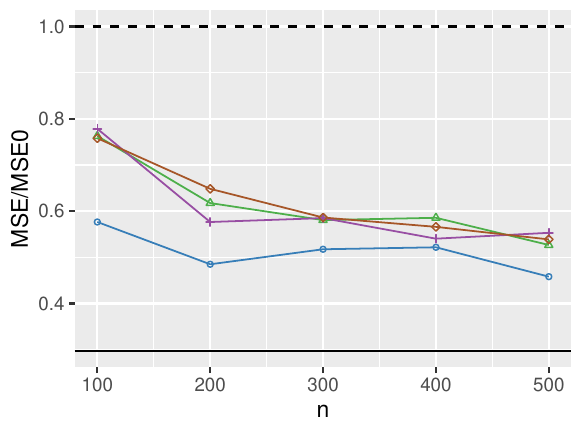}
    \caption{The ratio of mean squared errors. }\label{msem4}
  \end{subfigure}%
  ~ 
  \begin{subfigure}[t]{0.485\textwidth}
    \centering
    \includegraphics[height=0.65\linewidth,width=0.9\linewidth]{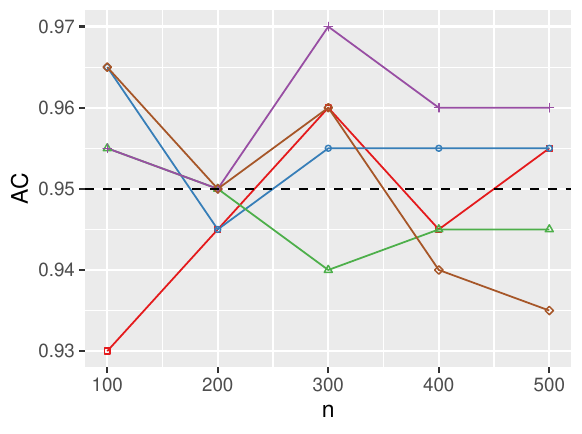}
    \caption{The average coverage. }\label{acm4}
  \end{subfigure}
\caption{Model 3: Comparison of SSL-method with the sample mean. The plot includes sample mean (red squares), SSL-Lasso (blue circles), SSL-Additive (green up triangles), SSL-XGBoost (purple pluses) and SSL-RF (brown diamonds) estimates.}
\end{figure*}

 {Model 3.} Let
$
X_i\overset{\mathrm{iid}}{\sim}N_{p-1}(0,C),
$ be equicorrelated with
  $C_{ij}=\{1-1/(2p)\}1_{\{i=j\}}+1/(2p)1_{\{i\neq j\}}$, with $p=1001$, $m=10n$. We consider a nonlinear additive outcome model
$
Y_i=\sum_{j=1}^{p-1}0.7^{j-1}\sin(X_{ij})+\delta_i,
$
where $\delta_i\overset{\mathrm{iid}}{\sim}N(0,0.25)$. Fig. \ref{msem4} demonstrates significant gain in reduction of MSE of the proposed method with the SSL-Lasso in the lead. Fig. \ref{acm4} presents strong finite sample coverage.

\begin{figure*}
  \centering
  \begin{subfigure}[t]{0.485\textwidth}
    \centering
    \includegraphics[height=0.65\linewidth,width=0.9\linewidth]{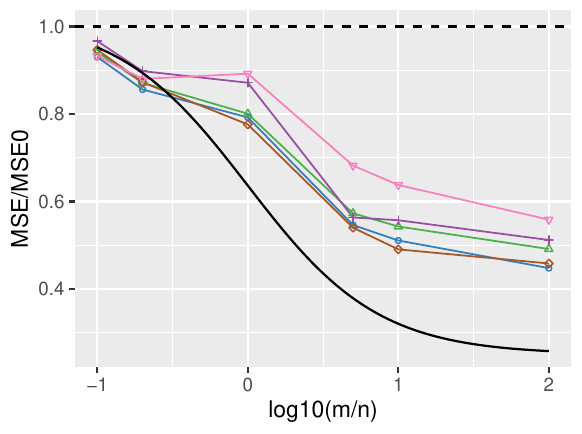}
    \caption{The ratio of mean squared errors. }\label{msem3}
  \end{subfigure}%
  ~ 
  \begin{subfigure}[t]{0.485\textwidth}
    \centering
    \includegraphics[height=0.65\linewidth,width=0.9\linewidth]{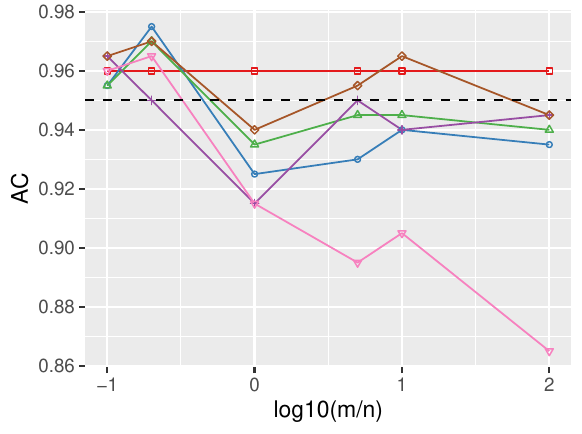}
    \caption{The average coverage.}\label{acm3}
  \end{subfigure}
  \caption{Model 4: Impact of the size of additional data. The plot shows sample mean (red squares), SSL-Lasso (blue circles), SSL-Additive (green up triangles), SSL-XGBoost (purple pluses), SSL-RF (brown diamonds) and SSLS (pink down triangles) estimates.}
\end{figure*}

{Model 4.} Here we observe behavior with varying $m$. Let $X_i$ and $\delta_i$ be as in {Model 1} and consider the nonlinear outcome of {Model 3}. Set 
 $p=201$, $n=500$ and let $m$ vary from $0.1n$ to $10n$. 
 We compare with $\bar Y$ and SSLS of \cite{zhang2019semi}. We see substantial gains in efficiency. SSL-RF dominates the other estimators, both in terms of MSE, Fig. \ref{msem3}, and coverage, Fig. \ref{acm3}. SSLS loses coverage with a larger $m$. 
 When $m$ is small, the ordinary least squares estimate's impact is not significant, and SSLS is similar to the sample mean $\bar Y$. As $m$ grows, the instability of least-squares and the unfitness of SSLS is exposed.

\begin{figure*}
  \centering
  \begin{subfigure}[t]{0.485\textwidth}
    \centering
    \includegraphics[height=0.65\linewidth,width=0.9\linewidth]{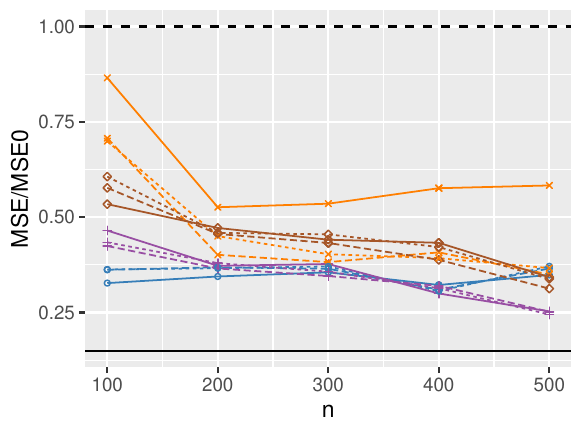}
    \caption{The ratio of mean squared errors. }\label{msem7}
  \end{subfigure}%
  ~ 
    \begin{subfigure}[t]{0.485\textwidth}
    \centering
    \includegraphics[height=0.65\linewidth,width=0.9\linewidth]{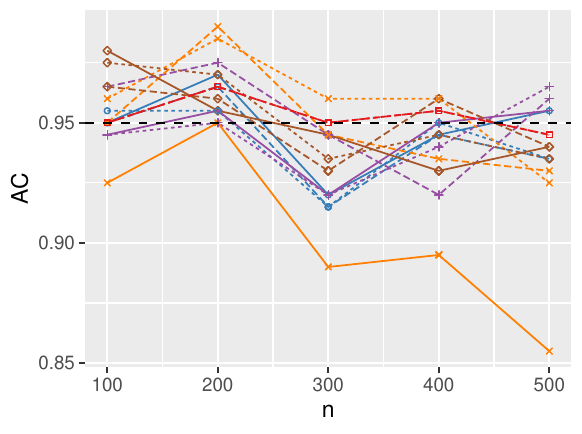}
    \caption{The average coverage. }\label{acm7}
  \end{subfigure}
\caption{Model 5: Is sample-splitting needed? The plot includes sample mean (red squares), SSL-Lasso (blue circles), SSL-XGBoost (purple pluses), SSL-MLP (orange crosses) and SSL-RF (brown diamonds) estimates. The number of folds, $K$, is denoted with solid, dashed and long dashed lines for $K=1$ (without cross-fitting), $K=5$ and $K=20$, respectively.}
\end{figure*}

{Model 5.} Is sample-splitting needed? Let $
X_i\overset{\mathrm{iid}}{\sim}\mathrm{Lognormal}_{p-1}(0,C),
$
with $C$ as in {Model 3} with $p=101$, $m=10n$.
Let $
Y_i=\sum_{j=1}^{3}\{\log(X_{ij}+1)^2+0.1\}+\delta_i,
$
where $\delta_i\overset{\mathrm{iid}}{\sim}N(0,0.25)$. We varied $K$ from $1$ to $5$ and then to $20$. 
 We observe that some methods, like SSL-MLP, benefit significantly from sample splitting: without it, they under-cover, Fig. \ref{acm7}, and have the largest MSE, Fig. \ref{msem7}.

\begin{figure*}
  \centering
  \begin{subfigure}[t]{0.485\textwidth}
    \centering
    \includegraphics[height=0.65\linewidth,width=0.9\linewidth]{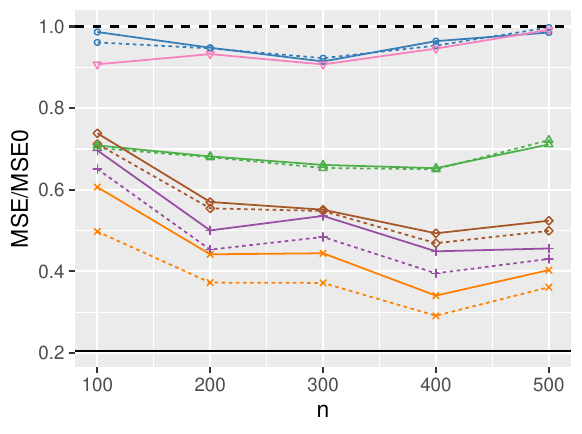}
    \caption{
    The ratio of mean squared errors. }\label{msem8}
  \end{subfigure}%
  ~ 
   \begin{subfigure}[t]{0.485\textwidth}
    \centering
    \includegraphics[height=0.65\linewidth,width=0.9\linewidth]{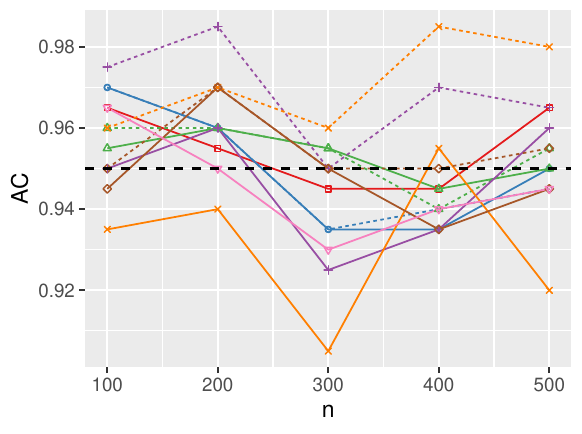}
    \caption{The average coverage.}\label{acm8}
  \end{subfigure}
\caption{Model 6: Does partitioning matter? The plot includes sample mean (red squares), SSL-OLS (blue circles), SSL-Additive (green up triangles), SSL-XGBoost (purple pluses), SSL-MLP (orange crosses), SSL-RF (brown diamonds) and SSL-SSLS (pink down triangles) estimates. The number of cross-fitting repetitions, $S$, is denoted with solid and dashed lines for $S=1$ and $S=5$, respectively.}
\end{figure*}

{Model 6.} In finite samples, the randomness from the $K$-partition creates an additional variance. We repeat the random $K$-partition for $S$ times, and for each time we obtain an estimate $\hat\theta^s$ and the corresponding estimated asymptotic variance $\hat V(\hat\theta^s)$. Here we compare $\hat\theta^1$ with the average $\tilde\theta=S^{-1}\sum_{s=1}^S\hat\theta^s$. An asymptotic confidence interval based on $\tilde\theta$ can be constructed using an estimated variance $\tilde V(\tilde\theta)=S^{-1}\sum_{s=1^S}\{\hat V(\hat\theta^s)+(\hat\theta^s-\tilde\theta)^2\}$. 
The outcome model is nonlinear with one interaction term
$
Y_i=X_{i1}X_{i2}+0.5(X_{i3}+0.5)^2+\delta_i,
$
and $X_i$ and $\delta_i$ are as in {Model 1} with $p=4$, $m=10n$. 
 Figures \ref{msem8} and  \ref{acm8} illustrate that partitions do not matter much for the least-squares procedure: SSL-Lasso,  SSL-Additive and SSL-RF do not vary much. However, highly nonlinear methods, such as SSL-MLP and SSL-XGBoost, benefit significantly from repeating the partitioning process.

{Model 7.} (Average treatment effect)  Consider
$
X_{ij}\overset{\mathrm{iid}}{\sim}\mathrm{Un}(-1,1)
$
with $p=11$, and $D_i\sim\mathrm{Bernoulli}[1/\{1+\exp(5^{1/2}\sum_{j=1}^5X_{ij}/2)\}]$. 
In the linear setting, the outcome model is 
$
Y_i=D_i(1+\beta_1^\T X_i)+(1-D_i)\beta_0^\T X_i+\delta_i,
$
where $\delta_i\overset{\mathrm{iid}}{\sim}N(0,0.2^2)$ and
$
\beta_0=-(0.5^{1/2},0.5,0.5^{3/2},0.5^2,0.5^2,0,0,0,0,0)$, $ \beta_1=- \beta_0$.
In the nonlinear setting, the outcome model is
$
Y_i=D_i\{X_{i1}X_{i2}+0.5(X_{i3}+0.5)^2\}+(1-D_i)\{X_{i1}X_{i2}-0.5(X_{i3}+0.5)^2\}+\delta_i.
$
 For the linear setting our proposed estimator and the estimator of \cite{chernozhukov2017double} estimate the propensity and the outcome model by cross-validated generalized and linear ridge regression. For the nonlinear setting, the outcome models are estimated by ridge regression, additive model and multilayer perceptron.
 Parameters $\alpha$ and $\beta$ of \cite{cheng2018efficient} are estimated by cross-validated adaptive lasso, where the initial weights are estimated by linear regression or generalized linear regression; the parameter $\gamma$ is estimated by cross-validated lasso; the kernel is chosen to be sixth-order Gaussian, and the bandwidth is estimated by the plug-in method.
Table \ref{ate4} contains all the results. 
We found that the biases of  our SSL and the supervised estimator of \cite{chernozhukov2017double} are not sensitive to the choice of the tuning parameters, while the bias of \cite{cheng2018efficient} is. Under the linear outcome models, the two SSL estimators have smaller mean squared errors than the supervised estimator; under nonlinear outcome models, our semi-supervised mlp+ridge estimator outperforms the others.

\begin{table}
\centering
\begin{threeparttable}
\caption{Experiments for the average treatment effect}\label{ate4}
\begin{tabular}{llllll}
Estimator&Bias&Emp SE&ASE&RMSE&AC\vspace{0.5em}\\
$n=100,m=200$& \multicolumn{5}{c}{ Linear Outcome}\\
Zhang $\&$ Bradic (ridge+ridge)&{0.0010}&{0.0881}&0.0812&{0.0879}&0.935\\
\cite{chernozhukov2017double} (ridge+ridge)&0.0097&0.1295&0.1238&0.1295&0.930\\
\cite{cheng2018efficient}&-0.0147&0.0885&0.0801&0.0895&0.925\vspace{0.5em}\\
$n=500,m=1000$& \multicolumn{5}{c}{ Linear Outcome}\\
Zhang $\&$ Bradic (ridge+ridge)&{{-0.0025}}&{0.0333}&0.0351&{0.0333}&0.945\\
\cite{chernozhukov2017double} (ridge+ridge)&-0.0052&0.0588&0.0546&0.0588&0.965\\
\cite{cheng2018efficient}&-0.0093&0.0329&0.0352&0.0341&0.940\vspace{0.5em}\\
$n=200,m=400$& \multicolumn{5}{c}{ Non-Linear Outcome}\\
Zhang $\&$ Bradic (ridge+ridge)&0.0031&0.0660&0.0672&0.0659&0.965\\
\cite{chernozhukov2017double} (ridge+ridge)&0.0051&0.0714&0.0737&0.0714&0.955\\
Zhang $\&$ Bradic (additive+ridge)&0.0027&0.0622&0.0638&0.0621&0.960\\
\cite{chernozhukov2017double} (additive+ridge)&0.0054&0.0705&0.0731&0.0706&0.960\\
Zhang $\&$ Bradic (mlp+ridge)&-0.0027&{0.0518}&0.0497&{0.0518}&0.935\\
\cite{chernozhukov2017double} (mlp+ridge)&{0.0015}&0.0570&0.0596&0.0569&0.960\\
\cite{cheng2018efficient}&-0.0209&0.0637&0.0655&0.0669&0.970\vspace{0.5em}\\
$n=500,m=1000$& \multicolumn{5}{c}{ Non-Linear Outcome}\\
Zhang $\&$ Bradic (ridge+ridge)&-0.0005&0.0384&0.0413&0.0383&0.970\\
\cite{chernozhukov2017double} (ridge+ridge)&-0.0014&0.0433&0.0457&0.0432&0.955\\
Zhang $\&$ Bradic (additive+ridge)&{-0.0001}&0.0385&0.0395&0.0383&0.975\\
\cite{chernozhukov2017double} (additive+ridge)&-0.0006&0.0436&0.0455&0.0435&0.960\\
Zhang $\&$ Bradic (mlp+ridge)&-0.0025&{0.0256}&0.0275&{0.0255}&0.975\\
\cite{chernozhukov2017double} (mlp+ridge)&-0.0017&0.0361&0.0354&0.0361&0.940\\
\cite{cheng2018efficient}&-0.0143&0.0377&0.0408&0.0402&0.945\vspace{0.5em}\\
\end{tabular}
\begin{tablenotes}
\footnotesize
\item Bias, average of the estimation biases; Emp SE, empirical standard error; ASE, average of estimated standard errors; RMSE, root-mean-square error; AC, average coverage of the $95\%$ confidence intervals.
\end{tablenotes}
\end{threeparttable}
\end{table}

 \subsection{HIV drug resistance}
 
\begin{figure*}
  \centering
  \begin{subfigure}[t]{0.495\textwidth}
    \centering
    \includegraphics[height= 0.65\linewidth,width=0.95\linewidth]{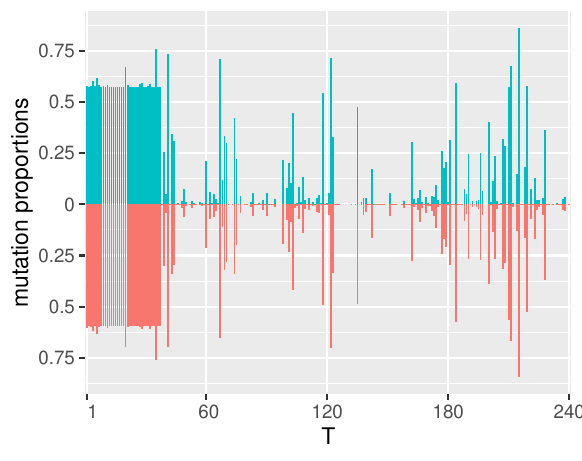}
    \caption{}\label{proportions}
  \end{subfigure}%
  ~ 
  \begin{subfigure}[t]{0.495\textwidth}
    \centering
    \includegraphics[height=0.65\linewidth,width=0.95\linewidth]{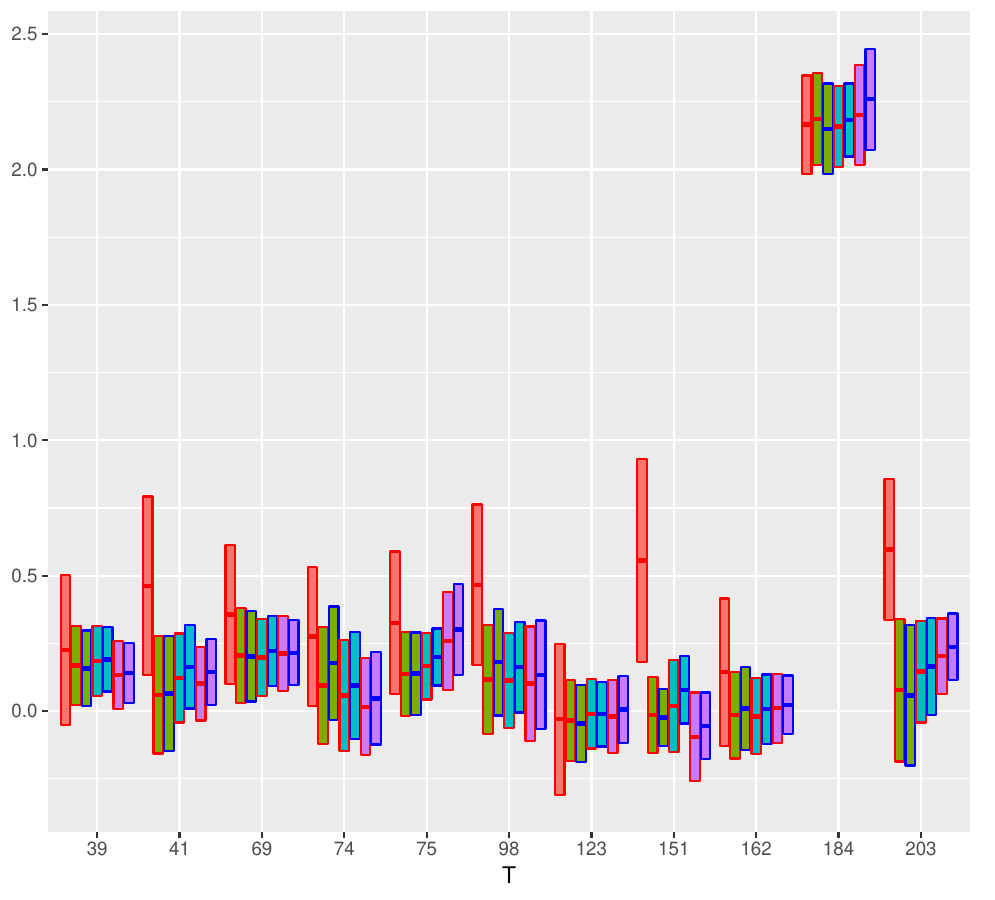}
    \caption{}\label{CIdeltaplot}
  \end{subfigure}
  \caption{Real data. (a) A back-to-back bar chart comparing the labeled and unlabeled group's mutation proportions on reverse transcriptase positions between 1 and 240. The blue color on the top denotes the unlabeled group, and the red color on the bottom denotes the labeled group.
  (b) Confidence intervals of the average treatment effect. We compare the sample mean of the labeled samples (red border and red fill), supervised \cite{chernozhukov2017double}  estimators  (red border), and our SSL-method estimators (blue border). Estimators of the propensity score and the outcome model are: logistic + Lasso (green fill), XGBoost + XGBoost (aqua fill), RF + RF (purple fill).}
\end{figure*}

We consider the dataset of \cite{baxter2006genotypic},  available at the Stanford University HIV Drug Resistance Database \citep{rhee2003human}, 
 \texttt{https://hivdb.stanford.edu}. It is known that mutations are common in HIV, and some of the mutations may affect HIV drug resistance. We provide estimation and inference for the average treatment effect of a specific mutation on the reverse transcriptase to the drug resistance. The outcome is lamivudine (3TC), a nucleoside reverse transcriptase inhibitor (NRTI), drug resistance. The treatment, $D$, denotes the existence of a mutation on the $T$th position of the HIV's reverse transcriptase. Explanatory variables $X_j$, where $j\in\{1,2,\ldots,240\}\setminus\{T\}$, denote existence of a mutation on the $j$th position. We  consider the subtype B sequence. Redundant viruses obtained from the same individuals were excluded.
We obtained $n=423$ pairs of supervised data $(D_{i,T},Y_i,\{X_{i,j}\}_{j\neq T})_{i=1}^n$ and $m=2458$ pairs of additional unlabeled covariates $(D_{i,T},\{X_{i,j}\}_{j\neq T})_{i=n+1}^{m+n}$. Fix $T\in\{1,2,\ldots,240\}$. Before we perform our semi-supervised methods, we first check whether there is a significant difference between the distribution of $X$ in the two groups; see the back-to-back bar chart of the labeled and unlabeled group's mutation proportions on different reverse transcriptase positions in Fig. \ref{proportions}.
The p-value based on Pearson Statistic was obtained using a permutation distribution \citep{agresti2005multivariate} and resulted in a value of $0.178$. We do not have any significant evidence that the covariates' distributions differ between the supervised and unlabeled groups.
 Estimators of the propensity score  and the outcome model are: (logistic) lasso  + lasso, XGBoost + XGBoost, and random forest + random forest.  In order to improve the stability of the estimator, we trim each $\hat e^{(-k)}(X_i)$ to $(0.01, 0.99)$. We  compare with the sample
 estimator 
 $(\sum_{i=1}^nD_i)^{-1}\sum_{i=1}^nD_iY_i-\{\sum_{i=1}^n(1-D_i)\}^{-1}\sum_{i=1}^n(1-D_i)Y_i$, suitable only for homogeneous effects.
Fig. \ref{CIdeltaplot} shows the confidence intervals for $\delta$ on several positions based on different estimators. We can see that there is a large average treatment effect on position 184, a small average treatment effect on positions 39, 69, and potentially a small average treatment effect on positions 41, 75 and 203. The sample estimator is most different from the rest on positions 41, 98, 151 and 203. The sample estimator is biased when the distribution of $X$ on treated and control is different. It implies that the mutations on positions 41, 98, 151 and 203 are significantly dependent on the other positions' mutations. Moreover, our confidence intervals are shorter than those of \cite{chernozhukov2017double}. It coincides with the fact that additional unlabeled data provide improved asymptotic efficiency.

\section*{Acknowledgement}

Zhang was previously at the Department of Mathematics, University of California San Diego during the initial preparation of this work. Bradic is also affiliated with the Halıcıo\v{g}lu Data Science Institute at the University of California San Diego.

\section*{Supplementary material}

Supplementary Material includes detailed proofs of all theoretical results and an additional section on the missing-at-random extension of our proposal.

\renewcommand{\thetheorem}{S.\arabic{theorem}}
\renewcommand{\thelemma}{S.\arabic{lemma}}
\renewcommand{\theequation}{\thesection.\arabic{equation}}

\appendix

\clearpage\newpage
\par\bigskip 
\begin{center}
\textbf{\uppercase{Supplementary Material to ``High-dimensional inference for dynamic treatment effects''}}
\end{center}

\par\medskip

This document contains additional sections and proofs of the main theoretical results. All results and notation are numbered and used, as in the main text unless stated otherwise. 

\section{Notation}

A  constant $c>0$, independent of $n,p,m$ may change value  from one line to  the other.  For any vector $a\in\mathbb R^p$ and $r>0$, $\|a\|_r=(\sum_{j=1}^pa_j^r)^{1/r},\ \|a\|_0=\#\{j\leq p:a_j\neq0\}$. For any matrix $A\in\mathbb R^{p\times p}$, we denote $\|A\|_2=\sup_{z\neq0}\|Az\|_2/\|z\|_2$.   
We define $\mu_r(f)=E\{f-E(f)\}^r$ being the $r$th central moment, and $\mu_{r,X}(f)=E_X\{f-E_X(f)\}^r$. Recall that $E_X(f)=\int fdP_X$ is the conditional expectation on the marginal distribution $P_X$. 
 With a slight abuse of notation, for any function $g$, 
\[
 E _{I_k^c}(g)= E \{g\mid (Y_i,X_i)_{i\in\{1,2,\dots,n\}\setminus I_k}\}
\]
and $(Y,X)\sim P_{Y,X}$ independent of $(Y_i,X_i)_{i\in\{1,2,\dots,n\}\setminus I_k}$ in the proof of Theorems \ref{c22}, \ref{t23}, \ref{t31}, \ref{c33}, \ref{gen}, \ref{genv}; 
\[
 E _{I_k^c}(g)= E \{g\mid (D_i,Y_i,X_i)_{i\in\{1,2,\dots,n\}\setminus I_k}\}
\]
and $(D,Y,X)\sim P_{D,Y,X}$ independent of $(D_i,Y_i,X_i)_{i\in\{1,2,\dots,n\}\setminus I_k}$ in the proof of Theorem \ref{t42}.
\[
 E _{J_k^c}(g)= E \{g\mid (T_i,Y_i,X_i)_{i\in\{1,2,\dots,m+n\}\setminus J_k}\}
\]
and $(T,Y,X)\sim P_{T,Y,X}$ independent of $(T_i,Y_i,X_i)_{i\in\{1,2,\dots,m+n\}\setminus J_k}$ in the proof of Theorem \ref{martheta}.
 A table of important notations is provided in Table \ref{notations}, the order of the table roughly follows the appearances of the notations in the mail file.

\begin{table}[h!]
\begin{center}
\caption{Table of notations}\label{notations}
\begin{tabular}{|l|l|}
\hline
Notation&Description\\
\hline
$Y_i,Y$&response variable or variable of interest\\
$X_i,X$&covariates or explanatory variables\\
$n$&number of labeled samples\\
$m$&number of unlabeled samples\\
$p$&dimension of the covariates containing the intercept\\
$\tau$&the limit of the proportion of the observed samples $\lim_{m,n\to\infty}n/(m+n)$\\
$\theta$&expectation of the response variable $\theta=E(Y)$\\
$\sigma_Y^2$&variance of the response variable $\sigma_Y^2=\mathrm{var}(Y)$\\
$\beta^*$&population slope\\
$\varepsilon_i,\varepsilon$&unexplained error of the response\\
$b^2$&explained variance\\
$\sigma_\varepsilon^2$&unexplained variance\\
$\{I_k\}_{k=1}^K$&a partition of $\{1,2,\dots,n\}$\\
$\{J_k\}_{k=1}^K$&a partition of $\{1,2,\dots,m+n\}$\\
$\hat\beta^{(-k)}$&an estimate of $\beta^*$ computed on all but the $k$th labeled observations\\
$\hat\mu^{(k)}$&empirical mean of the covariates in $J_k$\\
$\hat\theta$&proposed estimator of $\theta$\\
$\hat\sigma_\varepsilon^2$&proposed estimator of $\sigma_\varepsilon^2$\\
$\hat b^2$&proposed estimator of $b^2$\\
$\hat\sigma_Y^2$&proposed estimator of $\sigma_Y^2$\\
$\bar Y$&empirical sample mean of $Y$\\
$S_Y^2$&empirical sample variance of $Y$\\
$\alpha$&significance level\\
$z_{1-\alpha/2}$&the $(1-\alpha/2)$-quantile of a standard normal distribution\\
$\mu$&expectation of the covariates $\mu=E(X)$\\
$C$&covariance matrix of the covariates $C=\mathrm{cov}(X)$\\
$Z$&standardized covariates $Z=C^{-1/2}(X-\mu)$\\
$D_i,D$&treatment indicator\\
$\delta$&average treatment effect\\
$\hat\delta$&proposed estimator of $\delta$\\
$e(x)$&propensity score $e(x)=E(D\mid X=x)$\\
$\hat e^{(-k)}(x)$&an estimate of $E(D\mid X=x)$ computed on all but the $k$th observations\\
$\beta_w^*$&population slope of treatment group $D=w$\\
$\zeta_i,\zeta$&unexplained error of the propensity model\\
$g^0(x)$&underlying curve $g^0(x)=E(Y\mid X=x)$\\
$g^*(x)$&projection of $g^0$ to a functional class $\mathcal G_d$\\
$\hat g^{(-k)}(x)$&an estimate of $g^0(x)$ computed on all but the $k$th labeled observations\\
$T_i,T$&labeling indicator\\
$s(x)$&propensity score of labeling in the missing-at-random setting $s(x)=E(T\mid X=x)$\\
$\hat s^{(-k)}(x)$&an estimate of $E(T\mid X=x)$ computed on all but the $k$th observations\\
\hline
\end{tabular}
\end{center}
\end{table}

\section{Data missing-at-random}

Now, we turn to missing-at-random setting, where whether we observe $Y_i$ depends on $X_i$.  Suppose that we have $m+n$ independent and identically distributed samples $(T_i,Y_i,X_i)\sim P$, whose marginal distributions are $(P_T,P_Y,P_X)$. Here, $T_i\in\{0,1\}$ denotes the labeling: $Y_i$ is observable if and only if $T_i=1$. Assume the missing-at-random condition: $Y_i \perp T_i\mid  X_i$.
Let $Y_i^o=T_iY_i$. Let  $n=\sum_{i=1}^{m+n}T_i$ be the amount of labeled samples. Here, $n$ is a random variable.
Estimating the mean of $Y$ is equivalent to the estimation of the average treatment effect of the treated. 
 Let $\hat g^{(-k)}(x)$ and $\hat s^{(-k)}(x)$ be estimates of $g^0(x)=E(Y\mid X=x)$ and $s^0(x)=E(T\mid X=x)$ computed on all but the observations in $k$th fold, respectively. Then,
\begin{equation}\label{defmar}
\hat\theta_{\mbox{\tiny MAR}}=(m+n)^{-1}\sum_{k=1}^K\sum_{i\in J_k}\left[\hat g^{(-k)}(X_i)+\frac{T_i\{Y_i^o-\hat g^{(-k)}(X_i)\}}{\hat s^{(-k)}(X_i)}\right]
\end{equation}
is an estimate of the mean $\theta_{\mbox{\tiny MAR}}=E(Y)$ under the missing-at-random setting. 
Here, the mean estimator \eqref{defmar} is a special case of the double/debiased machine learning estimator of \cite{chernozhukov2018double}, where they require a positive overlap assumption $\mathrm{pr}\{s^0(X)>c\}=1$ for some constant $c>0$.
 In our  semi-supervised  setting, we do allow that $\tau=\lim_{m,n\to\infty}n/(m+n)=0$, i.e. $s^0(X)=E(T)\to0$. Hence, it is natural to ask  if we  can relax the positive overlap  to a more general condition on $s^0(x)$, rather than forcing $s^0(x)$ being a constant as in semi-supervised learning? In Theorem \ref{martheta} bellow, we showcase that only $\mathrm{pr}\{s^0(X)>c_1E(T)\}=1$ is needed, and that $E(T)\to 0$ is allowed.

\begin{theorem}\label{martheta}
Suppose that $E|Y|^{2+c}<c_1$, $E(\varepsilon^2\mid X)<c_1$ and $E(\varepsilon^2)>c_2$ for some $c,c_1,c_2>0$. Suppose the expected number of labeled samples grows to infinity, i.e. $(m+n)E(T)\to\infty$. Besides, for a given covariate, the ratio of the probability of observing the corresponding response to the overall labeling probability is bounded away from zero, i.e.
$
\mathrm{pr}\{s^0(X)>c_1E(T)\}=1,
$
for some $c_1>0$. Suppose $K<\infty$ and the estimators of the outcome and the propensity score model, $\hat g$ and $\hat s$, have estimation errors satisfying $E_X\{\hat g^{(-k)}(X)-g^0(X)\}^2=o_P(1),\  E_X\{1-s^0(X)/\hat s^{(-k)}(X)\}^2=o_P(1),$ and
\begin{align*}
&E_X\{\hat g^{(-k)}(X)-g^0(X)\}^2\cdot E_X\{1-s^0(X)/\hat s^{(-k)}(X)\}^2=o_P[(m+n)^{-1}\{E(T)\}^{-1}]
\end{align*}
as $m+n,p\to\infty$.  Then, the estimator $\hat\theta_{\mbox{\tiny \rm MAR}}$, \eqref{defmar}, is asymptotically normally distributed
\begin{equation}\label{asymar}
(m+n)^{1/2}\left[E\{g^0(X)\}^2+E\left\{T\varepsilon/s^0(X)\right\}^2\right]^{-1/2}(\hat\theta_{\mbox{\tiny \rm MAR}}-\theta)
{}\to N(0,1).
\end{equation}
Moreover, if
$
E_X[\{\hat g(X)-g^0(X)\}^2\{1-s^0(X)/\hat s^{(-k)}(X)\}^2]=o_P(1),
$
then,  
\begin{equation*}
\hat V_{\mbox{\tiny \rm MAR}}=(m+n)^{-1}\sum_{k=1}^K\sum_{i\in J_k}\left[\hat g^{(-k)}(X_i)+\frac{T_i\{Y_i^o-\hat g^{(-k)}(X_i)\}}{\hat s^{(-k)}(X_i)}-\hat\theta_\mathrm{MAR}\right]^2
\end{equation*}
is consistent, in that
as $m+n,p\to\infty$,
$
\hat V_{\mbox{\tiny \rm MAR}}(\theta)=[E\{g^0(X)\}^2+E\left\{T\varepsilon/s^0(X)\right\}^2]\{1+o_P(1)\}.
$ 
\end{theorem}
Hence, an asymptotic $(1-\alpha)$-level confidence interval for the mean $\theta_{\mbox{\tiny MAR}}$ could be defined as:
\begin{equation*}
\left(\hat\theta_{\mbox{\tiny MAR}}-z_{1-\alpha/2}\hat V_{\mbox{\tiny MAR}}(\theta)(m+n)^{-1/2},\ \hat\theta_{\mbox{\tiny MAR}}+z_{1-\alpha/2}\hat V_{\mbox{\tiny MAR}}(\theta)(m+n)^{-1/2}\right).
\end{equation*}
Observe that 
under the assumptions of Theorem \ref{martheta},
\begin{equation*}
(m+n)^{-1}[E\{g^0(X)\}^2+E\{T\varepsilon/s^0(X)\}^2]=O[\{(m+n)E(T)\}^{-1}],
\end{equation*}
which is of the same order as $n^{-1}$, since $n\sim\mathrm{Binomial}\{m+n,E(T)\}$. That is, the mean estimate $\hat\theta_{\mbox{\tiny MAR}}$ is $n^{1/2}$-consistent. Hence, the accuracy   depends on the number of labeled samples rather than the total number of samples. The consistency rate under the missing-at-random setting coincides with \cite{chakrabortty2018efficient}, see Section 2 of their Supplementary Material. 
Unlike assuming $s^0(X)$ to be known,   we consider the consistency rates of $E_X\{1-s^0(X)/\hat s^{(-k)}(X)\}^2$. 
Whenever, $E(T)\to0$, the rate of $1-s^0(X)/\hat s^{(-k)}(X)$ depends on $(m+n)E(T)$, rather than $m+n$ alone.
Hence the estimation error of $\hat s$ cannot be simply ignored for a large $m$. An illustrative  example is that of the case  of $T$ is independent of $X$ with the empirical mean $\bar T=(m+n)\sum_{i=1}^{m+n}T_i$ . One can easily check that, for $T_i\sim\mathrm{Bernoulli}\{E(T)\}$, we have $1-E(T)/\bar T=O_P[\{(m+n)E(T)\}^{-1/2}]$.

\section{ Further discussion on the variance}
\subsection{Asymptotic inference for the variance}

 When we are interested in estimating and perhaps constructing confidence intervals regarding the variance of $Y$, we require the same set of simple assumptions used in obtaining inferential statements regarding the mean of $Y$. Even when $\hat \beta$ is a biased estimate whose bias is bounded asymptotically (but is not diminishing) we are able to guarantee $n^{1/2}$ consistency of the estimate, \eqref{eq:var}. For consistent $\hat \beta$, even without specified rate assumptions, we can guarantee more, in distribution,
\begin{equation*}
n^{1/2}(\hat\sigma_Y^2-\sigma_Y^2)\rightarrow N \left \{0,\mathrm{var}\left(\varepsilon ^2+2{\beta^*}^\T\tilde V \varepsilon \right)+\tau\mathrm{var}\left({\beta^*}^\T\tilde V \right)^2\right\}.
\end{equation*}
The result above remains correct even when there is a large dependence of $\varepsilon_i$ on $\tilde V_i$. The result simplifies a lot if both the covariates $X_i$ and the errors $\varepsilon_i$ have Gaussian distribution; in that case, the asymptotic variance becomes $2 \sigma_\varepsilon^4 +4\sigma_\varepsilon^2b^2 + 2 \tau b^4$.
Moreover, under the same set of assumptions, we can consistently estimate the asymptotic variance of $\hat \sigma_Y^2$. To do so, we estimate the two components of the asymptotic variance separately.
Let's focus on estimating $\mbox{var} ({\beta^*}^\T\tilde V )^2$ first.
 To that end, we construct consistent estimates of $ ({\beta^*}^\T\tilde V_i)^2 - E ({\beta^*}^\T\tilde V)^2$, $\xi_i^{(k)}$, as follows
 \begin{equation}\label{xiv}
\xi_i ^{(k)}= { \ { \hat\beta^{(-k)}
 } }^\T\left( \hat V_i \hat V_i^\T - \hat C^{(k)}  \right)\hat\beta^{(-k)}.
\end{equation}
 Then we set  
 \begin{equation}\label{eq:sigmaxi}
\hat\sigma_\xi^{2}=N^{-1} \sum_{k=1}^K \sum_{i \in I_k} { \ \xi_i ^{(k)}} ^2 \approx \mbox{var} ({\beta^*}^\T\tilde V )^2.
\end{equation}
 Next, we estimate $\mathrm{var}(\varepsilon ^2+2{\beta^*}^\T\tilde V \varepsilon)$.
 To that end we define 
\begin{equation*}
 \eta_i^{(k)}
 = \hat \varepsilon_i^2 + 2 \hat\beta^{(-k)^\T} \hat V_i\hat\varepsilon_i + { \ { \hat\beta^{(-k)}
 } }^\T \hat C^{(k)} \hat\beta^{(-k)},
\end{equation*}
and observe that $\hat \sigma_Y^2$ is an average of $ \eta_i^{(k)}$. Then, 
we create a cross-fitted residuals of the following form 
\begin{equation}\label{nuv}
 \nu_i ^{(k)} =\eta_i^{(k)} - \hat\sigma_Y^{2} 
\end{equation}
 and show 
 \begin{equation}\label{eq:sigmanu}
\hat\sigma_\nu^{2}=N^{-1} \sum_{k=1}^K \sum_{i \in I_k} { \ \nu_i ^{(k)}} ^2 \approx \mathrm{var}(\varepsilon ^2+2{\beta^*}^\T\tilde V \varepsilon).
\end{equation}

In Theorem \ref{c33}, we showcase that as long as any consistent estimate of $\hat \beta$ is used, the confidence interval
\begin{equation}\label{ci4}
\mathrm{CI}(\sigma_Y^2)= \left (\hat\sigma_Y^2-z_{1-\alpha/2}\{\hat\sigma_\nu^2/n+\hat\sigma_\xi^2/(m+n)\}^{1/2}, \ \hat\sigma_Y^2+z_{1-\alpha/2}\{\hat\sigma_\nu^2/n+\hat\sigma_\xi^2/(m+n)\}^{1/2}\ \right)
\end{equation}
will be asymptotically correct.

\subsection{Variance estimation discussion}

Based on Theorem  \ref{c33},  when the   data  follows Gaussian distribution,    it is not difficult to see that  when $\tau\leq1$, i.e., $m \geq n$
\begin{equation*}
\mathrm{var}(\varepsilon^2+2{\beta^*}^\T\tilde V \varepsilon )+\tau\mathrm{var}({\beta^*}^\T\tilde V )^2 \leq  \mathrm{var}(Y-\theta)^2= \mathrm{var}(n^{1/2}S_Y^2) +o(1).
\end{equation*}
 Namely, the constructed confidence interval for $\sigma_Y^2$ as presented in \eqref{ci4} is asymptotically more accurate in the sense of having smaller width asymptotically, than the   interval that is solely based on $\{ Y_i\}_{i=1}^n$
 \begin{equation*}
\left(  {S _Y^2} \biggl/{ \left\{1-z_{\alpha/2} { (\hat\gamma-1)^{1/2}}/n^{1/2} \right\}}, {S _Y^2} \biggl/{ \left\{1+z_{\alpha/2} { (\hat\gamma-1)^{1/2}}/n^{1/2} \right\}} \right),
\end{equation*}
or its robust alternatives (see for example \cite{hummel2005better})
where $\hat\gamma$ is any consistent estimator for the kurtosis. One of the choices for $\hat\gamma$ can be
$$
\hat\gamma=\frac{n(n+1)}{(n-1)(n-2)(n-3)}\sum_i\frac{(Y_i-\bar Y)^4}{S_Y^4}-\frac{3(n-1)^2}{(n-2)(n-3)}+3.
$$

 This, in turn, implies that the proposed semi-supervised estimator, $\hat \sigma_Y^2$ is asymptotically more accurate than the sample variance, $S_ Y^2$.

In general, the efficiency of the proposed semi-supervised estimator $\hat \sigma^2_Y$ would depend on the particular model of nonlinearity, i.e., on the particular deviations from the linear model. 
We illustrate the discussion with two specific examples. To that end, we introduce a proportionality coefficient $r$ as the proportion of the decrease achieved by the semi-supervised estimator compared to $S_Y^2$. We define such coefficient with
\begin{equation}\label{dr}
r=\frac{\mathrm{var}(Y -\theta)^2-\mathrm{var}(\varepsilon ^2+2{\beta^*}^\T\tilde V\varepsilon)}{\mathrm{var}(Y -\theta)^2}.
\end{equation}
 Here, we have assumed that $m \gg n$ and the effect of $\tau$ is negligible.

The first example discusses a heteroscedastic linear model where the variance of the error depends quadratically on the covariates. The second discusses larger deviations from normality, where the response model is highly nonlinear. In particular, we consider 
\begin{align*}
Y_ i&= \sum_{j=1}^p X_{ij} + \left(a \sum_{j=1}^p X_{ij} ^2 + \sum_{j=1}^p X_{ij} \right) \eta_i \ & \mbox{(Example 1)} \\
  Y_i&=a \biggl| \log\Bigl( 0.8 \ \bigl|\sum_{j=1}^p X_{ij} \bigl|+0.01\Bigl)\biggl|+\sum_{j=1}^p X_{ij} +\eta_i \ \ & \mbox{(Example 2)} \ 
\end{align*}
where $a$ measures the size of the deviation from the linear model. In the above $X_i, \eta_i \sim N (0,1)$. When $r >0$ we see that the proposed estimator is more efficient than $S_Y^2$.

\begin{figure}[h]
    \centering
    \begin{subfigure}[b]{0.455\textwidth}
   \includegraphics[height=0.75\linewidth,width=\linewidth]{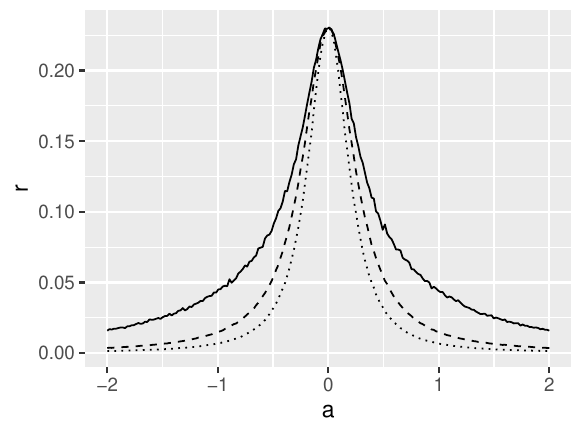}
        \caption{The proportion of decrease versus the size of the heteroscedasticity of the linear model}
        \label{fig:gull}
    \end{subfigure}
    \begin{subfigure}[b]{0.455\textwidth}
\includegraphics[height=0.75\linewidth,width=\linewidth]{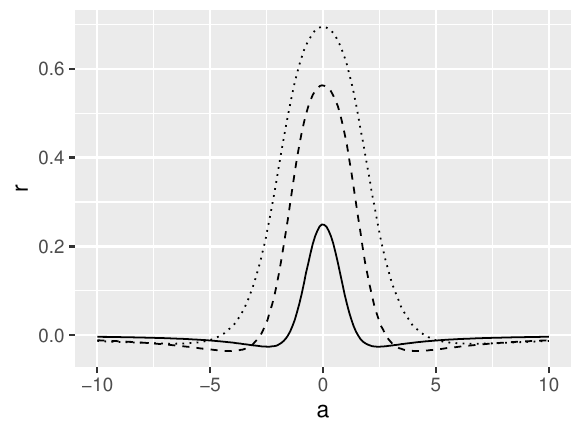}
        \caption{The proportion of decrease versus the size of the nonlinearity of Example 2}
        \label{fig:tiger}
    \end{subfigure}
    \begin{subfigure}[b]{0.455\textwidth}
\includegraphics[height=0.75\linewidth,width=\linewidth]{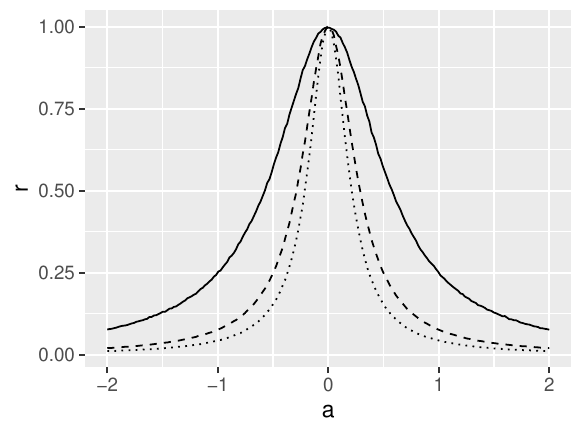}
        \caption{Signal to noise ratio versus the size of the heteroscedasticity of the  linear model}
        \label{fig:mouse}
    \end{subfigure}
        ~
    \begin{subfigure}[b]{0.455\textwidth}
\includegraphics[height=0.75\linewidth,width=\linewidth]{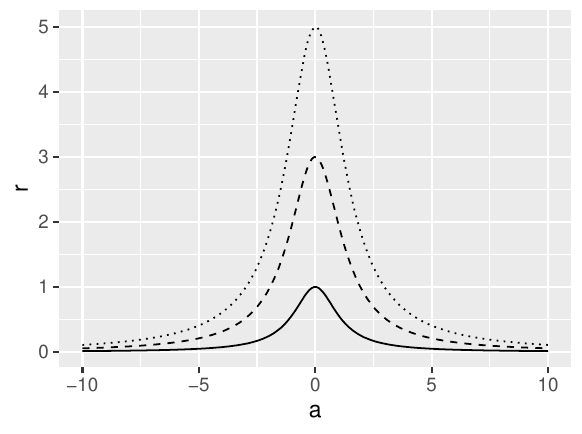}
        \caption{Signal to noise ratio versus the size of the nonlinearity of Example 2}
        \label{fig:mouse'}
        \end{subfigure}
    \caption{Proportion of the decrease in asymptotic variance achieved by the proposed semi-supervised estimator $\hat \sigma_Y^2$ as a function of the  coefficient $a$ representing the heteroscedasticity in (a) and (c) and nonlinearity in  (b) and (d)  subfigures. Different colors correspond to different dimensionality settings, in (a) and (c), we have $p=1$ (solid), $p=10$ (dashed) and $p=20$ (dotted); in (b) and (d), we have $p=1$ (solid), $p=3$ (dashed) and $p=5$ (dotted).}\label{fig:1'}
\end{figure}

From Example 1, we observe that efficiency persists over a broad range of heteroscedastic model specifications. We also observe that the larger the magnitude of $a$ is -- the more significant the effect of heteroscedasticity is --  the smaller the signal is in the linear model. This, in turn, results in smaller $r$ values. Example 2 is showing a more complex scenario; smaller magnitudes of $a$ indicate not too great deviations from the linear model and result in greater efficiency in $\hat \sigma_Y^2$. For larger $a$, the linear approximation is too far from the data generating process. 
Results are presented in Fig. \ref{fig:1'}, where we also showcase the Signal to Noise ratio corresponding to each setting.

\subsection{Inference of the variance by a general machine learning model}
In addition to the confidence interval of mean $E(Y)$, one may also be interested in inference towards the variance $\sigma_Y^2=\mathrm{var}(Y)$, the explained variance $b_\mathrm{gen}^2$ and the unexplained variance $\sigma_{\varepsilon,\mathrm{gen}}^2$.

The general semi-supervised estimators towards $\sigma_{\varepsilon,\mathrm{gen}}^2$ and $b_\mathrm{gen}^2$ are proposed in \eqref{genuev} and \eqref{genev} when we construct asymptotic confidence intervals of the mean. As for the variance of $Y$, recall that in our setting, $\sigma_Y^2=\sigma_{\varepsilon,\mathrm{gen}}^2+b_\mathrm{gen}^2$. Hence, the variance can be estimated by the sum of estimated explained variance and unexplained variance
\begin{equation}\label{sig2Ygen}
\hat\sigma_{Y,\mathrm{gen}}^2=\hat\sigma_{\varepsilon,\mathrm{gen}}^2+\hat b_\mathrm{gen}^2.
\end{equation}
The estimation of PVE can also be handled by the usage of a general machine learning model. An extension of $R^2$ can be defined as
\[
R_\mathrm{gen}^2=K^{-1}\sum_{k=1}^K\hat b_\mathrm{gen}^{2^{(k)}}/\hat\sigma_{Y,\mathrm{gen}}^{2^{(k)}}.
\]
Because of the limited length of the paper, here we only propose the asymptotic normality results of the generalized estimators. 

\begin{theorem}\label{genv}
Suppose that we have $n$ independent and identically distributed samples $(Y_i, X_i) \sim P$ whose marginal distributions are $(P_Y, P_X)$. In addition, suppose that we observe   a supplementary set of $m$ independent and identically distributed samples $X_i$ that are drawn from the same distribution $P_X$. Moreover, suppose that $E|Y|^{4+c}<C$, $E|g^*(X)|^{4+c}<C$ and the estimation error satisfies $\mu_{4,X}\{\hat g^{(-k)}(X)-g^*(X)\}=o_P(1)$, then
\begin{equation*}
\frac{n^{1/2}(\hat\sigma_{Y,\mathrm{gen}}^2-\sigma_Y^2)}{\{V(\sigma_Y^2)\}^{1/2}} \to  N (0,1),
\end{equation*}
in distribution, where
\begin{equation*}
V(\sigma_Y^2)=\mathrm{var}\left\{\varepsilon^2+2\varepsilon(g^*(X)-\theta)+\frac{n}{m+n}(g^*(X)-\theta)^2\right\}+\frac{mn}{(m+n)^2}\mathrm{var}\left\{(g^*(X)-\theta)^2\right\}.
\end{equation*}
The asymptotic variance $V(\sigma_Y^2)$ can be estimated by 
\begin{equation*}
\hat V(\sigma_Y^2)=n^{-1}\sum_{k=1}^K\sum_{i\in I_k}\left(\nu_i-N^{-1}\sum_{i\in I_k}\nu_i\right)^2+\frac{m}{(m+n)^2}\sum_{k=1}^K\sum_{i\in I_k}\left(\xi_i-N^{-1}\sum_{i\in I_k}\xi_i\right)^2,
\end{equation*}
with $\hat V(\sigma_Y^2)/V(\sigma_Y^2)=1+o_P(1)$, where
\begin{align*}
&\nu_i=\left(Y_i-\hat\theta\right)^2-\frac{m}{m+n}\left\{\hat g^{(-k)}(X_i)-M^{-1}\sum_{i\in J_k}\hat g^{(-k)}(X_i)\right\}^2,\\
&\xi_i=\left\{\hat g^{(-k)}(X_i)-M^{-1}\sum_{i\in J_k}\hat g^{(-k)}(X_i)\right\}^2.
\end{align*}
Moreover, if in addition that $\mu_{2,X}\{\hat g^{(-k)}(X)-g^*(X)\}=o_P(n^{-1/2})$ and either a) $g(x)$ is linear on $x$, or b) $\mu_{2,X}\{\hat g^{(-k)}(X)-g^*(X)\}\mu_{2,X}\{g^*(X)-g^0(X)\}=o_P(n^{-1})$, then
\begin{align}
&\frac{n^{1/2}(\hat\sigma_{\varepsilon,\mathrm{gen}}^2-\sigma_{\varepsilon,\mathrm{gen}}^2)}{\{V(\sigma_{\varepsilon,\mathrm{gen}}^2)\}^{1/2}} \to  N (0,1),\qquad\frac{n^{1/2}(\hat b_\mathrm{gen}^2-b_\mathrm{gen}^2)}{\{V(b_\mathrm{gen}^2)\}^{1/2}} \to  N (0,1),\label{gennormal3}\\
&n^{1/2}V^{-1/2}(R_\mathrm{gen}^2)(R_\mathrm{gen}^2-\mathrm{PVE})\to N(0,1),\nonumber
\end{align}
in distribution, provided that $V(\sigma_{\varepsilon,\mathrm{gen}}^2)>c$, $V(b_\mathrm{gen}^2)>c$ and $V(R_\mathrm{gen}^2)>c$, where $V(\sigma_{\varepsilon,\mathrm{gen}}^2)=\mathrm{var}(\varepsilon^2)$ and
\begin{align*}
V(b_\mathrm{gen}^2)&=\mathrm{var}\left[2\varepsilon\{g^*(X)-\theta\}+\frac{n}{m+n}\{g^*(X)-\theta\}^2\right]+\frac{mn}{(m+n)^2}\mathrm{var}\left[\{g^*(X)-\theta\}^2\right],\\
V(R_\mathrm{gen}^2)&=\mathrm{var}[\sigma_Y^{-4}b_\mathrm{gen}^2\varepsilon^2+\sigma_Y^{-4}\sigma_{\varepsilon,\mathrm{gen}}^2\{2\varepsilon(g^*(X)-\theta)+n(m+n)^{-1}(g^*(X)-\theta)^2\}]\\
&\qquad+n(m+n)^{-1}\sigma_Y^{-8}\sigma_{\varepsilon,\mathrm{gen}}^4\mathrm{var}\{(g^*(X)-\theta)^2\}.
\end{align*}
\end{theorem}

Based on Theorem \ref{genv}, an asymptotic confidence intervals for $\sigma_Y^2$, at significant level $\alpha$, is proposed as
\begin{equation}\label{CIsig2Ygen}
\mathrm{CI}_\mathrm{gen}(\sigma_Y^2)=\left(\hat\sigma_{Y,\mathrm{gen}}^2-z_{1-\alpha/2}\left\{\hat V(\sigma_Y^2)/n\right\}^{1/2},\ \hat\sigma_{Y,\mathrm{gen}}^2+z_{1-\alpha/2}\left\{\hat V(\sigma_Y^2)/n\right\}^{1/2}\right).
\end{equation}

For a general nonlinear model, as in Theorem \ref{genv}, we can see that the asymptotic normality results of estimating explained variance and unexplained variance \eqref{gennormal3} require rates on $\hat g-g^*$ and $g^*-g^0$. Here we provide some insights on the two rates. As the degree of freedom $d$ grows, the estimation error $\mathrm{err}_1=\mu_{2,X}\{\hat g^{(-k)}(X)-g^*(X)\}$ grows and the misspecification error $\mathrm{err}_2=\mu_{2,X}\{g^*(X)-g^0(X)\}$ decreases. To obtain \eqref{gennormal3}, we need $\mathrm{err}_1=o_P(n^{-1/2})$ and $\mathrm{err}_1\mathrm{err}_2=o_P(n^{-1})$, which is weaker than $\mathrm{err}_3=\mu_{2,X}\{\hat g^{(-k)}(X)-g^0(X)\}=o_P(n^{-1/2})$.

Here we take the ReLu network as an example and showcase the conditions when the asymptotic normalities \eqref{gennormal3} hold. Following the settings as in \cite{farrell2018deep}, let $W$ and $L$ being the number of parameters and the number of layers, respectively. Assume the conditions in Theorem 2 of \cite{farrell2018deep}. If $g^0(X)\in\mathcal W^{r,\infty}((-1,1)^p)=\{g:\max_{\alpha,|\alpha|\leq r}\mathrm{ess}\sup_{x\in(-1.1)^d}|D^\alpha g(x)|\leq1\}$. For $W\propto n^a$ with any $a>0$ and $L\propto\log n$, omitting the logarithm terms, we have $\mathrm{err}_1$, $\mathrm{err}_2$ are of the order $n^{-a}$ and $n^{-2ar/p}$ respectively. Hence, the asymptotic normalities \eqref{gennormal3} hold when $a\in(0,1/2)$ and $p<2r$. In other words, the degree of freedom $d$, or $W$ in the neural network example, is flexible, and we are able to obtain the asymptotic normalities for a wide range of $d$.

\section{Numerical experiments on the variance estimation}
\subsection{Supplementary views of the experiments in the main file}
In this section, we provide numerical experiment results of variance estimation under {Model 1} - {Model 6} (the results for mean estimation are provided in the main file), where all the settings are the same as in the main file. We compare estimation and inference results of our proposed generalized estimator $\hat\sigma_Y^2=\hat\sigma_{Y,\mathrm{gen}}^2$ \eqref{sig2Ygen} with the sample variance $S_Y^2$. The confidence interval based on our proposed variance estimator is defined as \eqref{CIsig2Ygen} and the significant level is set to be $\alpha=0.05$. Besides, we also investigate the estimation of the second moment $E(Y^2)$, and more generally, $E\{f(Y)\}$ with some function $f$, see {Model 8} and {Model 9}. 

\begin{figure*}[h!]
    \centering
    \begin{subfigure}[t]{0.485\textwidth}
        \centering
        \includegraphics[height=0.75\linewidth,width=\linewidth]{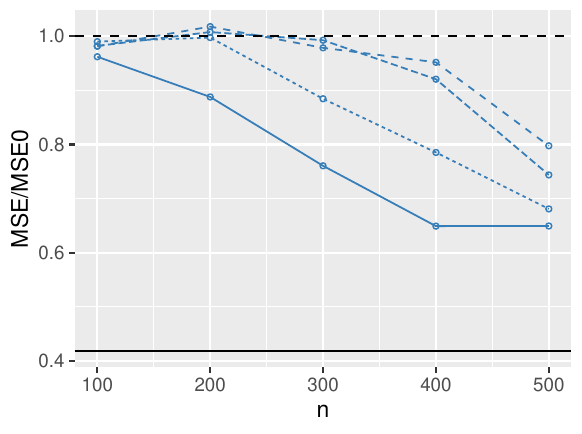}
        \caption{ The ratio of mean squared errors.  }\label{msev11}
    \end{subfigure}
        ~ 
        \begin{subfigure}[t]{0.485\textwidth}
        \centering
        \includegraphics[height=0.75\linewidth,width=\linewidth]{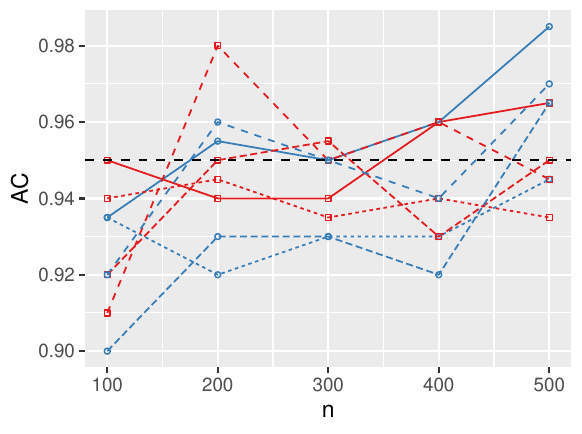}
        \caption{The average coverage of $S_Y^2$ and $\hat\sigma_Y^2$.}\label{acv11}
    \end{subfigure}
    \caption{Model 1: Comparison of  SSL-Lasso and the sample variance. The plot includes sample variance (red squares) and SSL-Lasso (blue circles) estimates. The sparsity level of the linear coefficients, $s$, is denoted with long dashed, dashed, dotted and solid line for $s=90$, $s=70$, $s=50$ and $s=30$, respectively.}
  \end{figure*}

{Model 1.} Let
$
X_i\overset{\mathrm{iid}}{\sim}N_{p-1}(0,I_{p-1}), 
$
with $p=500$, $m=10n$ and
$
Y_i=s^{-1/2}\sum_{j=1}^sX_{ij}+\delta_i, \ s\in\{30,50,70,90\},
$
 $\delta_i\overset{\mathrm{iid}}{\sim}N(0,0.25)$.
We compare our semi-supervised estimator based on lasso with the sample variance $S_Y^2$. Fig. \ref{msev11} shows the mean squared error (MSE) of $S_Y^2$ and $\hat\sigma_Y^2$ as $n$ varies, with $s\in\{30,50,70,90\}$. Fig. \ref{acv11} shows the average coverage (AC) of confidence intervals for $\sigma_Y^2$ estimated by $S_Y^2$ and $\hat\sigma_Y^2$. 
  
\begin{figure*}[h!]
    \centering
    \begin{subfigure}[t]{0.485\textwidth}
        \centering
        \includegraphics[height=0.75\linewidth,width=\linewidth]{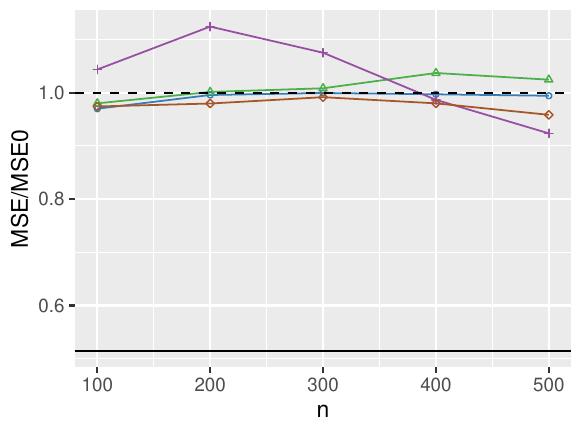}
        \caption{ The ratio of mean squared errors.  }\label{msev2}
    \end{subfigure}%
    ~ 
        \begin{subfigure}[t]{0.485\textwidth}
        \centering
        \includegraphics[height=0.75\linewidth,width=\linewidth]{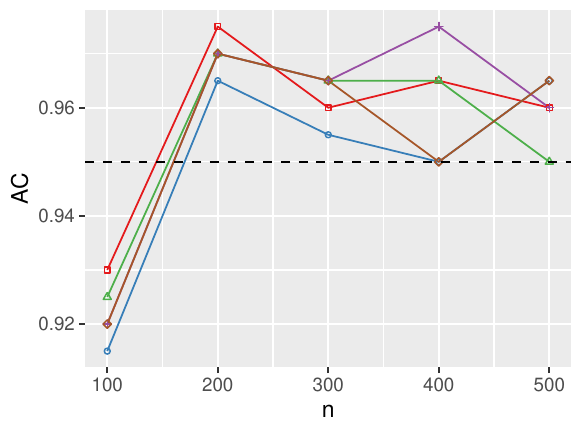}
        \caption{The average coverage.}\label{acv2}
    \end{subfigure}
\caption{Model 2: Comparison of our SSL-``method'' estimators and the sample variance. The plot includes sample variance (red squares), SSL-Lasso (blue circles), SSL-Additive (green up triangles), SSL-XGBoost (purple pluses), and SSL-RF (brown) estimates.}
\end{figure*}

{Model 2.} Let 
$X_i$ and $\delta_i$ be as in {Model 1.}  
   and consider a nonlinear model
$
Y_i=3\cos^0(X_{i1}+X_{i2}+X_{i3})+\delta_i,
$
with $p=51$, $m=10n$. 
We compare our semi-supervised estimators with the sample variance $S_Y^2$. Fig. \ref{msev2} shows the mean squared error (MSE) of $S_Y^2$ and $\hat\sigma_Y^2$ as $n$ varies. Fig. \ref{acv2} shows the average coverage (AC) of confidence intervals for $\sigma_Y^2$ estimated by $S_Y^2$ and $\hat\sigma_Y^2$.
  
\begin{figure*}[h!]
    \centering
    \begin{subfigure}[t]{0.465\textwidth}
        \centering
        \includegraphics[height=0.75\linewidth,width=\linewidth]{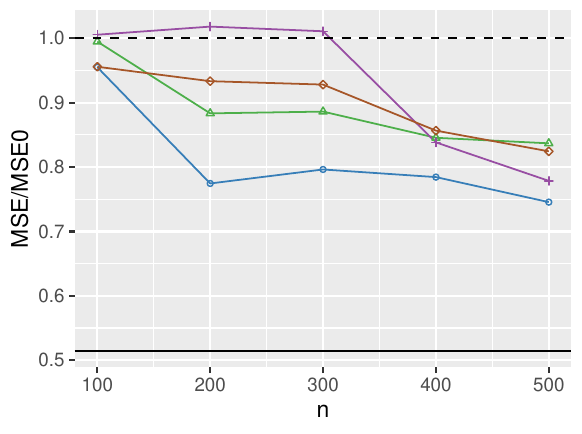}
        \caption{The ratio of mean squared errors. }\label{msev4}
    \end{subfigure}%
   ~ 
   \begin{subfigure}[t]{0.465\textwidth}
        \centering
        \includegraphics[height=0.75\linewidth,width=\linewidth]{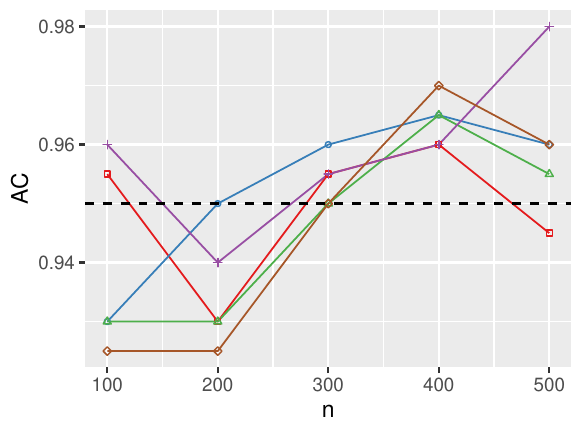}
        \caption{The average coverage. }\label{acv4}
    \end{subfigure}
\caption{Model 3: Comparison of SSL-``method'' with the sample variance. The plot includes sample variance (red squares), SSL-Lasso (blue circles), SSL-Additive (green up triangles), SSL-XGBoost (purple pluses), and SSL-RF (brown diamonds) estimates.}
\end{figure*}

{Model 3.} Let
$
X_i\overset{\mathrm{iid}}{\sim}N_{p-1}(0,C),
$ be equi-correlated with
   $C_{ij}=\{1-1/(2p)\}1_{\{i=j\}}+1/(2p)1_{\{i\neq j\}}$, with $p=1001$, $m=10n$. We consider a nonlinear additive outcome model
$
Y_i=\sum_{j=1}^{p-1}0.7^{j-1}\sin(X_{ij})+\delta_i,
$
where $\delta_i\overset{\mathrm{iid}}{\sim}N(0,0.25)$.
We compare our semi-supervised estimators with the sample variance $S_Y^2$. Fig. \ref{msev4} shows the mean squared error (MSE) of $S_Y^2$ and $\hat\sigma_Y^2$ as $n$ varies. Fig. \ref{acv4} shows the average coverage (AC) of confidence intervals for $\sigma_Y^2$ estimated by $S_Y^2$ and $\hat\sigma_Y^2$.
 
\begin{figure*}[h!]
    \centering
    \begin{subfigure}[t]{0.465\textwidth}
        \centering
        \includegraphics[height=0.75\linewidth,width=\linewidth]{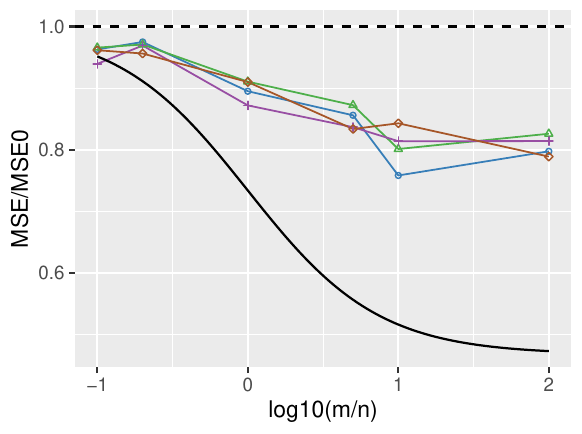}
        \caption{The ratio of mean squared errors. }\label{msev3}
    \end{subfigure}%
    ~ 
    \begin{subfigure}[t]{0.465\textwidth}
        \centering
        \includegraphics[height=0.75\linewidth,width=\linewidth]{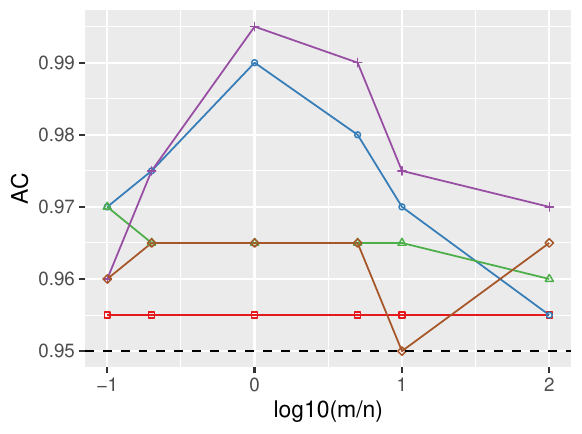}
        \caption{The average coverage.}\label{acv3}
    \end{subfigure}
    \caption{Model 4: Impact of the size of additional data. The plot includes sample variance (red squares), SSL-Lasso (blue circles), SSL-Additive (green up triangles), SSL-XGBoost (purple pluses), and SSL-RF (brown diamonds) estimates.}
\end{figure*}

{Model 4.} Here we observe behavior with varying $m$. Let $X_i$ and $\delta_i$ be as in {Model 1.} and consider the nonlinear outcome  of  {Model 3.}. Set 
 $p=201$, $n=500$ and let $m$ vary from $0.1n$ to $10n$.  
We compare our semi-supervised estimators with the sample variance $S_Y^2$. Fig. \ref{msev3} shows the mean squared error (MSE) of $S_Y^2$ and $\hat\sigma_Y^2$ as $m$ varies. Fig. \ref{acv3} shows the average coverage (AC) of confidence intervals for $\sigma_Y^2$ estimated by $S_Y^2$ and $\hat\sigma_Y^2$.
 
\begin{figure*}[h!]
    \centering
    \begin{subfigure}[t]{0.465\textwidth}
        \centering
        \includegraphics[height=0.75\linewidth,width=\linewidth]{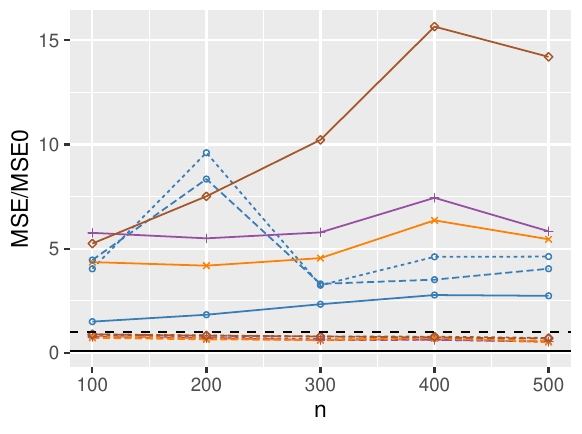}
        \caption{The ratio of mean squared errors. }\label{msev7}
    \end{subfigure}%
   ~ 
       \begin{subfigure}[t]{0.465\textwidth}
        \centering
        \includegraphics[height=0.75\linewidth,width=\linewidth]{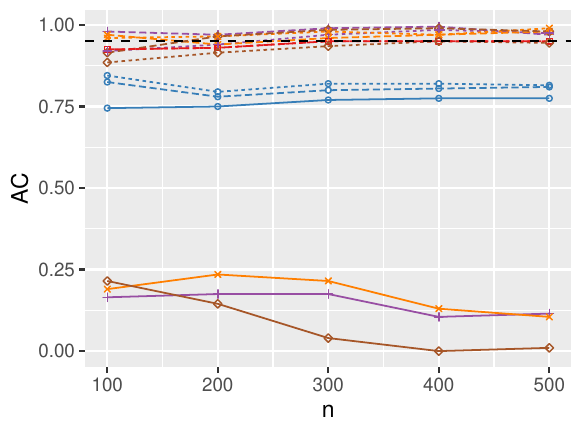}
        \caption{The average coverage. }\label{acv7}
    \end{subfigure}
\caption{Model 5: Is sample-splitting needed? The plot includes sample variance (red squares), SSL-Lasso (blue circles), SSL-XGBoost (purple pluses), SSL-MLP (orange crosses), and SSL-RF (brown diamonds) estimates. The number of folds, $K$, is denoted with solid, dashed, and long dashed line for $K=1$(without cross-fitting), $K=5$, and $K=20$, respectively.}
\end{figure*}

{Model 5.} Let $
X_i\overset{\mathrm{iid}}{\sim}\mathrm{Lognormal}_{p-1}(0,C),
$
with $C$ as in {Model 3.} with $p=101$ and $m=10n$.
Let $
Y_i=\sum_{j=1}^{3}\{\log(X_{ij}+1)^2+0.1\}+\delta_i,
$
where $\delta_i\overset{\mathrm{iid}}{\sim}N(0,0.25)$. We varied $K$ from $1$ to $5$ and then to $20$. 
We compare our semi-supervised estimators with the sample variance of $S_Y^2$. Fig. \ref{msev7} shows the mean squared error (MSE) of $S_Y^2$ and $\hat\sigma_Y^2$ as $n$ varies, with different choices of $K$ ($K=1$ stands for without sample splitting). Fig. \ref{acv7} shows the average coverage (AC) of confidence intervals for $\sigma_Y^2$ estimated by $S_Y^2$ and $\hat\sigma_Y^2$, with different choices of $K$.

\begin{figure*}[h!]
    \centering
    \begin{subfigure}[t]{0.465\textwidth}
        \centering
        \includegraphics[height=0.75\linewidth,width=\linewidth]{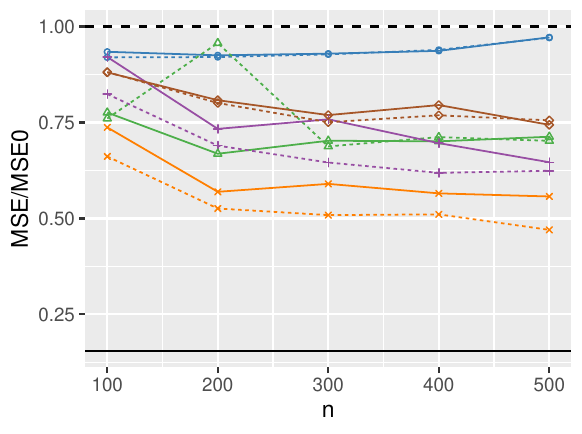}
        \caption{
        The ratio of mean squared errors. }\label{msev8}
    \end{subfigure}%
    ~ 
     \begin{subfigure}[t]{0.465\textwidth}
        \centering
        \includegraphics[height=0.75\linewidth,width=\linewidth]{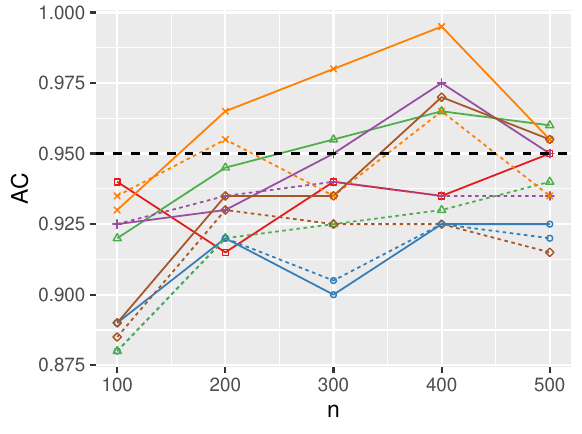}
        \caption{The average coverage.}\label{acv8}
    \end{subfigure}
\caption{Model 6: Does partitioning matter? The plot includes sample variance (red squares), SSL-OLS (blue circles), SSL-Additive (green up triangles), SSL-XGBoost (purple pluses), SSL-MLP (orange crosses), and SSL-RF (brown diamonds) estimates. The number of the repetitions of the cross-fitting, $S$, is denoted with a solid and dashed line for $S=1$ and $S=5$, respectively.}
\end{figure*}

{Model 6.} Let
$
Y_i=X_{i1}X_{i2}+0.5(X_{i3}+0.5)^2+\delta_i
$
and $X_i$ and $\delta_i$ are as in {Model 1.} with $p=4$, $m=10n$. 
We compare our semi-supervised estimators with the sample variance $S_Y^2$, with different choices of $S$, where $S$ denotes the amount of random $K$-partition. Fig. \ref{msev8} shows the mean squared error (MSE) of $S_Y^2$ and $\hat\sigma_Y^2$ as $n$ varies, with different choices of $S$. Fig. \ref{acv8} shows the average coverage (AC) of confidence intervals for $\sigma_Y^2$ estimated by $S_Y^2$ and $\hat\sigma_Y^2$, with different choices of $S$.
 
\subsection{New experiments}

{Model 8.} Instead of estimating $E(Y)$ and $\mathrm{var}(Y)$, we can also estimate the second moment $E(Y^2)$. Here we have two approaches.  {Method 1} is to simply create a new response $\tilde Y=Y^2$ and repeat our procedure on $(\tilde Y_i,X_i)$. This method can be generated to estimate $E\{f(Y)\}$ for any function $f$.  {Method 2} is to estimate $E(Y)$ and $\mathrm{var}(Y)$ by $\hat\theta$ and $\hat\sigma_Y^2$, respectively. Then, $E(Y^2)=\{E(Y)\}^2+\mathrm{var}(Y)$ can be estimated by $\hat\theta^2+\hat\sigma_Y^2$. Here, we compare the two approaches in two different settings. 

Setting {(a)} ``$Y$ is linear in $X$''. Generate
$
(X_{ij})_{(i,j)\in(1,2,\dots,m+n)\times(1,2,\dots,p-1)}\overset{\mathrm{iid}}{\sim}N(1,1)
$
with $p=100$, $m=10n$ and
\begin{equation*}
Y_i=5^{1/2}\sum_{j=1}^5X_{ij}+\delta_i,
\end{equation*}
where $\delta_1,\delta_2,\ldots,\delta_n\overset{\mathrm{iid}}{\sim}N(0,0.25)$.

Setting {(b)} ``$Y^2$ is nonlinear in $X$''.  Generate
$
(X_{ij})_{(i,j)\in(1,2,\dots,m+n)\times(1,2,\dots,p-1)}\overset{\mathrm{iid}}{\sim}\mathrm{Un}(-1,1)
$
with $p=101$, $m=10n$ and
\begin{equation*}
Y_i=r_i\left(5^{1/2}\sum_{j=1}^5X_{ij}+\delta_i+5^{1/2}+0.5\right)^{1/2},
\end{equation*}
where $\delta_1,\delta_2,\ldots,\delta_n\overset{\mathrm{iid}}{\sim}\mathrm{Un}(-1,1)$ and $r_1,r_2,\dots,r_n\in\{-1,1\}$ are independent and identical random variables which take the value $1$ of probability $0.99$ and the value $-1$ with probability $0.01$.
 
\begin{figure*}[h]
    \centering
    \begin{subfigure}[t]{0.465\textwidth}
        \centering
        \includegraphics[height=0.75\linewidth,width=\linewidth]{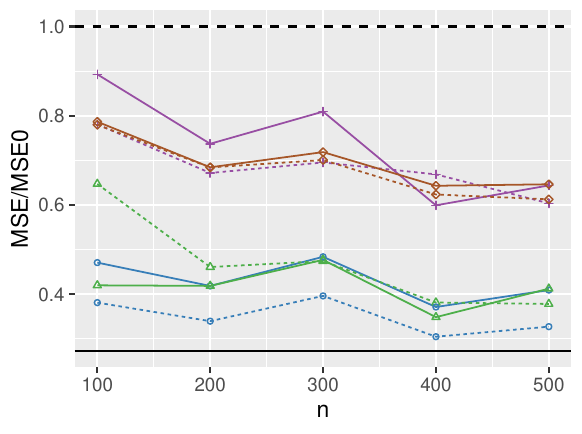}
        \caption{The ratio of mean squared errors:  setting a.}\label{msem12}
    \end{subfigure}%
    ~ 
    \begin{subfigure}[t]{0.465\textwidth}
        \centering
        \includegraphics[height=0.75\linewidth,width=\linewidth]{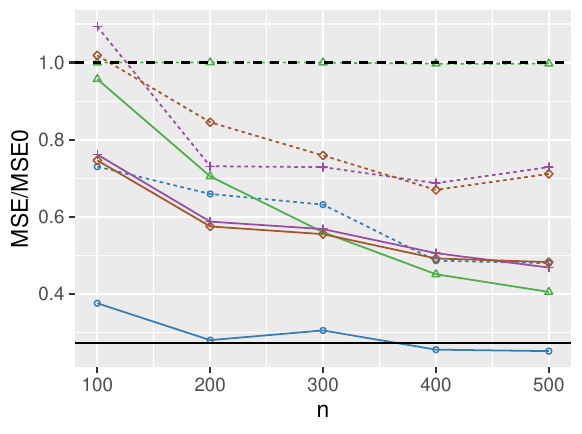}
        \caption{The ratio of mean squared errors:  setting b.}\label{msem13}
    \end{subfigure}
    \caption{Model 8: Comparison between the two proposed estimates of the second moment $E(Y^2)$, solid line denotes method 1 and dashed line denotes method 2. The plot includes SSL-Lasso (blue circles), SSL-additive (green up triangles), SSL-XGBoost (purple pluses), and SSL-RF (brown diamonds). }
\end{figure*}

The simulation is repeated for 200 times. We compare the semi-supervised estimators of $E(Y^2)$ derived from the two approaches, with the empirical mean estimator $\overline{Y^2}=n^{-1}\sum_{i=1^n}Y_i^2$.  Fig. \ref{msem12} shows the mean squared error (MSE) of $\overline{Y^2}$ and $\hat\theta_1$, $\hat\theta_2$ as $n$ varies, in setting {(a)}. Here, $\hat\theta_1$ is the semi-supervised mean estimator \eqref{thetagen} performed on the transformed sample $(\tilde Y_i,X_i)$, following the method 1; and $\hat\theta_2=\hat\theta^2+\hat\sigma_Y^2$ follows the method 2. Fig. \ref{msem13} shows the mean squared error (MSE) of $\overline{Y^2}$ and $\hat\theta_1$, $\hat\theta_2$ as $n$ varies, on setting {(b)}. We observe that {Method 1} is slightly more efficient than {Method 2} when the outcome model is linear. However, when the outcome is a nonlinear function of covariates, our proposed {Method 2} is significantly more efficient.
 
\vskip 10pt 

{Model 9.} Transforming the response?
Instead of estimating $E(Y^2)$ as we proposed we can also estimate $E\{f(Y)\}$, for some function $f$, by simply creating a new response $\tilde Y=f(Y)$ and repeat our procedure on $(X_i,\tilde Y_i)$. 
 We generate $X_i$ and $\delta_i$ as in {Model 1.}
with $p=4$, $m=10n$ and a linear outcome
$
Y_i=X_{i1}+X_{i2}+X_{i3}+\delta_i.
$
we consider: $f(y)=y^2$, $f(y)=\sin(y)$ and $f(y)=\log(y^2+1)$.
 
\begin{figure*}
    \centering
    \begin{subfigure}[t]{0.465\textwidth}
        \centering
        \includegraphics[height=0.75\linewidth,width=\linewidth]{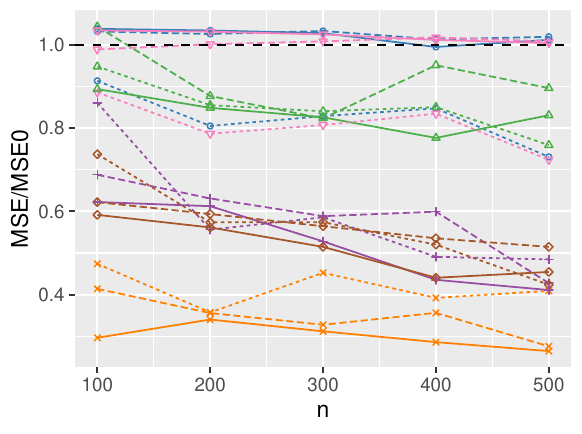}
        \caption{The ratio of mean squared errors.}\label{msem9}
    \end{subfigure}%
    ~ 
      \begin{subfigure}[t]{0.465\textwidth}
        \centering
        \includegraphics[height=0.75\linewidth,width=\linewidth]{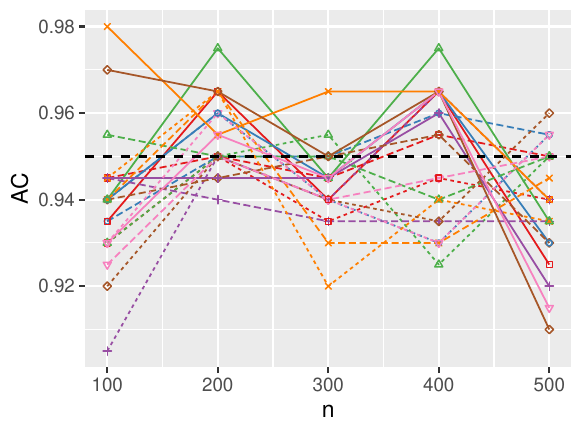}
        \caption{The average coverage.}\label{acm9}
    \end{subfigure} 
\caption{Model 9: Transforming the response? The plot includes empirical average (red squares), SSL-Lasso (blue circles), SSL-additive (green up triangles), SSL-XGBoost (purple pluses), and SSL-RF (brown diamonds). The transformation on the response, $f(y)$, is denoted with solid, dashed and long dashed line for $f(y)=\log(y^2+1)$, $f(y)=\sin(y)$ and $f(y)=y^2$, respectively.}
\end{figure*}

 Fig. \ref{msem9} illustrates the benefit of our estimator of $E(Y^2)$. A simple outcome transformation, leads to no improvements over a naive sample average of the outcomes
  $\overline{f(Y)}$. However, for nonlinear outcome transformation, one can see that even a simple outcome transformation may lead to more efficient estimation.  Fig. \ref{acm9} shows the average coverage (AC) of confidence intervals for $\theta=E\{f(Y)\}$ estimated by $\overline{f(Y)}$, $\hat\theta_\mathrm{SSLS}$ and $\hat\theta$.
 
\section{Proofs of main results}

\subsection{ Auxiliary Lemmas}
We begin by presenting three simple results that will be useful throughout the document.

\begin{lemma}[Lemma B.1 in \cite{chernozhukov2017double}]\label{l1}
Let $\{X_n\}$ and $\{Y_n\}$ be sequences of random variables. If for any $c>0$, $\mathrm{pr} (|X_n|>c|Y_n)=o_P(1)$. Then, $X_n=o_P(1)$. In particular, this occurs if $ E (|X_n|^q\mid Y_n)=o_P(1)$ for any $q\geq1$. Typical examples we used in our proofs are a) $ E (X_n^2\mid Y_n)=o_P(1)$, b) $X_n=\sum_{i=1}^nZ_{n,i}/n$, where $(Z_{n,i})$ is a row-wise independent and identically distributed triangular array, conditional on $Y_n$, with $ E (|Z_{n,1}|\mid Y_n)=o_P(1)$.
\end{lemma}

\begin{lemma}\label{l2}
Let $\{X_n\}$ and $\{Y_n\}$ be sequences of random variables. If $ E (X_n^2\mid Y_n)= O_P(1)$, then $X_n= O_P(1)$. Consequently, if $(Z_{n,i})$ is a row-wise independent and identically distributed triangular array conditional on $Y_n$, with $\mathrm{var}(Z_{n,1}\mid Y_n)=O_P(1)$, or a stronger condition that $E(Z_{n,1}^2\mid Y_n)=O_P(1)$. Then, $\sum_{i=1}^nZ_{n,i}/n=E(Z_{n,1})+O_P(n^{-1/2})$.
\end{lemma}

\begin{proof}[of Lemma~\ref{l2}]
For any $c>0$, since $ E (X_n^2\mid Y_n)= O_P(1)$, there exists $C>0$ such that, for all $n\geq1$,
\[
\mathrm{pr} \{ E (X_n^2\mid Y_n)>C\}<c_1/2.
\]
Hence,
\begin{align*}
& \mathrm{pr}\{|X_n|>(2C/c)^{1/2}\}
= E[1_{\{|X_n|>(2C/c)^{1/2}\}}]\\
&=  E \left[1_{\{E (X_n^2\mid Y_n)\leq C\}} E (1_{\{|X_n|>(2C/c)^{1/2}\}}\mid Y_n)\right]  + E \left(1_{\{ E (X_n^2\mid Y_n)>C\}} E [1_{\{|X_n|>(2C/c)^{1/2}\}}\mid Y_n]\right)\\
&<  E \left[1_{\{ E (X_n^2\mid Y_n)\leq C\}} E \{cX_n^2/(2C)\mid Y_n\}\right]+ E[1_{\{ E (X_n^2\mid Y_n)>C\}}]\\
&=  cE \left[1_{\{ E (X_n^2\mid Y_n)\leq C\}} E (X_n^2\mid Y_n)\right]/(2C)+\mathrm{pr}\{E (X_n^2\mid Y_n)>C\}\\
&\leq c/2+c/2=c.
\end{align*}
That is, $X_n= O_P(1)$.
It follows that $\sum_{i=1}^nZ_{n,i}/n=E(Z_{n,1})+O_P(n^{-1/2})$ since
$$
E\left[n\left\{n^{-1}\sum_{i=1}^nZ_{n,i}-E(Z_{n,1})\right\}^2\mid Y_n\right]=\mathrm{var}(Z_{n,1}\mid Y_n)=O_P(1).
$$
\end{proof}

\begin{lemma}\label{l3}
Let $(Z_{n,i})$ be a row-wise independent and identically distributed triangular array with $ E(Z_{n,1})=0$ and $ E |Z_{n,1}|^q<c_1$ for $q>1$ and $C<\infty$. Let $X_n=\sum_{i=1}^nZ_{n,i}/n$. Then, $X_n=o_P(1)$.
\end{lemma}

\begin{proof}[of Lemma~\ref{l3}]
Let $Y_{n,i}=Z_{n,i}1_{\{|Z_{n,i}|<n\}}$. For any $c>0$, 
\[
\mathrm{pr} (|X_n|\geq c)\leq\mathrm{pr} \{\cup_{i=1}^n(Z_{n,i}\neq Y_{n,i})\}+\mathrm{pr}\left(\left|\sum_{i=1}^nY_{n,i}\right|\geq nc\right)
\]
Let $r\in(1,q\land2)$, then $ E |Z_{n,1}|^r\leq( E |Z_{n,1}|^q)^{r/q}<c_1^{r/q}$. By Markov's Inequality,
\[
\mathrm{pr} \{\cup_{i=1}^n(Z_{n,i}\neq Y_{n,i})\}\leq n\mathrm{pr} (|Z_{n,1}|\geq n)\leq n E |Z_{n,1}|^q/n^q=n^{1-q} E |Z_{n,1}|^q=o(1)
\]
and
\begin{align*}
\mathrm{pr}\left(\left|\sum_{i=1}^nY_{n,i}\right|\geq nc\right)&\leq E\left|\sum_{i=1}^nY_{n,i}\right|^2/(nc)^2=n E\left[Z_{n,1}^21_{\{|Z_{n,i}|<n\}}/(nc)^2\right]\\
&= E \left[|Z_{n,1}|^r|Z_{n,1}|^{2-r}1_{\{|Z_{n,i}|<n\}}/(nc^2)\right]\leq n^{1-r} E |Z_{n,1}|^r/c^2=o(1).
\end{align*}
Hence, $\mathrm{pr} (|X_n|\geq c)=o_P(1)$. That is, $X_n=o_P(1)$.
\end{proof}

\subsection{Proofs of the main results}

\begin{proof}[of Theorem~\ref{c22}] This proof provides $ n^{1/2}$ consistencies of $\hat\theta$ and $\hat\sigma_Y^2$.

{Part {1}.} We first assume Condition \ref{a2} and \ref{a3} and show that $\hat\theta-\theta= O_P(n^{-1/2})$. By the definition of $\beta^*$, as in Lemma 1 of \cite{zhang2019semi}, 
\[
 E (\varepsilon)=0,\  E (\tilde X\varepsilon)=0,\ \theta={\ {\beta^*} }^\T\tilde\mu,\ \sigma_Y^2=b^2+\sigma_\varepsilon^2.
\]
By the definition of $\hat\theta^{(k)}$ as in \eqref{eEY}, 
\begin{equation}\label{p21}
\hat\theta^{(k)}-\theta=N^{-1}\sum_{i\in I_k}(Y_i-\theta)-N^{-1}\sum_{i\in I_k}{\ {\hat\beta^{(-k)}} }^\T\tilde V_i+M^{-1}\sum_{i\in J_k}{\ {\hat\beta^{(-k)}} }^\T\tilde V_i.
\end{equation}
Now we will show that each of the terms on the RHS is of the order $O_P(n^{-1/2})$. From Condition \ref{a2}, \ref{a3} and recall that $\tilde V=\tilde X-\tilde\mu$ is independent of $(Y_i,X_i)_{i\in\{1,2,\dots,n\}\setminus I_k}$, while $\hat\beta^{(-k)}$ is a function of $(Y_i,X_i)_{i\in\{1,2,\dots,n\}\setminus I_k}$, 
\begin{align*}
 E (Y-\theta)^2&\leq( E |Y-\theta|^{2+c})^{2/(2+c)}<c_1,\\
 E ({{\beta^*} }^\T\tilde V)^2&= E (Y-\theta)^2-\sigma_\varepsilon^2\leq E (Y-\theta)^2<c_1,\\
 E _{I_k^c} \left\{(\hat\beta^{(-k)}-\beta^*)^\T\tilde V\right\}^2&=(\hat\beta^{(-k)}-\beta^*)^\T\tilde C(\hat\beta^{(-k)}-\beta^*)\leq\|\hat\beta^{(-k)}-\beta^*\|_2^2\|\tilde C\|_2= O_P(1),
\end{align*} 
and by triangle inequality, $E _{I_k^c}({\ {\hat\beta^{(-k)}} }^\T\tilde V)^2= O_P(1)$. Then, by Lemma \ref{l2},
\begin{align}
 N^{-1}\sum_{i\in I_k}(Y_i-\theta)= O_P(N^{-1/2}),   \label{p251}\\
 N^{-1}\sum_{i\in I_k}{ \ { \hat\beta^{(-k)}} }^\T\tilde V_i= O_P(N^{-1/2}),\label{p252}\\
 M^{-1}\sum_{i\in J_k}{\ {\hat\beta^{(-k)}} }^\T\tilde V_i= O_P(M^{-1/2}).\label{p253}
\end{align}
Therefore, $\hat\theta^{(k)}-\theta= O_P(N^{-1/2})+ O_P(N^{-1/2})+ O_P(M^{-1/2})= O_P(N^{-1/2})$, since $M\geq N$. When $K<\infty$,
\begin{equation}\label{p24}
\hat\theta=K^{-1}\sum_{k=1}^K\hat\theta^{(k)}=\theta+ O_P(n^{-1/2}).
\end{equation}

{Part {2}.} Now we assume Condition \ref{a6} and \ref{a3} and show that $\hat\sigma_Y^2-\sigma_Y^2= O_P(n^{-1/2})$. Recall the definition of $\hat\sigma_Y^{2^{(k)}}$ in Section \ref{s22},
\[
\hat\sigma_Y^{2^{(k)}}=N^{-1}\sum_{i\in I_k} (Y_i-\hat\theta)^2   + M^{-1}\sum_{i\in J_k} \left({ \ { \hat\beta^{(-k)}} }^\T \hat V_i\right)^2- N^{-1}\sum_{i\in I_k} \left({ \ { \hat\beta^{(-k)}} }^\T \hat V_i\right)^2,
\] 
We first approximate the terms on the RHS by replacing $\hat\theta$ and $\hat V_i$ by $\theta$ and $\tilde V_i$, respectively. Recall \eqref{p251} and \eqref{p24},
\begin{align*}
N^{-1}\sum_{i\in I_k} (Y_i-\hat\theta)^2&=N^{-1}\sum_{i\in I_k} (Y_i-\theta)^2+(\hat\theta-\theta)^2-2(\hat\theta-\theta)N^{-1}\sum_{i\in I_k}(Y_i-\theta)\\
&=N^{-1}\sum_{i\in I_k} (Y_i-\theta)^2+ O_P(N^{-1})
\end{align*}
Besides, by definition, $\hat V_i=\tilde V_i-(\hat\mu^{(k)}-\tilde\mu)$, where $\hat\mu^{(k)}-\tilde\mu=M^{-1}\sum_{i\in J_k}{\ {\hat\beta^{(-k)}} }^\T\tilde V_i$. Recall \eqref{p252} and \eqref{p253}, 
\begin{align}
M^{-1}\sum_{i\in J_k} \left({ \ { \hat\beta^{(-k)}} }^\T \hat V_i\right)^2&=M^{-1}\sum_{i\in J_k}({ \ { \hat\beta^{(-k)}} }^\T\tilde V_i)^2-\{{ \ { \hat\beta^{(-k)}} }^\T(\hat\mu^{(k)}-\tilde\mu)\}^2
\nonumber\\
&=M^{-1}\sum_{i\in J_k}({ \ { \hat\beta^{(-k)}} }^\T\tilde V_i)^2+ O_P(M^{-1}),\label{p23}
\end{align}
and
\begin{align}
N^{-1}\sum_{i\in I_k}({ \ { \hat\beta^{(-k)}} }^\T\hat V_i)^2&=N^{-1}\sum_{i\in I_k}({ \ { \hat\beta^{(-k)}} }^\T\tilde V_i)^2+\{{ \ { \hat\beta^{(-k)}} }^\T(\hat\mu^{(k)}-\tilde\mu)\}^2
\nonumber
\\
&\qquad-2{ \ { \hat\beta^{(-k)}} }^\T(\hat\mu^{(k)}-\tilde\mu)N^{-1}\sum_{i\in I_k}{ \ { \hat\beta^{(-k)}} }^\T\tilde V_i
\nonumber
\\
&=N^{-1}\sum_{i\in I_k}({ \ { \hat\beta^{(-k)}} }^\T\tilde V_i)^2+ O_P(M^{-1}+N^{-1/2}M^{-1/2}).\label{p23'}
\end{align}
Hence,
\begin{equation}\label{p22}
\hat\sigma_Y^{2^{(k)}}=N^{-1}\sum_{i\in I_k} (Y_i-\theta)^2-N^{-1}\sum_{i\in I_k}({ \ { \hat\beta^{(-k)}} }^\T\tilde V_i)^2+M^{-1}\sum_{i\in J_k}({ \ { \hat\beta^{(-k)}} }^\T\tilde V_i)^2+ O_P(N^{-1}).
\end{equation}
Now we will show that each of the terms on the RHS of \eqref{p22} is of the order $O_P(N^{-1/2})$. By Lemma \ref{l2}, it suffices to show
\begin{align}
&E (Y-\theta)^4=O(1),\label{align1}\\
&E _{I_k^c}({\ {\hat\beta^{(-k)}} }^\T\tilde V)^4= O_P(1)\label{align2}.
\end{align}
Here, \eqref{align1} follows by the assumption that $E|Y|^{4+c}<c_1$. Besides, recall that $\tilde V=\tilde X-\tilde\mu=(0,V^\T)^\T$, where $V=C^{1/2}Z$. By Condition \ref{a6}, we have bounded 4th moments
\begin{align*}
&E ({{\beta^*} }^\T\tilde V)^4= E ({\ \beta_{-1}^*}^\T C^{1/2}Z)^4=b^4 E ({\ \beta_{-1}^*}^\T C^{1/2}Z/b)^4\leq b^4\sup_{\|a\|_2=1} E (a^\T Z)^4=O(1),\\
&E _{I_k^c} \left\{(\hat\beta^{(-k)}-\beta^*)^\T\tilde V\right\}^4\leq\|\hat\beta^{(-k)}-\beta^*\|_{\tilde C}^4\sup_{\|a\|_2=1} E (a^\T Z)^4= O_P(1),
\end{align*}
and hence \eqref{align2} follows. Now, we obtain
\begin{equation}\label{hsigmaY2k}
\hat\sigma_Y^{2^{(k)}}-\sigma_Y^2= O_P(N^{-1/2})
\end{equation}
and the proof is finalized by noticing that for finite $K$, the rate above is inherited for the averaged estimator
\begin{equation}\label{hsigmaY2}
\hat\sigma_Y^2=\sigma_Y^2+ O_P(n^{-1/2}).
\end{equation}
\end{proof}

\begin{proof}[of Theorem~\ref{t23}] This proof provides an asymptotic normal result for $ n^{1/2}(\hat\theta-\theta)$ by relying on Conditions \ref{a2} and \ref{a4}. Recall from \eqref{p21},
\begin{align*}
\hat\theta^{(k)}-\theta&=N^{-1}\sum_{i\in I_k}(Y_i-\theta)-N^{-1}\sum_{i\in I_k}{\ {\hat\beta^{(-k)}} }^\T\tilde V_i+M^{-1}\sum_{i\in J_k}{\ {\hat\beta^{(-k)}} }^\T\tilde V_i\\
&=N^{-1}\sum_{i\in I_k}\varepsilon_i+M^{-1}\sum_{i\in J_k}{\beta^*}^\T\tilde V_i-N^{-1}\sum_{i\in I_k}(\hat\beta^{(-k)}-\beta^*)^\T\tilde V_i+M^{-1}\sum_{i\in J_k}(\hat\beta^{(-k)}-\beta^*)^\T\tilde V_i
\end{align*}
Since
\begin{align*}
E_{I_k^c}\left\{(\hat\beta^{(-k)}-\beta^*)^\T\tilde V\right\}^2&=\|\hat\beta^{(-k)}-\beta^*\|_{\tilde C}^2\leq\|\hat\beta^{(-k)}-\beta^*\|_2^2\|\tilde C\|_2=o_P(1),
\end{align*}
and by Lemma \ref{l1}, 
\[
N^{-1}\sum_{i\in I_k}(\hat\beta^{(-k)}-\beta^*)^\T\tilde V_i=o_P(N^{-1/2}),\qquad M^{-1}\sum_{i\in J_k}(\hat\beta^{(-k)}-\beta^*)^\T\tilde V_i=o_P(M^{-1/2}).
\]
Therefore,
\[
\hat\theta^{(k)}-\theta=N^{-1}\sum_{i\in I_k}\varepsilon_i+M^{-1}\sum_{i\in J_k}{\ {\beta^*} }^\T\tilde V_i+o_P(N^{-1/2}).
\]
When $K<\infty,$
\[
 n^{1/2}(\hat\theta-\theta)=n^{-1/2}\sum_{i=1}^n\varepsilon_i+n^{1/2}M^{-1}\sum_{i\in J_k}{ \ {\beta^*} }^\T\tilde V_i+o_P(1).
\]
By Condition \ref{a2}, $E |\varepsilon|^{2+c}<c_1,\  E |{\ {\beta^*} }^\T\tilde V|^{2+c}<c_1$. With a slight abuse of notation, assume that $\sigma_\varepsilon^2=\lim_{n\rightarrow\infty} E( \varepsilon^2),\quad \tau b^2=\lim_{n\rightarrow\infty}n E ({\ {\beta^*} }^\T\tilde V)^2/(m+n)$ both exists. We continue the analysis by analyzing three separate cases.\\
a) When $\sigma_\varepsilon^2>0$ and $\tau b^2>0$,
\[
\frac{ E |\varepsilon|^{2+c}}{\{E(\varepsilon^2)\}^{1+c/2}}<c_1,\quad\frac{ E |(n/(m+n))^{1/2}{\ {\beta^*} }^\T\tilde V|^{2+c}}{\{n E ({\ {\beta^*} }^\T\tilde V)^2/(m+n)\}^{1+c/2}}<c_1,
\]
i.e. the Lyapunov condition holds.  By Lindeberg-Feller Central Limit Theorem, 
\[
n^{-1/2}\sum_{i=1}^n\varepsilon_i\rightarrow N\left(0,\sigma_\varepsilon^2\right),\quad n^{1/2}M^{-1}\sum_{i\in J_k}{ \ {\beta^*} }^\T\tilde V_i\rightarrow N\left(0,\tau b^2\right).
\]
in distribution. By Slutsky's Theorem and multivariate delta method,
\begin{equation}\label{p31}
 n^{1/2}(\hat\theta-\theta)=n^{-1/2}\sum_{i=1}^n\varepsilon_i+n^{1/2}M^{-1}\sum_{i\in J_k}{ \ {\beta^*} }^\T\tilde V_i+o_P(1)\rightarrow N\left(0,\sigma_\varepsilon^2+\tau b^2\right).
\end{equation}
b) When $\sigma_\varepsilon^2=0$, recall the assumption that $\sigma_\varepsilon^2+\tau b^2>0$, we have $\tau b^2>0$. In this case, by Lemma \ref{l1} and Lindeberg-Feller Central Limit Theorem,
\[
n^{-1/2}\sum_{i=1}^n\varepsilon_i=o_P(1),\quad n^{1/2}M^{-1}\sum_{i\in J_k}{ \ {\beta^*} }^\T\tilde V_i\rightarrow N\left(0,\tau b^2\right).
\]
By Slutsky's Theorem, \eqref{p31} holds.\\
c) When $\tau b^2=0$, similarly as in b), \eqref{p31} holds.
\end{proof}

 \begin{proof}[of Theorem~\ref{t31}] This proof provides consistency results for $\hat\sigma_\varepsilon^2$ and $\hat b^2$ by assuming Conditions \ref{a2} and \ref{a4}. 

{Part {1}.} We first show that $\hat\sigma_\varepsilon^2=\sigma_\varepsilon^2+o_P(1)$. Recall the definition of $\hat\sigma_\varepsilon^{2^{(k)}}$,
\[
\hat\sigma_\varepsilon^{2^{(k)}}=N^{-1}\sum_{i\in I_k}(Y_i-\hat\theta-{ \ { \hat\beta^{(-k)}} }^\T \hat V_i)^2.
\]
Now we first approximate the RHS by replacing $\hat\theta$ and $\hat V_i$ by $\theta$ and $\tilde V_i$, respectively. Recall from \eqref{p251} -- \eqref{p24},
\begin{align}
\hat\sigma_\varepsilon^{2^{(k)}}&=N^{-1}\sum_{i\in I_k}(Y_i-\theta-{ \ { \hat\beta^{(-k)}} }^\T\tilde V_i)^2+\{\hat\theta-\theta-{ \ { \hat\beta^{(-k)}} }^\T (\hat\mu^{(k)}-\tilde\mu)\}^2\nonumber\\
&\qquad-2\{\hat\theta-\theta-{ \ { \hat\beta^{(-k)}} }^\T (\hat\mu^{(k)}-\tilde\mu)\}N^{-1}\sum_{i\in I_k}(Y_i-\theta-{ \ { \hat\beta^{(-k)}} }^\T\tilde V_i)\nonumber\\
&=N^{-1}\sum_{i\in I_k}\varepsilon_i^2+N^{-1}\sum_{i\in I_k}\{(\hat\beta^{(-k)}-\beta^*)^\T\tilde V_i\}^2\nonumber\\
&\qquad-2N^{-1}\sum_{i\in I_k}\varepsilon_i(\hat\beta^{(-k)}-\beta^*)^\T\tilde V_i+ O_P(N^{-1}).\label{sigeps2k}
\end{align}
 Remember the definition that $ E _{I_k^c}(g)= E \{g\mid (Y_i,X_i)_{i\in\{1,2,\dots,m+n\}\setminus I_k}\}$.  By Condition \ref{a2}, $E |\varepsilon|^{2+c}<c_1,\ E _{I_k^c}\{(\hat\beta^{(-k)}-\beta^*)^\T\tilde V\}^2=o_P(1)$ and hence $E _{I_k^c}|\varepsilon(\hat\beta^{(-k)}-\beta^*)^\T\tilde V|=o_P(1)$. By Lemma \ref{l1}, 
\[
N^{-1}\sum_{i\in I_k}\{(\hat\beta^{(-k)}-\beta^*)^\T\tilde V_i\}^2=o_P(1),\qquad N^{-1}\sum_{i\in I_k}\varepsilon_i(\hat\beta^{(-k)}-\beta^*)^\T\tilde V_i=o_P(1).
\]
By Lemma \ref{l3},
\[
N^{-1}\sum_{i\in I_k}\varepsilon_i^2=\sigma_\varepsilon^2+o_P(1).
\]
Hence, $\hat\sigma_\varepsilon^{2^{(k)}}=\sigma_\varepsilon^2+o_P(1)$. When $K<\infty$, $\hat\sigma_\varepsilon^2=\sigma_\varepsilon^2+o_P(1)$.

{Part {2}.} Now we show that $\hat b^2=b^2+o_P(1)$. By the definition of $\hat b^{2^{(k)}}$,
\[
\hat b^{2^{(k)}}=M^{-1}\sum_{i\in J_k}(\hat\beta^{(-k)^\T} \hat V_i)^2+2N^{-1}\sum_{i\in I_k}\hat\beta^{(-k)^\T} \hat V_i(Y_i-\hat\theta)-2N^{-1}\sum_{i\in I_k}(\hat\beta^{(-k)^\T} \hat V_i)^2.
\]
We first approximate the terms of RHS by replacing $\hat\theta$ and $\hat V_i$ by $\theta$ and $\tilde V_i$, respectively. Recall from \eqref{p23} and \eqref{p23'},
\begin{align}
{ \ { \hat\beta^{(-k)}} }^\T\hat C^{(k)}\hat\beta^{(-k)}&=M^{-1}\sum_{i\in J_k}({ \ { \hat\beta^{(-k)}} }^\T\tilde V_i)^2+ O_P(M^{-1}),\label{bcb1}\\
N^{-1}\sum_{i\in I_k}(\hat\beta^{(-k)^\T} \hat V_i)^2&=N^{-1}\sum_{i\in I_k}({ \ { \hat\beta^{(-k)}} }^\T\tilde V_i)^2+ O_P(N^{-1}).\nonumber
\end{align}
Recall from \eqref{p251} -- \eqref{p24},
\begin{align*}
&N^{-1}\sum_{i\in I_k}\hat\beta^{(-k)^\T} \hat V_i(Y_i-\hat\theta)
\\
&=N^{-1}\sum_{i\in I_k}\hat\beta^{(-k)^\T}\tilde V_i(Y_i-\theta)+{ \ { \hat\beta^{(-k)}} }^\T(\hat\mu^{(k)}-\tilde\mu)(\hat\theta-\theta)\\
&\qquad-\hat\beta^{(-k)^\T}(\hat\mu^{(k)}-\tilde\mu)N^{-1}\sum_{i\in I_k}(Y_i-\theta)-(\hat\theta-\theta)N^{-1}\sum_{i\in I_k}\hat\beta^{(-k)^\T}\tilde V_i\\
&=  N^{-1}\sum_{i\in I_k}{ \ { \hat\beta^{(-k)}} }^\T\tilde V_i(Y_i-\theta)+ O_P(N^{-1}).
\end{align*}
Hence,
\begin{align*}
\hat b^{2^{(k)}}&=M^{-1}\sum_{i\in J_k}({ \ { \hat\beta^{(-k)}} }^\T\tilde V_i)^2+2N^{-1}\sum_{i\in I_k}{ \ { \hat\beta^{(-k)}} }^\T\tilde V_i(Y_i-\theta)\\
&\qquad-2N^{-1}\sum_{i\in I_k}({ \ { \hat\beta^{(-k)}} }^\T\tilde V_i)^2+ O_P(N^{-1}).
\end{align*}
The first term on the RHS can be expressed as
\begin{align*}
M^{-1}\sum_{i\in J_k}({ \ { \hat\beta^{(-k)}} }^\T\tilde V_i)^2&=M^{-1}\sum_{i\in J_k}({\ {\beta^*}}^\T\tilde V_i)^2+2M^{-1}\sum_{i\in J_k}(\hat\beta^{(-k)}-\beta^*)^\T\tilde V_i{\ {\beta^*}}^\T\tilde V_i\\
&\qquad+M^{-1}\sum_{i\in J_k}\{(\hat\beta^{(-k)}-\beta^*)^\T\tilde V_i\}^2.
\end{align*}
By Condition \ref{a2} and \ref{a4}, $ E |{\ \beta^*}^\T\tilde V|^{2+c}<c_1,\quad E _{I_k^c}\{(\hat\beta^{(-k)}-\beta^*)^\T\tilde V\}^2=o_P(1)$, which implies that $E _{I_k^c}|(\hat\beta^{(-k)}-\beta^*)^\T\tilde V{\ {\beta^*}}^\T\tilde V|=o_P(1)$. By Lemma \ref{l3}, $M^{-1}\sum_{i\in J_k}({\ {\beta^*}}^\T\tilde V_i)^2=b^2+o_P(1)$. By Lemma \ref{l1},
\[
M^{-1}\sum_{i\in J_k}(\hat\beta^{(-k)}-\beta^*)^\T\tilde V_i{\ {\beta^*}}^\T\tilde V_i=o_P(1),\quad M^{-1}\sum_{i\in J_k}\{(\hat\beta^{(-k)}-\beta^*)^\T\tilde V_i\}^2=o_P(1).
\]
Hence,
\begin{equation}\label{bcb2}
M^{-1}\sum_{i\in J_k}({ \ { \hat\beta^{(-k)}} }^\T\tilde V_i)^2=b^2+o_P(1).
\end{equation}
Similarly, $N^{-1}\sum_{i\in I_k}({ \ { \hat\beta^{(-k)}} }^\T\tilde V_i)^2=b^2+o_P(1)$. Recall from \eqref{p251} -- \eqref{p24},
\begin{align}
& N^{-1}\sum_{i\in I_k}\hat\beta^{(-k)^\T} \hat V_i(Y_i-\hat\theta)\nonumber\\
&=N^{-1}\sum_{i\in I_k}\hat\beta^{(-k)^\T}\tilde V_i(Y_i-\theta)+{ \ { \hat\beta^{(-k)}} }^\T(\hat\mu^{(k)}-\tilde\mu)(\hat\theta-\theta)\nonumber\\
&\qquad-\hat\beta^{(-k)^\T}(\hat\mu^{(k)}-\tilde\mu)N^{-1}\sum_{i\in I_k}(Y_i-\theta)-(\hat\theta-\theta)N^{-1}\sum_{i\in I_k}\hat\beta^{(-k)^\T}\tilde V_i\nonumber\\
&=  N^{-1}\sum_{i\in I_k}{\ \beta^*}^\T\tilde V_i(Y_i-\theta)+N^{-1}\sum_{i\in I_k}(\hat\beta^{(-k)}-\beta^*)^\T\tilde V_i(Y_i-\theta)+ O_P(N^{-1}).\label{sth1}
\end{align}
By Condition \ref{a2} and \ref{a4}, $ E |{\ \beta^*}^\T\tilde V(Y-\theta)|^{2+c}<c_1$ and $E _{I_k^c}|(\hat\beta^{(-k)}-\beta^*)^\T\tilde V(Y-\theta)|=o_P(1)$. By Lemma \ref{l3}, $N^{-1}\sum_{i\in I_k}{\ \beta^*}^\T\tilde V_i(Y_i-\theta)=b^2+o_P(1)$,
and by Lemma \ref{l1}, $N^{-1}\sum_{i\in I_k}(\hat\beta^{(-k)}-\beta^*)^\T\tilde V_i(Y_i-\theta)=o_P(1)$.
Hence, $N^{-1}\sum_{i\in I_k}\hat\beta^{(-k)^\T} \hat V_i(Y_i-\hat\theta)=b^2+o_P(1)$. Combining all the previous results, 
\begin{align}
\hat b^{2^{(k)}}&=M^{-1}\sum_{i\in J_k}({ \ { \hat\beta^{(-k)}} }^\T\tilde V_i)^2+2N^{-1}\sum_{i\in I_k}{ \ { \hat\beta^{(-k)}} }^\T\tilde V_i(Y_i-\theta)\nonumber\\
&\qquad-2N^{-1}\sum_{i\in I_k}({ \ { \hat\beta^{(-k)}} }^\T\tilde V_i)^2+ O_P(N^{-1})\label{b2k}\\
&=b^2+o_P(1)+2\{b^2+o_P(1)\}-2\{b^2+o_P(1)\}=b^2+o_P(1).\label{hb2k}
\end{align}
When $K<\infty$, $\hat b^2=b^2+o_P(1)$.

{Part {3}.} Now assume Conditions \ref{a6} and \ref{a4}, we provide consistency rate results for $\hat\sigma_\varepsilon^2$ and $\hat b^2$. We first consider $\hat\sigma_\varepsilon^2$, recall \eqref{sigeps2k},
\begin{align*}
\hat\sigma_\varepsilon^{2^{(k)}}&=N^{-1}\sum_{i\in I_k}\varepsilon_i^2+N^{-1}\sum_{i\in I_k}\{(\hat\beta^{(-k)}-\beta^*)^\T\tilde V_i\}^2-2N^{-1}\sum_{i\in I_k}\varepsilon_i(\hat\beta^{(-k)}-\beta^*)^\T\tilde V_i+ O_P(N^{-1}).
\end{align*}
By Conditions \ref{a6} and \ref{a4}, $E(\varepsilon^4)<c_1$,
\[
E _{I_k^c}\{(\hat\beta^{(-k)}-\beta^*)^\T\tilde V\}^4\leq\|\hat\beta^{(-k)}-\beta^*\|_{\tilde C}^4\sup_{\|a\|_2=1} E (a^\T Z)^4=O(\|\hat\beta^{(-k)}-\beta^*\|_{\tilde C}^4)
\]
and hence $E _{I_k^c}\{\varepsilon(\hat\beta^{(-k)}-\beta^*)^\T\tilde V\}^2=O(\|\hat\beta^{(-k)}-\beta^*\|_{\tilde C}^2)$.
By Lemma \ref{l2},
\begin{align*}
&N^{-1}\sum_{i\in I_k}\varepsilon_i^2=\sigma_\varepsilon^2+ O_P(N^{-1/2}),\\
&N^{-1}\sum_{i\in I_k}\{(\hat\beta^{(-k)}-\beta^*)^\T\tilde V_i\}^2=\|\hat\beta^{(-k)}-\beta^*\|_{\tilde C}^2+ O_P(\|\hat\beta^{(-k)}-\beta^*\|_{\tilde C}^2N^{-1/2}),\\
&N^{-1}\sum_{i\in I_k}\varepsilon_i(\hat\beta^{(-k)}-\beta^*)^\T\tilde V_i=O_P(\|\hat\beta^{(-k)}-\beta^*\|_{\tilde C}N^{-1/2}).
\end{align*}
Hence,
\begin{align*}
\hat\sigma_\varepsilon^{2^{(k)}}&=N^{-1}\sum_{i\in I_k}\varepsilon_i^2+N^{-1}\sum_{i\in I_k}\{(\hat\beta^{(-k)}-\beta^*)^\T\tilde V_i\}^2\\
&\qquad-2N^{-1}\sum_{i\in I_k}\varepsilon_i(\hat\beta^{(-k)}-\beta^*)^\T\tilde V_i+ O_P(N^{-1})\\
&=\sigma_\varepsilon^2+ O_P(\|\hat\beta^{(-k)}-\beta^*\|_{\tilde C}^2+N^{-1/2}).
\end{align*}
When $K<\infty$, $\hat\sigma_\varepsilon^2=\sigma_\varepsilon^2+ O_P(\|\hat\beta^{(-k)}-\beta^*\|_{\tilde C}^2+N^{-1/2})=\sigma_\varepsilon^2+ O_P(\|\hat\beta^{(-k)}-\beta^*\|_2^2+n^{-1/2})$.

By the same strategy, now we show the consistency result for $\hat b^2$. Recall \eqref{b2k},
\begin{align*}
\hat b^{2^{(k)}}&=M^{-1}\sum_{i\in J_k}({ \ { \hat\beta^{(-k)}} }^\T\tilde V_i)^2+2N^{-1}\sum_{i\in I_k}{ \ { \hat\beta^{(-k)}} }^\T\tilde V_i(Y_i-\theta)\\
&\qquad-2N^{-1}\sum_{i\in I_k}({ \ { \hat\beta^{(-k)}} }^\T\tilde V_i)^2+ O_P(N^{-1}),
\end{align*}
where
\begin{align*}
M^{-1}\sum_{i\in J_k}({ \ { \hat\beta^{(-k)}} }^\T\tilde V_i)^2&=M^{-1}\sum_{i\in J_k}({\ {\beta^*}}^\T\tilde V_i)^2+2M^{-1}\sum_{i\in J_k}(\hat\beta^{(-k)}-\beta^*)^\T\tilde V_i{\ {\beta^*}}^\T\tilde V_i\\
&\qquad+M^{-1}\sum_{i\in J_k}\{(\hat\beta^{(-k)}-\beta^*)^\T\tilde V_i\}^2.
\end{align*}
By Conditions \ref{a2} and \ref{a6}, $E({\ {\beta^*}}^\T\tilde V)^4<c_1$, and recall that  $E _{I_k^c}\{(\hat\beta^{(-k)}-\beta^*)^\T\tilde V\}^4=O(\|\hat\beta^{(-k)}-\beta^*\|_{\tilde C}^4)$, hence
\[
E _{I_k^c}\{(\hat\beta^{(-k)}-\beta^*)^\T\tilde V{\ {\beta^*}}^\T\tilde V\}^2=O(\|\hat\beta^{(-k)}-\beta^*\|_{\tilde C}^2)
\]
By Lemma \ref{l2},
\begin{align}
&M^{-1}\sum_{i\in J_k}({\ {\beta^*}}^\T\tilde V_i)^2=b^2+ O_P(M^{-1}),\nonumber
\\
&M^{-1}\sum_{i\in J_k}\{(\hat\beta^{(-k)}-\beta^*)^\T\tilde V_i\}^2=\|\hat\beta^{(-k)}-\beta^*\|_{\tilde C}^2+ O_P(\|\hat\beta^{(-k)}-\beta^*\|_{\tilde C}^2M^{-1/2}),\label{line2}\\
&M^{-1}\sum_{i\in J_k}(\hat\beta^{(-k)}-\beta^*)^\T\tilde V_i{\ {\beta^*}}^\T\tilde V_i=(\hat\beta^{(-k)}-\beta^*)^\T\tilde C\beta^*+ O_P(\|\hat\beta^{(-k)}-\beta^*\|_{\tilde C}M^{-1/2}).\label{line3}
\end{align}
Hence, $M^{-1}\sum_{i\in J_k}({ \ { \hat\beta^{(-k)}} }^\T\tilde V_i)^2=b^2+\|\hat\beta^{(-k)}-\beta^*\|_{\tilde C}^2+2(\hat\beta^{(-k)}-\beta^*)^\T\tilde C\beta^*+ O_P(M^{-1})$. Similarly, $N^{-1}\sum_{i\in I_k}({ \ { \hat\beta^{(-k)}} }^\T\tilde V_i)^2=b^2+\|\hat\beta^{(-k)}-\beta^*\|_{\tilde C}^2+2(\hat\beta^{(-k)}-\beta^*)^\T\tilde C\beta^*+ O_P(N^{-1})$. Besides, recall \eqref{sth1}. Then, simple algebra concludes
\begin{align*}
&N^{-1}\sum_{i\in I_k}\hat\beta^{(-k)^\T} \hat V_i(Y_i-\hat\theta)\\
&=N^{-1}\sum_{i\in I_k}{\ \beta^*}^\T\tilde V_i(Y_i-\theta)+N^{-1}\sum_{i\in I_k}(\hat\beta^{(-k)}-\beta^*)^\T\tilde V_i(Y_i-\theta)+O_P(N^{-1}).
\end{align*}
By Conditions \ref{a6} and \ref{a4},
\[
E \{{\ \beta^*}^\T\tilde V(Y-\theta)\}^{2+c}=O(1),\qquad E _{I_k^c}\{(\hat\beta^{(-k)}-\beta^*)^\T\tilde V(Y-\theta)\}^2=O(\|\hat\beta^{(-k)}-\beta^*\|_{\tilde C}^2).
\]
By Lemma \ref{l2},
\begin{align*}
&N^{-1}\sum_{i\in I_k}{\ \beta^*}^\T\tilde V_i(Y_i-\theta)=b^2+ O_P(N^{-1/2}),\\
&N^{-1}\sum_{i\in I_k}(\hat\beta^{(-k)}-\beta^*)^\T\tilde V_i(Y_i-\theta)=(\hat\beta^{(-k)}-\beta^*)^\T\tilde C\beta^*+ O_P(\|\hat\beta^{(-k)}-\beta^*\|_{\tilde C}N^{-1/2}).
\end{align*}
Hence, $N^{-1}\sum_{i\in I_k}\hat\beta^{(-k)^\T} \hat V_i(Y_i-\hat\theta)=b^2+(\hat\beta^{(-k)}-\beta^*)^\T\tilde C\beta^*+ O_P(N^{-1/2})$. Combining all previous results,
\begin{align*}
\hat b^{2^{(k)}}&=b^2+\|\hat\beta^{(-k)}-\beta^*\|_{\tilde C}^2+2(\hat\beta^{(-k)}-\beta^*)^\T\tilde C\beta^*+2\{b^2+(\hat\beta^{(-k)}-\beta^*)^\T\tilde C\beta^*\}\\
&\qquad-2\{b^2+\|\hat\beta^{(-k)}-\beta^*\|_{\tilde C}^2+2(\hat\beta^{(-k)}-\beta^*)^\T\tilde C\beta^*\}+ O_P(N^{-1/2})\\
&=b^2+ O_P(\|\hat\beta^{(-k)}-\beta^*\|_{\tilde C}^2+N^{-1/2}).
\end{align*}
When $K<\infty$, $\hat b^2=b^2+ O_P(\|\hat\beta^{(-k)}-\beta^*\|_{\tilde C}^2+N^{-1/2})=b^2+ O_P(\|\hat\beta^{(-k)}-\beta^*\|_2^2+n^{-1/2})$.
\end{proof}

\begin{proof}[of Theorem~\ref{c33}] This proof provides an asymptotic normal result for $ n^{1/2}(\hat\sigma_Y^2-\sigma_Y^2)$ and a consistent estimate for the asymptotic variance.

{Part {1}.} We first show $ n^{1/2}(\hat\sigma_Y^2-\sigma_Y^2)\rightarrow N\left(0,\mathrm{var}(\varepsilon^2+2{ \ {\beta^*} }^\T\tilde V \varepsilon )+\tau\mathrm{var}({ \ {\beta^*} }^\T\tilde V )^2 \right)$. Recall from \eqref{p22},
\[
\hat\sigma_Y^{2^{(k)}}=N^{-1}\sum_{i\in I_k} (Y_i-\theta)^2-N^{-1}\sum_{i\in I_k}({ \ { \hat\beta^{(-k)}} }^\T\tilde V_i)^2+M^{-1}\sum_{i\in J_k}({ \ { \hat\beta^{(-k)}} }^\T\tilde V_i)^2+ O_P(N^{-1}).
\]
By \eqref{line2} and \eqref{line3},
\[
M^{-1}\sum_{i\in J_k}({ \ { \hat\beta^{(-k)}} }^\T\tilde V_i)^2=M^{-1}\sum_{i\in J_k}({\ \beta^*}^\T\tilde V_i)^2+\|\hat\beta^{(-k)}-\beta^*\|_{\tilde C}+2(\hat\beta^{(-k)}-\beta^*)^\T\tilde C\beta^*+o_P(M^{-1/2}).
\]
Similarly,
\[
N^{-1}\sum_{i\in I_k}({ \ { \hat\beta^{(-k)}} }^\T\tilde V_i)^2=N^{-1}\sum_{i\in I_k}({\ \beta^*}^\T\tilde V_i)^2+\|\hat\beta^{(-k)}-\beta^*\|_{\tilde C}+2(\hat\beta^{(-k)}-\beta^*)^\T\tilde C\beta^*+o_P(N^{-1/2}).
\]
Hence,
\[
\hat\sigma_Y^{2^{(k)}}=M^{-1}\sum_{i\in J_k}({\ \beta^*}^\T\tilde V_i)^2+N^{-1}\sum_{i\in I_k} (Y_i-\theta)^2-N^{-1}\sum_{i\in I_k}({\ \beta^*}^\T\tilde V_i)^2+o_P(N^{-1/2}).
\]
When $K<\infty$, by the independency between $(Y_i,X_i)_{i=1}^n$ and $\{X_i\}_{i=n+1}^{m+n}$, and similarly as \eqref{p31},
\begin{align}
 n^{1/2}(\hat\sigma_Y^2-\sigma_Y^2)&=n^{-1/2}\sum_{i\in I_k}(\varepsilon_i^2+2\varepsilon_i{ \ { \beta^*} }^\T\tilde V_i-\sigma_\varepsilon^2)
 \\
 &\  \ \ \  +n^{1/2}(m+n)^{-1}\sum_{i=1}^{m+n}\left\{({ \ { \beta^*} }^\T\tilde V_i)^2-b^2\right\}+o_P(1)\label{result:thm5}\\
&\rightarrow N\left\{0,\mathrm{var}(\varepsilon^2+2{ \ {\beta^*} }^\T\tilde V \varepsilon )+\tau\mathrm{var}({ \ {\beta^*} }^\T\tilde V )^2 \right\},\nonumber
\end{align}
in distribution, provided that $\mathrm{var}(\varepsilon^2+2{ \ {\beta^*} }^\T\tilde V \varepsilon )+\tau\mathrm{var}({ \ {\beta^*} }^\T\tilde V )^2>0$.

{Part {2}.} Now we prove the consistency of $\hat\sigma_\nu^2+n\hat\sigma_\xi^2/(m+n)$. It suffices to show 
$$\hat\sigma_\nu^2= E (\varepsilon^2+2{\beta^*}^\T\tilde V \varepsilon-\sigma_\varepsilon^2)^2+o_P(1)$$ and $\hat\sigma_\xi^2= E \{{\ \beta^*}^\T(\tilde V\tilde V^\T-\tilde C)\beta^*\}^2+o_P(1)$. Recall \eqref{xiv}, $\xi_i ^{(k)}= { \ { \hat\beta^{(-k)} } }^\T\left( \hat V_i \hat V_i^\T - \hat C^{(k)}    \right)\hat\beta^{(-k)}$. Now define $
\xi_i={\ \beta^*}^\T(\tilde V_i\tilde V_i^\T-\tilde C)\beta^*$. Observe that by algebraic manipulation followed by a Cauchy - Schwarz inequality
\begin{align}
 \left|N^{-1}\sum_{i\in I_k}\xi_i^{(k)^2}-N^{-1}\sum_{i\in I_k}\xi_i^2\right| 
& =\left|N^{-1}\sum_{i\in I_k}(\xi_i^{(k)}-\xi_i)^2+2N^{-1}\sum_{i\in I_k}\xi_i(\xi_i^{(k)}-\xi_i)\right|\nonumber\\
\leq\left|N^{-1}\sum_{i\in I_k}(\xi_i^{(k)}-\xi_i)^2\right|+&2\left\{N^{-1}\sum_{i\in I_k}\xi_i^2 N^{-1}\sum_{i\in I_k}(\xi_i^{(k)}-\xi_i)^2\right\}^{1/2}.\label{xi0}
\end{align}
By Condition \ref{a6}, $ E |{\ \beta^*}^\T(\tilde V\tilde V^\T-\tilde C)\beta^*|^{2+c}<c_1$. By Lemma \ref{l3}, $N^{-1}\sum_{i\in I_k}\xi_i^2= E\{{\ \beta^*}^\T(\tilde V\tilde V^\T-\tilde C)\beta^*\}^2+o_P(1)$. Now, we need to prove $N^{-1}\sum_{i\in I_k}(\xi_i^{(k)}-\xi_i)^2=o_P(1)$. Observe that
\[
\xi_i^{(k)}-\xi_i=\left\{({\ {\hat\beta^{(-k)}}}^\T\hat V_i)^2-({\ \beta^*}^\T\tilde V_i)^2\right\}-\left[{\ {\hat\beta^{(-k)}}}^\T\hat C^{(k)}\hat\beta^{(-k)}-b^2\right].
\]
It suffices to show
\[
N^{-1}\sum_{i\in I_k}\left\{({\ {\hat\beta^{(-k)}}}^\T\hat V)^2-({\ \beta^*}^\T\tilde V)^2\right\}^2=o_P(1),\quad{\ {\hat\beta^{(-k)}}}^\T\hat C^{(k)}\hat\beta^{(-k)}-b^2=o_P(1).
\]
By \eqref{bcb1} and \eqref{bcb2}, we can see that ${\ {\hat\beta^{(-k)}}}^\T\hat C^{(k)}\hat\beta^{(-k)}-b^2=o_P(1)$. 

New, we consider $a_i={\ \beta^*}^\T\tilde V_i$ and $\Delta_i={\ {\hat\beta^{(-k)}}}^\T\hat V_i-{\ \beta^*}^\T\tilde V_i$.  Then
\begin{align*}
N^{-1}\sum_{i\in I_k}\left[({\ {\hat\beta^{(-k)}}}^\T\hat V_i)^2-({\ \beta^*}^\T\tilde V_i)^2\right]^2&=N^{-1}\sum_{i\in I_k}\Delta_i^2(2a_i+\Delta_i)^2
\\
=N^{-1}\sum_{i\in I_k}(\Delta_i^4+4a_i\Delta_i^3+4a_i^2\Delta_i^2).
\end{align*}
By Condition \ref{a6} and \ref{a4}, $N^{-1}\sum_{i\in I_k}a_i^4= E ({\ \beta^*}^\T\tilde V)^4+o_P(1)$ with $ E ({\ \beta^*}^\T\tilde V)^4<c_1$ and by the fact that $\{(a+b+c)/3\}^4\leq(a^4+b^4+c^4)/3$,
\begin{align}
&N^{-1}\sum_{i\in I_k}\Delta_i^4\\
&=N^{-1}\sum_{i\in I_k}\left\{(\hat\beta^{(-k)}-\beta^*)^\T\tilde V_i-{\ \beta^*}^\T(\hat\mu^{(k)}-\tilde\mu)-(\hat\beta^{(-k)}-\beta^*)^\T(\hat\mu^{(k)}-\tilde\mu)\right\}^4\nonumber\\
&\leq27N^{-1}\sum_{i\in I_k}\left[\left\{(\hat\beta^{(-k)}-\beta^*)^\T\tilde V_i\right\}^4+27\left\{{\ \beta^*}^\T(\hat\mu^{(k)}-\tilde\mu)\right\}^4+27\left\{(\hat\beta^{(-k)}-\beta^*)^\T(\hat\mu^{(k)}-\tilde\mu)\right\}^4\right]\nonumber\\
&=o_P(1),\label{Delta4}
\end{align}
here $N^{-1}\sum_{i\in I_k}\{(\hat\beta^{(-k)}-\beta^*)^\T\tilde V_i\}^4=o(1)$ results from $ E _{I_k^c}\{(\hat\beta^{(-k)}-\beta^*)^\T\tilde V\}^4=o(1)$ and Lemma \ref{l1}. Hence, by Holder's Inequality,
\[
N^{-1}\sum_{i\in I_k}\left\{({\ {\hat\beta^{(-k)}}}^\T\hat V_i)^2-({\ \beta^*}^\T\tilde V_i)^2\right\}^2=o_P(1).
\]
Therefore, $N^{-1}\sum_{i\in I_k}(\xi_i^{(k)}-\xi_i)^2=o_P(1)$ and
\[
N^{-1}\sum_{i\in I_k}\xi_i^{(k)^2}=N^{-1}\sum_{i\in I_k}\xi_i^2+o_P(1)= E \{{\ \beta^*}^\T(\tilde V\tilde V^\T-\tilde C)\beta^*\}^2+o_P(1).
\]
When $K<\infty$,
\[
\hat\sigma_\xi^2=K^{-1}\sum_{k=1}^KN^{-1}\sum_{i\in I_k}\xi_i^{(k)^2}= E \{{\ \beta^*}^\T(\tilde V\tilde V^\T-\tilde C)\beta^*\}^2+o_P(1).
\]
To show $\hat\sigma_\nu^2= E (\varepsilon^2+2{\beta^*}^\T\tilde V \varepsilon-\sigma_\varepsilon^2)^2+o_P(1)$, recall \eqref{nuv}, $\nu_i^{(k)}=\hat \varepsilon_i^2 + 2 \hat\beta^{(-k)^\T} \hat V_i\hat\varepsilon_i + { \ { \hat\beta^{(-k)}
 } }^\T  \hat C^{(k)}  \hat\beta^{(-k)}-\hat\sigma_Y^2,$ where $\hat\varepsilon_i=Y_i-\hat\theta-\hat\beta^{(-k)^\T} \hat V_i$. Define $\nu_i=\varepsilon_i^2+2{\beta^*}^\T\tilde V_i\varepsilon_i-\sigma_\varepsilon^2$. Similarly as in \eqref{xi0},
\begin{align*}
&\left|N^{-1}\sum_{i\in I_k}\nu_i^{(k)^2}-N^{-1}\sum_{i\in I_k}\nu_i^2\right|\\
\leq&\left|N^{-1}\sum_{i\in I_k}(\nu_i^{(k)}-\nu_i)^2\right|+2\left\{N^{-1}\sum_{i\in I_k}\nu_i^2  N^{-1}\sum_{i\in I_k}(\nu_i^{(k)}-\nu_i)^2\right\}^{1/2}.
\end{align*}
By Condition \ref{a6}, $E |\varepsilon^2+2{\beta^*}^\T\tilde V \varepsilon-\sigma_\varepsilon^2|^{2+c}<c_1$,
and by Lemma \ref{l3},
\[
N^{-1}\sum_{i\in I_k}\nu_i^2= E \left(\varepsilon^2+2{\beta^*}^\T\tilde V \varepsilon-\sigma_\varepsilon^2\right)^2+o_P(1).
\]
Now it remains to prove $N^{-1}\sum_{i\in I_k}(\nu_i^{(k)}-\nu_i)^2=o_P(1)$. It suffices to show
\begin{align}
&N^{-1}\sum_{i\in I_k}(\hat\varepsilon_i^2-\varepsilon_i^2)^2=o_P(1),\label{nu1}\\
&N^{-1}\sum_{i\in I_k}(\hat\beta^{(-k)^\T} \hat V_i\hat\varepsilon_i-{\beta^*}^\T\tilde V_i\varepsilon_i)^2=o_P(1),\label{nu2}\\
&{ \ { \hat\beta^{(-k)}
 } }^\T  \hat C^{(k)}  \hat\beta^{(-k)}-\hat\sigma_Y^2+\sigma_\varepsilon^2=o_P(1).\label{nu3}
\end{align}
Recall that from \eqref{hsigmaY2}, \eqref{bcb1} and \eqref{bcb2}, we have ${\ {\hat\beta^{(-k)}}}^\T\hat C^{(k)}\hat\beta^{(-k)}-b^2=o_P(1),\ \hat\sigma_Y^2=\sigma_Y^2+o_P(1)$ and hence \eqref{nu3} holds. As for \eqref{nu1},
\begin{equation}\label{Holder1}
N^{-1}\sum_{i\in I_k}(\hat\varepsilon_i^2-\varepsilon_i^2)^2=N^{-1}\sum_{i\in I_k}(\hat\varepsilon_i-\varepsilon_i)^2\{(\hat\varepsilon_i-\varepsilon_i)+2\varepsilon_i\}^2.
\end{equation}
By Condition \ref{a6}, $ E |\varepsilon|^{4+c}<c_1$. By Lemma \ref{l3},
\begin{equation}\label{Holder2}
N^{-1}\sum_{i\in I_k}\varepsilon_i^4= E (\varepsilon^4)+o_P(1)
\end{equation}
with $ E (\varepsilon^4)<c_1$. Besides,
\begin{align}
N^{-1}\sum_{i\in I_k}(\hat\varepsilon_i-\varepsilon_i)^4&=N^{-1}\sum_{i\in I_k}(\hat\theta-\theta+\hat\beta^{(-k)^\T} \hat V_i-{\ \beta^*}^\T\tilde V_i)^4\nonumber\\
&\leq 8(\hat\theta-\theta)^4+8N^{-1}\sum_{i\in I_k}(\hat\beta^{(-k)^\T} \hat V_i-{\ \beta^*}^\T\tilde V_i)^4=o_P(1),\label{Holder3}
\end{align}
where the last equality results from \eqref{p24} and \eqref{Delta4}. By \eqref{Holder1}, \eqref{Holder2}, \eqref{Holder3} and Holder's Inequality, \eqref{nu1} holds. Now, for \eqref{nu2},
\begin{align*}
&N^{-1}\sum_{i\in I_k}(\hat\beta^{(-k)^\T} \hat V_i\hat\varepsilon_i-{\beta^*}^\T\tilde V_i\varepsilon_i)^2\\
&=N^{-1}\sum_{i\in I_k}\biggr\{(\hat\beta^{(-k)^\T} \hat V_i-{\ \beta^*}^\T\tilde V_i)\varepsilon_i+{\ \beta^*}^\T\tilde V_i(\hat\varepsilon_i-\varepsilon_i)+(\hat\beta^{(-k)^\T} \hat V_i-{\ \beta^*}^\T\tilde V_i)(\hat\varepsilon_i-\varepsilon_i)\biggr\}^2.
\end{align*}
Since
\begin{align*}
N^{-1}\sum_{i\in I_k}\varepsilon_i^4= E \varepsilon^4+o_P(1),\quad N^{-1}\sum_{i\in I_k}({\ \beta^*}^\T\tilde V_i)^4= E ({\ \beta^*}^\T\tilde V)^4+o_P(1),\\
N^{-1}\sum_{i\in I_k}(\hat\varepsilon_i-\varepsilon_i)^4=o_P(1),\quad N^{-1}\sum_{i\in I_k}(\hat\beta^{(-k)^\T} \hat V_i-{\ \beta^*}^\T\tilde V_i)^4=o_P(1),
\end{align*}
by Holder's Inequality, \eqref{nu2} holds. Now combining \eqref{nu1}, \eqref{nu2} and \eqref{nu3}, we have $N^{-1}\sum_{i\in I_k}(\nu_i^{(k)}-\nu_i)^2=o_P(1)$ and hence
\[
N^{-1}\sum_{i\in I_k}\nu_i^{(k)^2}=N^{-1}\sum_{i\in I_k}\nu_i^2+o_P(1)= E \left(\varepsilon^2+2{\beta^*}^\T\tilde V \varepsilon-\sigma_\varepsilon^2\right)^2+o_P(1).
\]
When $K<\infty$,
\[
\hat\sigma_\nu^2=K^{-1}\sum_{k=1}^KN^{-1}\sum_{i\in I_k}\nu_i^{(k)^2}= E \left(\varepsilon^2+2{\beta^*}^\T\tilde V \varepsilon-\sigma_\varepsilon^2\right)^2+o_P(1).
\]
Therefore, $\hat\sigma_\nu^2+n\hat\sigma_\xi^2/(m+n) = \mathrm{var}(\varepsilon ^2+2{ \ {\beta^*} }^\T\tilde V \varepsilon )+\tau\mathrm{var}({ \ {\beta^*} }^\T\tilde V )^2 + o_P(1)$.
\end{proof}

\begin{proof}[of Corollary~\ref{cor}]
When Conditions \ref{a2} and \ref{a4} hold, we have consistency results \eqref{hsigmaY2k} and \eqref{hb2k}. By Slutsky's Theorem, for each $k\leq K$,
\[
\hat b^{2^{(k)}}/\hat\sigma_Y^{2^{(k)}}=b^2/\sigma_Y^2+o_p(1)=PVE+o_p(1)
\]
and hence $R^2=PVE+o_p(1)$. 

The asymptotic normality result holds as a consequence of Theorem \ref{genv}.
\end{proof}

\begin{proof}[of Theorem~\ref{gen}]
 
This proof provides an asymptotic normal result for $n^{1/2}(\hat\theta_\mathrm{gen}-\theta)$ and the consistency of the asymptotic variance estimator. We first show the asymptotic normality. Let 
\[
\hat\theta_\mathrm{gen}^{(k)}=M^{-1}\sum_{i\in J_k}\hat g^{(-k)}(X_i)+N^{-1}\sum_{i\in I_k}\left\{Y_i-\hat g^{(-k)}(X_i)\right\},
\]
where $M=(m+n)/K$ and $N=n/K$, then $\hat\theta_\mathrm{gen}=K^{-1}\sum_{k=1}^K\hat\theta_\mathrm{gen}^{(k)}$. Observe that
\begin{align*}
\hat\theta_\mathrm{gen}^{(k)}&=M^{-1}\sum_{i\in J_k}g^*(X_i)+N^{-1}\sum_{i\in I_k}\varepsilon_i+M^{-1}\sum_{i\in J_k}\left\{\hat g^{(-k)}(X_i)-g^*(X_i)\right\}\\
&\qquad-N^{-1}\sum_{i\in I_k}\left\{\hat g^{(-k)}(X_i)-g^*(X_i)\right\}.
\end{align*}
By Lemma \ref{l1},
\begin{align*}
&M^{-1}\sum_{i\in J_k}\left\{\hat g^{(-k)}(X_i)-g^*(X_i)\right\}=o_P(M^{-1/2}),\\
&N^{-1}\sum_{i\in I_k}\left\{\hat g^{(-k)}(X_i)-g^*(X_i)\right\}=o_P(N^{-1/2}).
\end{align*}
Hence, when $K<\infty$,
\[
\hat\theta_\mathrm{gen}=(m+n)^{-1}\sum_{i=1}^{m+n}g^*(X_i)+n^{-1}\sum_{i=1}^n\varepsilon_i+o_P(n^{-1/2}).
\]
By Lindeberg-Feller Central Limit Theorem, as $m,n,p\to\infty$,
\[
\frac{n^{1/2}\left\{(m+n)^{-1}\sum_{i=1}^{m+n}g^*(X_i)+n^{-1}\sum_{i=1}^n\varepsilon_i\right\}}{\mathrm{var}(\varepsilon)+\tau\mathrm{var}\{g^*(X)\}}\to N(0,1)
\]
and hence
\[
\frac{n^{1/2}(\hat\theta_\mathrm{gen}-\theta)}{\sigma_{\varepsilon,\mathrm{gen}}^2+\frac{n}{m+n} b_\mathrm{gen}^2} \to  N (0,1).
\]
Now, we showcase that $\hat b_\mathrm{gen}^2$ and $\hat\sigma_{\varepsilon,\mathrm{gen}}^2$ are consistent estimators of $b_\mathrm{gen}^2$ and $\sigma_{\varepsilon,\mathrm{gen}}^2$, respectively. For $k\in K$ and $i\in I_k$, let 
\[
\nu_{\varepsilon,\mathrm{gen},i}=Y_i-\hat\theta_\mathrm{gen}-\hat g^{(-k)}(X_i)+M^{-1}\sum_{i\in J_k}\hat g^{(-k)}(X_i).
\]
Then,
\begin{align*}
&\left|N^{-1}\sum_{i\in I_k}\nu_{\varepsilon,\mathrm{gen},i}^2-N^{-1}\sum_{i\in I_k}\varepsilon_i^2\right|\\
&\leq  N^{-1}\sum_{i\in I_k}(\nu_{\varepsilon,\mathrm{gen},i}-\varepsilon_i)^2+2\left\{N^{-1}\sum_{i\in I_k}\varepsilon_i^2N^{-1}\sum_{i\in I_k}(\nu_{\varepsilon,\mathrm{gen},i}-\varepsilon_i)^2\right\}^{1/2},
\end{align*}
where $N^{-1}\sum_{i\in I_k}\varepsilon_i^2=\sigma_{\varepsilon,\mathrm{gen}}^2+o_P(1)$. Besides,
\[
\nu_{\varepsilon,\mathrm{gen},i}-\varepsilon_i=-(\hat\theta_\mathrm{gen}-\theta)-\left\{\hat g^{(-k)}(X_i)-M^{-1}\sum_{i\in J_k}\hat g^{(-k)}(X_i)-g^*(X_i)+\theta\right\},
\]
where $\hat\theta_\mathrm{gen}-\theta=o_P(1)$ and by Lemma \ref{l1},
\[
\sum_{i\in I_k}\left\{\hat g^{(-k)}(X_i)-M^{-1}\sum_{i\in J_k}\hat g^{(-k)}(X_i)-g^*(X_i)+\theta\right\}=o_P(1).
\]
Hence,
\[
N^{-1}\sum_{i\in I_k}\nu_{\varepsilon,\mathrm{gen},i}^2=N^{-1}\sum_{i\in I_k}\varepsilon_i^2+o_P(1)=\sigma_{\varepsilon,\mathrm{gen}}^2+o_P(1)
\]
and therefore, $\sigma_{\varepsilon,\mathrm{gen}}^2=\sigma_{\varepsilon,\mathrm{gen}}^2+o_P(1)$. Similarly, $b_\mathrm{gen}^2=b_\mathrm{gen}^2+o_P(1)$.
\end{proof}

\begin{proof}[of Theorem~\ref{t42}]

{Part {1}.} We first provide consistency rates of $\hat\delta$ and an asymptotic normal result for $ n^{1/2}(\hat\delta-\delta)$ when some specific rates are satisfied. Recall that, by definition,
\[
Y=D{\ \beta_1^*}^\T\tilde X+(1-D){\ \beta_0^*}^\T\tilde X+\varepsilon,\quad D=e(X)+\zeta,\quad E (\zeta\mid X)=0.
\]
By the definitions of $\beta_1^*$ and $\beta_0^*$ and Lemma 2.1 in \cite{zhang2019semi},
\begin{align*}
&E(Y-{\ \beta_1^*}^\T\tilde X\mid D=1)=0,\ E(Y-{\ \beta_0^*}^\T\tilde X\mid D=0)=0,\\
&E\{(Y-{\ \beta_1^*}^\T\tilde X)\tilde X\mid D=1\}=0,\ E\{(Y-{\ \beta_0^*}^\T\tilde X)\tilde X\mid D=0\}=0.
\end{align*}
Hence, $E(D{\ \beta_1^*}^\T\tilde X\varepsilon)=E\{{\ \beta_1^*}^\T\tilde X(Y-{\ \beta_1^*}^\T\tilde X)\mid D=1\}E(D)=0$. Similarly, $E\{(1-D){\ \beta_0^*}^\T\tilde X\varepsilon\}=0$. Besides,
\begin{align*}
E(\varepsilon)&=E\{Y-D{\ \beta_1^*}^\T\tilde X-(1-D){\ \beta_0^*}^\T\tilde X\}\\
&=E(Y-{\ \beta_1^*}^\T\tilde X\mid D=1)E(D)-E(Y-{\ \beta_0^*}^\T\tilde X\mid D=0)\{1-E(D)\}=0.
\end{align*}
Therefore,
\[
 E(Y^2)= E\{D({\ \beta_1^*}^\T\tilde X)^2\}+ E\{(1-D)({\ \beta_0^*}^\T\tilde X)^2\}+\sigma_\varepsilon^2,
\]
where $\sigma_\varepsilon=\mathrm{var}(\varepsilon)$. Since $\mathrm{pr} \{c\leq e(X)\leq1-c\}=1$,
\begin{equation}\label{beta1V}
 E ({\ \beta_1^*}^\T\tilde V)^2\leq E ({\ \beta_1^*}^\T\tilde X)^2\leq c^{-1} E\{e(X)({\ \beta_1^*}^\T\tilde X)^2\}=c^{-1} E\{D({\ \beta_1^*}^\T\tilde X)^2\}\leq c^{-1} E(Y^2).
\end{equation}
Let
\[
r_i^{(-k)} = {D_i}/{ \hat e^{(-k)}(X_i)},\ r^{(-k)} = {D}/{ \hat e^{(-k)}(X)},\ r_i = {D_i}/{e(X_i)},\ r=D/e(X).
\]
Since both $e(X)$ and $\hat e^{(-k)}(X)$ are bounded away from 0 uniformly with probability 1,
\begin{equation}\label{rk}
 E _{I_k^c}(r^{(-k)}-r)^2= E _{I_k^c}\frac{D\{\hat e^{(-k)}(X)-e(X)\}^2}{\{\hat e^{(-k)}(X)e(X)\}^2}= O_P(b_{m+n,p}^2).
\end{equation}
Here, recall that $ E _{I_k^c}g= E \{g\mid (D_i,Y_i,X_i)_{i\in\{1,2,\dots,n\}\setminus I_k}\}$ and $(D,Y,X)\sim P_{D,Y,X}$ independent of $(D_i,Y_i,X_i)_{i\in\{1,2,\dots,n\}\setminus I_k}$. By the definition of $\hat\tau_1^{(k)}$, we can obtain the following formula
\begin{align}\
\hat\tau_1^{(k)}&={\ \beta_1^*}^\T\hat\mu^{(k)} + N^{-1} \sum_{i\in I_k} r_i (Y_i-{\ {\beta_1^*} }^\T\tilde X_i)+(\hat\beta_1^{(-k)}-\beta_1^*)^\T\hat\mu^{(k)} \nonumber\\
&\qquad+N^{-1}\sum_{i\in I_k}(r_i^{(-k)}-r_i)(Y_i-{\ {\beta_1^*} }^\T\tilde X_i)\ - N^{-1}\sum_{i\in I_k}r_i(\hat\beta_1^{(-k)}-\beta_1^*)^\T\tilde X_i \nonumber
\\
&\qquad-N^{-1}\sum_{i\in I_k}(r_i^{(-k)}-r_i)(\hat\beta_1^{(-k)}-\beta_1^*)^\T\tilde X_i,\label{sth2}
\end{align}
where recall that $\hat\mu^{(k)}=M^{-1}\sum_{i\in J_k}\tilde X_i$.
Observe that each term of the RHS in \eqref{sth2} is an average of (conditional) independent and identically distributed random variables. Hence, by Lemma \ref{l2}, we can obtain the rates of each of the terms by looking at the first and second moments. For the first moments, recall that $r=D/e(X)$ and $E(r\mid X)=E(D\mid X)/e(X)=1$, we have
\begin{align*}
&E({\beta_1^*}^\T\tilde X)={\beta_1^*}^\T\tilde\mu,\\
&E\{r(Y-{\ \beta_1^*}^\T\tilde X)\}=\tau_1-{\ \beta_1^*}^\T\tilde\mu,\\
&E\{(\hat\beta_1^{(-k)}-\beta_1^*)^\T\tilde X\}=(\hat\beta_1^{(-k)}-\beta_1^*)^\T\tilde\mu,\\
&E_{I_k^c}\{r(\hat\beta_1^{(-k)}-\beta_1^*)^\T\tilde X\}=(\hat\beta_1^{(-k)}-\beta_1^*)^\T\tilde\mu,
\end{align*}
and by the Holder's inequality,
\begin{align}
&E_{I_k^c}\{(r^{(-k)}-r)(Y-{\ \beta_1^*}^\T\tilde X)\}=E_{I_k^c}\{(r^{(-k)}-r)(E(Y\mid X)-{\ \beta_1^*}^\T\tilde X)\}=O_P(b_{m+n,p}c_p),\label{contribution1}\\
&E _{I_k^c}\{(r^{(-k)}-r)(\hat\beta_1^{(-k)}-\beta_1^*)^\T\tilde X\}=O_P(a_{n,p}b_{m+n,p}).\label{contribution2}
\end{align}
We can see that the terms ${\beta_1^*}^\T\tilde\mu$ and $(\hat\beta_1^{(-k)}-\beta_1^*)^\T\tilde\mu$ will cancel out, and the terms \eqref{contribution1} and \eqref{contribution2} will be the main contributions of the first moment.

As for the second moments, we have
\begin{align}
&\mathrm{var}({\beta_1^*}^\T\tilde X)=E({\beta_1^*}^\T\tilde V)^2\leq c^{-1}E(Y^2)=O(1)\label{beta1X},\\
&E _{I_k^c}\{(\hat\beta_1^{(-k)}-\beta_1^*)^\T\tilde V\}^2\leq E _{I_k^c}\{(\hat\beta_1^{(-k)}-\beta_1^*)^\T\tilde X\}^2=O_P(a_{n,p}^2),\nonumber
\end{align}
where \eqref{beta1X} results from \eqref{beta1V}. By Condition \ref{cond:4}, $r=D/e(X)\leq c^{-1}$ and $|r^{(-k)}-r|=|D/\hat e^{(-k)}(X)-D/e(X)|\leq c^{-1}$ with probability 1. Hence, we have following results for the variance (or second moments) of the terms in \eqref{sth2},
\begin{align}
&E\{r^2(Y-{\ \beta_1^*}^\T\tilde X)^2\}\leq c^{-2}\sigma_\varepsilon^2=O(1),\nonumber\\
&E _{I_k^c}r^2\{(\hat\beta_1^{(-k)}-\beta_1^*)^\T\tilde V\}^2\leq c_1^{-2}E _{I_k^c}\{(\hat\beta_1^{(-k)}-\beta_1^*)^\T\tilde X\}^2=O_P(a_{n,p}^2),\label{A12}\\
&E _{I_k^c}(r^{(-k)}-r)^2\{(\hat\beta_1^{(-k)}-\beta_1^*)^\T\tilde X\}^2\leq c_1^{-2}E _{I_k^c}\{(\hat\beta_1^{(-k)}-\beta_1^*)^\T\tilde X\}^2=O_P(a_{n,p}^2).\label{A13}
\end{align}
Besides, by the assumption that $\mathrm{pr}\{E (\varepsilon^2\mid X)<C\}=1$, we have
\begin{equation}\label{A11}
E _{I_k^c}(r^{(-k)}-r)^2(Y-{\ {\beta_1^*} }^\T\tilde X)^2\leq CE _{I_k^c}(r^{(-k)}-r)^2=O_P(b_{m+n,p}^2).
\end{equation}
Now, by Lemma \ref{l2}, we have asymptotic results for each of the terms in \eqref{sth2}. The terms ${\ \beta_1^*}^\T\hat\mu^{(k)}$ and $(\hat\beta_1^{(-k)}-\beta_1^*)^\T\hat\mu^{(k)}$ are averages of $M$ (conditional) independent and identically distributed random variables, we have
\begin{align*}
{\ \beta_1^*}^\T\hat\mu^{(k)}&={\ \beta_1^*}^\T\tilde\mu+ O_P(M^{-1/2}),\\
(\hat\beta_1^{(-k)}-\beta_1^*)^\T\hat\mu^{(k)}&=(\hat\beta_1^{(-k)}-\beta_1^*)^\T\tilde\mu+O_P(a_{n,p}M^{-1/2}).
\end{align*}
The other terms in \eqref{sth2} are averages of $N$ (conditional) independent and identically distributed random variables, we have
\begin{align*}
N^{-1} \sum_{i\in I_k} r_i (Y_i-{\ {\beta_1^*} }^\T\tilde X_i)&=\tau_1-{\ \beta_1^*}^\T\tilde\mu+O_P(N^{-1/2}),\\
N^{-1}\sum_{i\in I_k}r_i(\hat\beta_1^{(-k)}-\beta_1^*)^\T\tilde X_i&=(\hat\beta_1^{(-k)}-\beta_1^*)^\T\tilde\mu+O_P(a_{n,p}N^{-1/2}),\\
\end{align*}
and
\begin{align*}
N^{-1}\sum_{i\in I_k}(r_i^{(-k)}-r_i)(Y_i-{\ {\beta_1^*} }^\T\tilde X_i)&=O_P(b_{m+n,p}c_p+b_{m+n,p}N^{-1/2}),\\
N^{-1}\sum_{i\in I_k}(r^{(-k)}-r)(\hat\beta_1^{(-k)}-\beta_1^*)^\T\tilde X&=O_P(a_{n,p}b_{m+n,p}+a_{n,p}N^{-1/2}).
\end{align*} 
Combining the previous results, we have
\begin{align*}
\hat\tau_1^{(k)}&=\tau_1+ O_P(a_{n,p}b_{m+n,p}+b_{m+n,p}c_p+(1+a_{n,p}+b_{m+n,p})N^{-1/2}),
\end{align*}
Similarly,
\[
\hat\tau_2^{(k)}=\tau_2+ O_P(a_{n,p}b_{m+n,p}+b_{m+n,p}c_p+(1+a_{n,p}+b_{m+n,p})N^{-1/2}).
\]
When $K<\infty$, $a_{n,p}= O(1)$, $a_{n,p}b_{m+n,p}= O(n^{-1/2})$ and $b_{m+n,p}c_p= O(n^{-1/2})$,
\begin{align}
\hat\delta&=\hat\tau_1-\hat\tau_2=\delta+ O_P(a_{n,p}b_{m+n,p}+b_{m+n,p}c_p+(1+a_{n,p}+b_{m+n,p})n^{-1/2})\nonumber\\
&=\delta+ O_P(n^{-1/2}).\label{cons:delta}
\end{align}

Moreover, if $a_{n,p}=o_P(1),\ b_{m+n,p}=o_P(1),\ a_{n,p}b_{m+n,p}=o_P(n^{-1/2})$ and $b_{m+n,p}c_p=o_P(n^{-1/2})$. Then, by the previous results and Lindeberg-Feller Central Limit Theorem,
\begin{align}
 n^{1/2}(\hat\delta-\delta)&= n^{1/2}(m+n)^{-1}\sum_{i=1}^{m+n}(\beta_1^*-\beta_0^*)^\T\tilde V_i+n^{-1/2} \sum_{i=1}^n\varepsilon_i\zeta_i/[e(X_i)\{1-e(X_i)\}]\nonumber\\
 &\qquad-E[\varepsilon\zeta/e(X)\{1-e(X)\}]+o_P(1)\label{result:thm6}\\
&\rightarrow N(0,V_\delta),\nonumber
\end{align}
in distribution, provided that
\[
V_\delta=\mathrm{var} \left[\frac{\varepsilon \zeta }{e(X)\{1-e(X)\}}\right]+\tau(\beta_1^*-\beta_0^*)^\T\tilde C(\beta_1^*-\beta_0^*)>c>0.
\]
\vskip 10pt 

{Part {2}.} Now we provide a consistency result for $\hat V_\delta$. Recall the definition of $\nu_{\delta,i}$,
\[
\nu_{\delta,i}= r_i^{(-k)}(Y_i-\hat\beta_1^{(-k)^\T}\tilde X_i) - \rho_i^{(-k)} (Y_i-\hat\beta_0^{(-k)^\T}\tilde X_i)  -\hat\delta +(\hat\beta_1^{(-k)}-\hat\beta_0^{(-k)})^\T\hat\mu^{(k)},
\]
where $r_i^{(-k)}=D_i/\hat e^{(-k)}(X_i)$ and $\rho_i^{(-k)}=(1-D_i)/\{1-\hat e^{(-k)}(X_i)\}$. Define $\nu_{\delta,i}^*= r_i(Y_i-{\ \beta_1^*}^\T\tilde X_i) - \rho_i(Y_i-{\ \beta_0^*}^\T\tilde X_i)$, where $r_i=D_i/e(X_i)$ and $\rho_i=(1-D_i)/\{1-e(X_i)\}$. Similarly as in \eqref{xi0},
\begin{align*}
&\left|N^{-1}\sum_{i\in I_k}\nu_{\delta,i}^2-N^{-1}\sum_{i\in I_k}{\nu_{\delta,i}^*}^2\right|\\
\leq&\left|N^{-1}\sum_{i\in I_k}(\nu_{\delta,i}-\nu_{\delta,i}^*)^2\right|+2\left\{N^{-1}\sum_{i\in I_k}{\nu_{\delta,i}^*}^2 N^{-1}\sum_{i\in I_k}(\nu_{\delta,i}-\nu_{\delta,i}^*)^2\right\}^{1/2}.
\end{align*}
By Conditions \ref{a2} and \ref{cond:4}, $ E |r(Y-{\ \beta_1^*}^\T\tilde X) - \rho(Y-{\ \beta_0^*}^\T\tilde X)|^{2+c}<c_1$, where $r=D/e(X)$ and $\rho=(1-D)/\{1-e(X)\}$. By Lemma \ref{l2},
\[
N^{-1}\sum_{i\in I_k}{\nu_{\delta,i}^*}^2=V_1+o_P(1),
\]
where $V_1=\mathrm{var}\{r(Y-{\beta_1^*}^\T\tilde X)-\rho(Y-{\beta_0^*}^\T\tilde X)\}$. Now it remains to show $N^{-1}\sum_{i\in I_k}(\nu_{\delta,i}-\nu_{\delta,i}^*)^2=o_P(1)$. Observe that $\nu_{\delta,i}-\nu_{\delta,i}^*=A_{1,i}+A_{2,i}+A_3$, where
\begin{align*}
 A_{1,i}&=r_i^{(-k)}(Y_i-\hat\beta_1^{(-k)^\T}\tilde X_i)-r_i(Y_i-{\ \beta_1^*}^\T\tilde X_i),\\
 A_{2,i}&=-\rho_i^{(-k)} (Y_i-\hat\beta_0^{(-k)^\T}\tilde X_i)+\rho_i(Y_i-{\ \beta_0^*}^\T\tilde X_i),\\
 A_3&=(\hat\beta_1^{(-k)}-\hat\beta_0^{(-k)})^\T\hat\mu^{(k)}-\hat\delta.
\end{align*}
Hence, it suffices to show 
\[
N^{-1}\sum_{i\in I_k}A_{1,i}^2=o_P(1),\quad N^{-1}\sum_{i\in I_k}A_{2,i}^2=o_P(1),\quad A_3=o_P(1).
\]
Observe that
\[
A_{1,i}=(r_i^{(-k)}-r_i)\varepsilon_i-r_i(\hat\beta_1^{(-k)}-\beta_1^*)^\T\tilde X_i-(r_i^{(-k)}-r_i)(\hat\beta_1^{(-k)}-\beta_1^*)^\T\tilde X_i
\]
From \eqref{A11}, \eqref{A12} and \eqref{A13},
\begin{align*}
& E _{I_k^c}\{(r^{(-k)}-r)^2\varepsilon^2\}=o_P(1),\quad E _{I_k^c}[r^2\{(\hat\beta_1^{(-k)}-\beta_1^*)^\T\tilde X\}^2]=o_P(1),\\
& E _{I_k^c}[(r^{(-k)}-r)^2\{(\hat\beta_1^{(-k)}-\beta_1^*)^\T\tilde X\}^2]=o_P(1).
\end{align*}
Hence,
\[
 E _{I_k^c}\{(r^{(-k)}-r)\varepsilon-r(\hat\beta_1^{(-k)}-\beta_1^*)^\T\tilde X-(r^{(-k)}-r)(\hat\beta_1^{(-k)}-\beta_1^*)^\T\tilde X\}^2=o_P(1).
\]
By Lemma \ref{l1}, $N^{-1}\sum_{i\in I_k}A_{1,i}^2=o_P(1)$. Similarly, we have $N^{-1}\sum_{i\in I_k}A_{2,i}^2=o_P(1)$. Besides,
\begin{align*}
A_3&=(\hat\beta_1^{(-k)}-\beta_1^*)^\T\tilde\mu-(\hat\beta_0^{(-k)}-\beta_0^*)^\T\tilde\mu+(\beta_1^*-\beta_0^*)^\T(\hat\mu^{(k)}-\tilde\mu)\\
&\qquad+(\hat\beta_1^{(-k)}-\beta_1^*)^\T(\hat\mu^{(k)}-\tilde\mu)-(\hat\beta_0^{(-k)}-\beta_0^*)^\T(\hat\mu^{(k)}-\tilde\mu)-(\hat\delta-\delta).
\end{align*}
Under the condition $a_{n,p}=o(1)$,  we have $(\hat\beta_1^{(-k)}-\beta_1^*)^\T\tilde\mu=o_P(1)$ and $(\hat\beta_0^{(-k)}-\beta_0^*)^\T\tilde\mu=o_P(1)$. By Lemma \ref{l2}, $(\beta_1^*-\beta_0^*)^\T(\hat\mu^{(k)}-\tilde\mu)=o_P(1)$. By Lemma \ref{l1}, $(\hat\beta_1^{(-k)}-\beta_1^*)^\T(\hat\mu^{(k)}-\tilde\mu)=o_P(1)$ and $(\hat\beta_0^{(-k)}-\beta_0^*)^\T(\hat\mu^{(k)}-\tilde\mu)=o_P(1)$. Recall \eqref{cons:delta}, $\hat\delta-\delta=o_P(1)$. Therefore, $A_3=o_P(1)$. Now, combining all the previous results,
\begin{equation}\label{nudeltai}
N^{-1}\sum_{i\in I_k}(\nu_{\delta,i}-\nu_{\delta,i}^*)^2=o_P(1),
\end{equation}
and hence
\[
N^{-1}\sum_{i\in I_k}\nu_{\delta,i}^2=N^{-1}\sum_{i\in I_k}{\nu_{\delta,i}^*}^2+o_P(1)=V_1+o_P(1).
\]
Now recall $\xi_{\delta,i}=(\hat\beta_1^{(-k)}-\hat\beta_0^{(-k)})^\T(\tilde X_i-\hat\mu^{(k)})$. Define $\xi_{\delta,i}^*=(\beta_1^*-\beta_0^*)^\T\tilde V_i$. Similarly as in \eqref{xi0},
\begin{align*}
&\left|N^{-1}\sum_{i\in I_k}\xi_{\delta,i}^2-N^{-1}\sum_{i\in I_k}{\xi_{\delta,i}^*}^2\right|\\
\leq& \left|N^{-1}\sum_{i\in I_k}(\xi_{\delta,i}-\xi_{\delta,i}^*)^2\right|+2\left\{N^{-1}\sum_{i\in I_k}{\xi_{\delta,i}^*}^2\cdot N^{-1}\sum_{i\in I_k}(\xi_{\delta,i}-\xi_{\delta,i}^*)^2\right\}^{1/2},
\end{align*}
By Condition \ref{a2}, $ E |(\beta_1^*-\beta_0^*)^\T\tilde V|^{2+c}<c_1$. By Lemma \ref{l3},
\[
N^{-1}\sum_{i\in I_k}{\xi_{\delta,i}^*}^2=V_2+o_P(1),
\]
where $V_2=(\beta_1^*-\beta_0^*)^\T\tilde C(\beta_1^*-\beta_0^*)$. Now it remains to show $N^{-1}\sum_{i\in I_k}(\xi_{\delta,i}-\xi_{\delta,i}^*)^2=o_P(1)$. Observe that
\begin{align*}
\xi_{\delta,i}-\xi_{\delta,i}^*&=(\hat\beta_1^{(-k)}-\beta_1^*)^\T\tilde V_i-(\hat\beta_0^{(-k)}-\beta_0^*)^\T\tilde V_i\\
&\qquad-(\beta_1^*-\beta_0^*)^\T(\hat\mu^{(k)}-\tilde\mu)-(\hat\beta_1^{(-k)}-\beta_1^*)^\T(\hat\mu^{(k)}-\tilde\mu)+(\hat\beta_0^{(-k)}-\beta_0^*)^\T(\hat\mu^{(k)}-\tilde\mu).
\end{align*}
By Lemma \ref{l1}, $N^{-1}\sum_{i\in I_k}(\hat\beta_1^{(-k)}-\beta_1^*)^\T\tilde V_i=o_P(1)$, $N^{-1}\sum_{i\in I_k}(\hat\beta_0^{(-k)}-\beta_0^*)^\T\tilde V_i=o_P(1)$, $(\hat\beta_1^{(-k)}-\beta_1^*)^\T(\hat\mu^{(k)}-\tilde\mu)=o_P(1)$ and $(\hat\beta_0^{(-k)}-\beta_0^*)^\T(\hat\mu^{(k)}-\tilde\mu)=o_P(1)$. By Lemma \ref{l3}, $(\beta_1^*-\beta_0^*)^\T(\hat\mu^{(k)}-\tilde\mu)=o_P(1)$. Therefore,
\[
N^{-1}\sum_{i\in I_k}(\xi_{\delta,i}-\xi_{\delta,i}^*)^2=o_P(1),
\]
and hence $N^{-1}\sum_{i\in I_k}\xi_{\delta,i}^2=N^{-1}\sum_{i\in I_k}{\xi_{\delta,i}^*}^2+o_P(1)=V_2+o_P(1)$. When $K<\infty$,
\[
\hat V_\delta=K^{-1}\sum_{k=1}^K\left\{N^{-1}\sum_{i\in I_k}\nu_{\delta,i}^2+nN^{-1}\sum_{i\in I_k}\xi_{\delta,i}^2/(m+n)\right\}=V_\delta+o_P(1).
\]
\end{proof}

\begin{proof}[of Theorem~\ref{martheta}]

This proof provides an asymptotic normal result for $n^{1/2}(\hat\theta_\mathrm{MAR}-\theta_\mathrm{MAR})$. Assume the following rates 
\[
E_{J_k^c}\{\hat g^{(-k)}(X)-g^0(X)\}^2=O_P(a_{m+n,p}),\qquad E_{J_k^c}\{1-s^*(X)/\hat s^{(-k)}(X)\}^2=O_P(b_{m+n,p}).
\]
By definition, the proposed estimator $\hat\theta_\mathrm{MAR}$ can be rewritten as
\[
\hat\theta_\mathrm{MAR}=K^{-1}\sum_{k=1}^K\hat\theta_\mathrm{MAR}^{(k)},
\]
with
\[
\hat\theta_\mathrm{MAR}^{(k)}=M^{-1}\sum_{i\in J_k}\left[g^{(-k)}(X_i)+\frac{T_i\{Y_i^o-g^{(-k)}(X_i)\}}{\hat s^{(-k)}(X_i)}\right],
\]
where $M=|J_k|=(m+n)/K$. Recall that for each $i$,
\[
Y_i=g^0(X_i)+\varepsilon_i,\qquad T_i=s^*(X_i)+r_i.
\]
Hence,
\begin{align*}
&\hat\theta_\mathrm{MAR}^{(k)}-\theta\\
&\qquad=M^{-1}\sum_{i\in J_k}\biggr[g^0(X_i)+\frac{T_i\varepsilon_i}{s^*(X_i)}+T_i\varepsilon_i\left\{\frac{1}{\hat s^{(-k)}(X_i)}-\frac{1}{s^*(X_i)}\right\}\\
&\qquad\qquad-\frac{r_i\left\{\hat g^{(-k)}(X_i)-g^0(X_i)\right\}}{s^*(X_i)}-T_i\left\{\hat g^{(-k)}(X_i)-g^0(X_i)\right\}\left\{\frac{1}{\hat s^{(-k)}(X_i)}-\frac{1}{s^*(X_i)}\right\}\biggr].
\end{align*}
Since $Y_i$ and $T_i$ are independent conditional on $X_i$, the expectations of the terms on RHS are
\begin{align*}
&E\left\{g^0(X)+\frac{T\varepsilon}{s^*(X)}\right\}=\theta,\qquad E_{J_k^c}\left[T\varepsilon\left\{\frac{1}{\hat s^{(-k)}(X)}-\frac{1}{s^*(X)}\right\}\right]=0,\\
&E_{J_k^c}\left[\frac{r\left\{\hat g^{(-k)}(X)-g^0(X)\right\}}{s^*(X)}\right]=0.
\end{align*}
Now, since $E_{J_k^c}(T\mid X) =s^*(X) $ and by the tower property of the conditional expectations, we have
\begin{align*}
&E_{J_k^c}\left|T\left\{\hat g^{(-k)}(X)-g^0(X)\right\}\left\{\frac{1}{\hat s^{(-k)}(X)}-\frac{1}{s^*(X)}\right\}\right|
\\
&=E_{J_k^c}\left\{\left|\hat g^{(-k)}(X)-g^0(X)\right|\cdot\left|\frac{1}{\hat s^{(-k)}(X)}-\frac{1}{s^*(X)}\right|E_{J_k^c}(T\mid X)\right\}\\
&=E_{J_k^c}\left|\left\{\hat g^{(-k)}(X)-g^0(X)\right\}\frac{s^*(X)-\hat s^{(-k)}(X)}{\hat s^{(-k)}(X)}\right|.
\end{align*}
Now, by simple Holder's inequality and following the assumptions,  the above is of the order of 
$
 O_P(a_{m+n,p}b_{m+n,p}).
$

As for the the second moments, with similar reasoning, we have
\begin{align*}
&E_{J_k^c}\left[T\varepsilon\left\{\frac{1}{\hat s^{(-k)}(X)}-\frac{1}{s^*(X)}\right\}\right]^2  =E_{J_k^c}\left[\left\{\frac{\hat s^{(-k)}(X)-s^*(X)}{\hat s^{(-k)}(X)s^*(X)}\right\}^2E_{J_k^c}(T\varepsilon^2\mid X)\right]\\
& =E_{J_k^c}\left[\left\{\frac{\hat s^{(-k)}(X)-s^*(X)}{\hat s^{(-k)}(X)s^*(X)}\right\}^2E_{J_k^c}(T\mid X)E_{J_k^c}(\varepsilon^2\mid X)\right]
\\
&=E_{J_k^c}\left[\frac{\left\{\hat s^{(-k)}(X)-s^*(X)\right\}^2}{\left\{\hat s^{(-k)}(X)\right\}^2s^*(X)}E_{J_k^c}(\varepsilon^2\mid X)\right]=O_P(b_{m+n,p}^2/E(T)),
\end{align*}
as well as
\begin{align*}
&E_{J_k^c}\left[\frac{r_i\left\{\hat g^{(-k)}(X)-g^0(X)\right\}}{s^*(X)}\right]^2=E_{J_k^c}\left[\frac{\left\{\hat g^{(-k)}(X)-g^0(X)\right\}^2}{\{s^*(X)\}^2}E_{J_k^c}(r^2\mid X)\right]\\
&=E_{J_k^c}\left[\frac{\left\{\hat g^{(-k)}(X)-g^0(X)\right\}^2\{1-s^*(X)\}}{s^*(X)}\right]=O_P(a_{m+n,p}^2/E(T)).
\end{align*}
By Lemma \ref{l2}, we have
\begin{align*}
M^{-1}\sum_{i\in J_k}T_i\varepsilon_i\left\{\frac{1}{\hat s^{(-k)}(X_i)}-\frac{1}{s^*(X_i)}\right\}&=O_P(M^{-1/2}b_{m+n,p}\{E(T)\}^{-1/2})\\
&=o_P(M^{-1/2}\{E(T)\}^{-1/2}),\\
M^{-1}\sum_{i\in J_k}\frac{r_i\left\{\hat g^{(-k)}(X_i)-g^0(X_i)\right\}}{s^*(X_i)}&=O_P(M^{-1/2}a_{m+n,p}\{E(T)\}^{-1/2})\\
&=o_P(M^{-1/2}\{E(T)\}^{-1/2}).
\end{align*}
By Lemma \ref{l1}, and that $a_{m+n,p}b_{m+n,p}=o_P((m+n)^{-1/2}\{E(T)\}^{-1/2})$,
\[
M^{-1}\sum_{i\in J_k}T_i\left\{\hat g^{(-k)}(X_i)-g^0(X_i)\right\}\left\{\frac{1}{\hat s^{(-k)}(X_i)}-\frac{1}{s^*(X_i)}\right\}=o_P(M^{-1/2}\{E(T)\}^{-1/2}).
\]
Therefore,
\[
\hat\theta_\mathrm{MAR}^{(k)}=\sum_{i\in J_k}\left\{g^0(X_i)+\frac{T_i\varepsilon_i}{s^*(X_i)}\right\}+o_P\left(M^{-1/2}\{E(T)\}^{-1/2}\right).
\]
For $K<\infty$, we have
\[
\hat\theta_\mathrm{MAR}=\sum_{i=1}^{m+n}\left\{g^0(X_i)+\frac{T_i\varepsilon_i}{s^*(X_i)}\right\}+o_P\left((m+n)^{-1/2}\{E(T)\}^{-1/2}\right).
\]
Let $V_T=E\{T\varepsilon/s^*(X)\}^2$, and recall that $E(\varepsilon^2\mid X)<c_1$,
\begin{align*}
V_T&=E\left[\frac{1}{\{s^*(X)\}^2}E(T\varepsilon^2\mid X)\right]=E\left[\frac{1}{\{s^*(X)\}^2}E(T\mid X)E(\varepsilon^2\mid X)\right]= E\left\{\frac{1}{s^*(X)}E(\varepsilon^2\mid X)\right\}\\
&=E\left\{\frac{\varepsilon^2}{s^*(X)}\right\}\geq \frac{\left\{E(\varepsilon^2)\right\}^2}{E\{s^*(X)\varepsilon^2\}}>\frac{\{E(\varepsilon^2)\}^2}{C_1E(T)}.
\end{align*}
which implies that
\begin{align*}
 &1_{\{|\frac{T\varepsilon}{s^*(X)}|>\delta (m+n)V_T^{1/2}\}}\leq1_{\{\frac{\varepsilon^2}{c_1^2\{E(T)\}^2}>\frac{\delta^2(m+n)^2\{E(\varepsilon^2)\}^2}{C_1E(T)}\}}=1_{\{C_1\varepsilon^2>c_1^2\delta^2(m+n)^2\left\{E(\varepsilon^2)\right\}^2E(T)\}}\\
 &\leq\frac{C_1\varepsilon^2}{c_1^2\delta^2(m+n)^2\left\{E(\varepsilon^2)\right\}^2E(T)}.
\end{align*}
Hence, for any $\delta>0$,
\begin{align*}
&V_T^{-1}E\left[\left\{\frac{T\varepsilon}{s^*(X)}\right\}^21_{\{|\frac{T\varepsilon}{s^*(X)}|>\delta (m+n)V^{1/2}\}}\right] 
\\
&\leq \frac{C_1E(T)}{\{E(\varepsilon^2)\}^2}E\left[\frac{T\varepsilon^2}{\{s^*(X)\}^2}\cdot\left|\frac{C_1\varepsilon^2}{c_1^2\delta^2(m+n)^2\left\{E(\varepsilon^2)\right\}^2E(T)}\right|^{c/2}\right]\\
&=\frac{C_1^{1+c/2}\{E(T)\}^{1-c/2}}{c_1^c\delta^c(m+n)^c\{E(\varepsilon^2)\}^{2+c}}E\left[\frac{|\varepsilon|^{2+c}}{\{s^*(X)\}^2}E(T\mid X)\right]
\\
&=\frac{C_1^{1+c/2}\{E(T)\}^{1-c/2}}{c_1^c\delta^c(m+n)^c\{E(\varepsilon^2)\}^{2+c}}E\left\{\frac{|\varepsilon|^{2+c}}{s^*(X)}\right\}\\
&=\frac{C_1^{1+c/2}E|\varepsilon|^{2+c}}{c_1^{1+c}\delta^c(m+n)^c\{E(\varepsilon^2)\}^{2+c}\{E(T)\}^{c/2}}\to0,
\end{align*}
since $(m+n)^2E(T)\to\infty$. Therefore, by the Lindeberg Central Limit Theorem, as $m+n,p\to\infty$,
\[
\{(m+n)V_T\}^{-1/2}\sum_{i=1}^{m+n}\frac{T_i\varepsilon_i}{s^*(X_i)}\to N(0,1)
\]
in distribution. Besides, when $E\{g^0(X)\}^2>C>0$, we have
\[
\frac{E|g^0(X)|^{2+c}}{\left[E\{g^0(X)\}^2\right]^{1+c/2}}<\infty,
\]
by the Lindeberg-Feller Central Limit Theorem, as $m+n,p\to\infty$,
\[
\left[(m+n)E\{g^0(X)\}^2\right]^{1/2}\sum_{i=1}^{m+n}\{g^0(X_i)-\theta\}\to N(0,1)
\]
in distribution. Observe that
\[
\mathrm{cov}\left\{g^0(X),\frac{T\varepsilon}{s^*(X)}\right\}=0.
\]
Then, by the 
 delta method, as $m+n,p\to\infty$, we obtain
\begin{equation}\label{asyvar}
\left[\frac{m+n}{V_T+E\{g^0(X)\}^2}\right]^{1/2}\sum_{i=1}^{m+n}\left\{g^0(X_i)+\frac{T_i\varepsilon_i}{s^*(X_i)}-\theta\right\}\to N(0,1),
\end{equation}
in distribution. When $E\{g^0(X)\}^2\to0$, by Lemma \ref{l1}, $\sum_{i=1}^{m+n}\{g^0(X_i)-\theta\}=o_P((m+n)^{-1/2})$. Since $[V_T+E\{g^0(X)\}^2]/V_T\to1$, by the Slutsky's Theorem, \eqref{asyvar} still holds. Now, recall that 
\[
\hat\theta_\mathrm{MAR}=\sum_{i=1}^{m+n}\left\{g^0(X_i)+\frac{T_i\varepsilon_i}{s^*(X_i)}\right\}+R
\]
where 
\[
R=o_P\left((m+n)^{-1/2}\{E(T)\}^{-1/2}\right).
\]
Hence,
\[
\frac{m+n}{V_T+E\{g^0(X)\}^2}R^2=o_P\left(\frac{1}{E(T)[V_T+E\{g^0(X)\}^2]}\right)=o_P(1).
\]
Therefore, as $m+n,p\to\infty$, the estimator $\hat\theta_\mathrm{MAR}$ is asymptotically normal
\[
\left[\frac{m+n}{V_T+E\{g^0(X)\}^2}\right]^{1/2}(\hat\theta_\mathrm{MAR}-\theta)\to N(0,1).
\]
Here,
\[
(m+n)^{-1}[V_T+E\{g^0(X)\}^2]\leq(m+n)^{-1}\left[\frac{\{E(\varepsilon^2)\}^2}{C_1E(T)}+E\{g^0(X)\}^2\right]=O_P((m+n)^{-1}/E(T)).
\]
Now we showcase that $\hat V_\mathrm{MAR}(\theta)$ is a consistent estimator of $V_\mathrm{MAR}(\theta)=V_T+E\{g^0(X)^2\}$. Let 
\[
\nu_{\theta,i}=g^{(-k)}(X_i)+\frac{T_i\{Y_i-g^{(-k)}(X_i)\}}{s^{(-k)}(X_i)}-\hat\theta_\mathrm{MAR},\qquad\nu_{\theta,i}^*=g^0(X_i)+\frac{T_i\{Y_i-g^0(X_i)\}}{s^*(X_i)}-\theta.
\]
Then, similarly as in \eqref{xi0},
\begin{align*}
 \left|M^{-1}\sum_{i\in J_k}\nu_{\theta,i}^2-M^{-1}\sum_{i\in J_k}{\nu_{\theta,i}^*}^2\right|
 &\leq M^{-1}\sum_{i\in J_k}\left(\nu_{\theta,i}-\nu_{\theta,i}^*\right)^2
 \\
 &+2\left\{M^{-1}\sum_{i\in J_k}{\nu_{\theta,i}^*}^2M^{-1}\sum_{i\in J_k}\left(\nu_{\theta,i}-\nu_{\theta,i}^*\right)^2\right\}^{1/2}.
\end{align*}
We first consider the term $M^{-1}\sum_{i\in J_k}{\nu_{\theta,i}^*}^2$. Let $W_{n,i}={\nu_{\theta,i}^*}^2/V_\mathrm{MAR}(\theta)$. Then,
\begin{align*}
n\mathrm{pr}(|W_{n,1}|>n)&\leq E\left[|W_{n,1}|1_{\{|W_{n,1}|>n\}}\right]\leq M^{-c/2}E|W_{n,1}|^{1+c/2}\\
&\leq M^{-c/2}\frac{E|g^0(X)-\theta+T\varepsilon/s^*(X)|^{2+c}}{\{V_\mathrm{MAR}(\theta)\}^{1+c/2}}\\
&\leq M^{-c/2}\frac{\left\{\left(E|g^0(X)-\theta|^{2+c}\right)^{1/(2+c)}+\left(E|T\varepsilon/s^*(X)|^{2+c}\right)^{1/(2+c)}\right\}^{2+c}}{\{V_\mathrm{MAR}(\theta)\}^{1+c/2}}\\
&\leq M^{-c/2}\frac{E\left[|\varepsilon|^{2+c}/\{s^*(X)\}^{1+c}\right]}{\{V_\mathrm{MAR}(\theta)\}^{1+c/2}}+O(M^{-c/2})\\
&\leq \frac{C_1^{1+c/2}E|\varepsilon|^{2+c}}{c_1^{1+c}M^{c/2}\{E(\varepsilon^2)\}^{2+c}\{E(T)\}^{c/2}}+O(M^{-c/2})=o(1),
\end{align*}
since $ME(T)\to\infty$. Besides, for any $0<c_1<2$, similarly,
\begin{align*}
M^{-1}E[W_{n,1}^21_{\{|W_{n,1}|\leq M\}}]&\leq M^{-1}E(W_{n,1}^2|M/W_{n,1}|^{1-c/2})=M^{-c/2}E|W_{n,1}|^{1+c/2}=o(1).
\end{align*}
By general weak law of large numbers,
\[
\frac{M^{-1}\sum_{i\in J_k}{\nu_{\theta,i}^*}^2}{V_\mathrm{MAR}(\theta)}=1+o_P(1).
\]
Now, consider the term $M^{-1}\sum_{i\in J_k}(\nu_{\theta,i}-\nu_{\theta,i}^*)^2$. Observe that
\begin{align*}
\nu_{\theta,i}-\nu_{\theta,i}^*&=T_i\varepsilon_i\left\{\frac{1}{\hat s^{(-k)}(X_i)}-\frac{1}{s^*(X_i)}\right\}-\frac{r_i\{\hat g^{(-k)}(X_i)-g^0(X_i)\}}{s^*(X_i)}\\
&\qquad-T_i\left\{\hat g^{(-k)}(X_i)-g^0(X_i)\right\}\left\{\frac{1}{\hat s^{(-k)}(X_i)}-\frac{1}{s^*(X_i)}\right\}-(\hat\theta_\mathrm{MAR}-\theta).
\end{align*}
Recall that 
\begin{align*}
&E_{J_k^c}\left[T\varepsilon\left\{\frac{1}{\hat s^{(-k)}(X)}-\frac{1}{s^*(X)}\right\}\right]^2=o_P\{1/E(T)\}, \\
&E_{J_k^c}\left[\frac{r_i\left\{\hat g^{(-k)}(X)-g^0(X)\right\}}{s^*(X)}\right]^2=o_P\{1/E(T)\}.
\end{align*}
Besides,
\begin{align*}
&E_{J_k^c}\left[T\left\{\hat g^{(-k)}(X)-g^0(X)\right\}\left\{\frac{1}{\hat s^{(-k)}(X)}-\frac{1}{s^*(X)}\right\}\right]^2\\
&=E_{J_k^c}\left[\left\{\hat g^{(-k)}(X)-g^0(X)\right\}^2\frac{\left\{\hat s^{(-k)}(X)-s^*(X)\right\}^2}{\left\{\hat s^{(-k)}(X)\right\}^2s^*(X)}\right]=o_P\{1/E(T)\}.
\end{align*}
By Lemma \ref{l1},
\begin{align*}
&M^{-1}\sum_{i\in J_k}\left[T_i\varepsilon_i\left\{\frac{1}{\hat s^{(-k)}(X_i)}-\frac{1}{s^*(X_i)}\right\}\right]^2=o_P\{1/E(T)\},\\
&M^{-1}\sum_{i\in J_k}\left[\frac{r_i\{\hat g^{(-k)}(X_i)-g^0(X_i)\}}{s^*(X_i)}\right]^2=o_P\{1/E(T)\},\\
&M^{-1}\sum_{i\in J_k}\left[T_i\left\{\hat g^{(-k)}(X_i)-g^0(X_i)\right\}\left\{\frac{1}{\hat s^{(-k)}(X_i)}-\frac{1}{s^*(X_i)}\right\}\right]^2=o_P\{1/E(T)\}.
\end{align*}
Combining with the fact that $(\hat\theta_\mathrm{MAR}-\theta)^2=o_P\{1/E(T)\}$, we have
\[
M^{-1}\sum_{i\in J_k}(\nu_{\theta,i}-\nu_{\theta,i}^*)^2=o_P\{1/E(T)\}.
\]
Therefore,
\begin{align*}
M^{-1}\sum_{i\in J_k}\nu_{\theta,i}^2&=M^{-1}\sum_{i\in J_k}{\nu_{\theta,i}^*}^2+o_P\{1/E(T)\}+2\left[V_\mathrm{MAR}\{1+o_P(1)\}o_P\{1/E(T)\}\right]^{1/2}\\
&=V_\mathrm{MAR}\{1+o_P(1)\}+o_P\{1/E(T)\}
\end{align*}
and hence
\[
\frac{\hat V_\mathrm{MAR}}{V_\mathrm{MAR}}=\frac{K^{-1}\sum_{k=1}^KM^{-1}\sum_{i\in J_k}\nu_{\theta,i}^2}{V_\mathrm{MAR}}=1+o_P(1).
\]
\end{proof}

\begin{proof}[of Theorem~\ref{genv}]

This proof provides asymptotic normalities for the variance, explained variance and unexplained variance estimators. We first work on the explained variance and the unexplained variance. With a slight abuse of notation, define
\begin{align*}
\hat b^{2^{(k)}}&=M^{-1}\sum_{i\in J_k}\{\hat h^{(-k)}(X_i)\}^2+2N^{-1}\sum_{i\in I_k}\hat h^{(-k)}(X_i)\{Y_i-\hat\theta_\mathrm{gen}-\hat h^{(-k)}(X_i)\},\\
\tilde b^{2^{(k)}}&=M^{-1}\sum_{i\in J_k}\{\hat g^{(-k)}(X_i)-\mu^{(-k)}\}^2\\
&\qquad+2N^{-1}\sum_{i\in I_k}\{\hat g^{(-k)}(X_i)-\mu^{(-k)}\}\{Y_i-\theta-\hat g^{(-k)}(X_i)+\mu^{(-k)}\},\\
\check b^{2^{(k)}}&=M^{-1}\sum_{i\in J_k}\{\tilde g^*(X_i)\}^2+2N^{-1}\sum_{i\in I_k}\varepsilon_i\tilde g^*(X_i),\\
\hat\sigma_\varepsilon^{2^{(k)}}&=N^{-1}\sum_{i\in I_k}\{Y_i-\hat\theta_\mathrm{gen}-\hat h^{(-k)}(X_i)\}^2,\\
\tilde\sigma_\varepsilon^{2^{(k)}}&=N^{-1}\sum_{i\in I_k}\{Y_i-\theta-\hat g^{(-k)}(X_i)+\mu^{(-k)}\}^2,\\
\check\sigma_\varepsilon^{2^{(k)}}&=N^{-1}\sum_{i\in I_k}\varepsilon_i^2.
\end{align*}
where $\hat h^{(-k)}(X_i)=\hat g^{(-k)}(X_i)-M^{-1}\sum_{i\in J_k}\hat g^{(-k)}(X_i)$, $\mu^{(-k)}= E_{I_k^c}\{\hat g^{(-k)}(X)\}$ and $\tilde g^*(X_i)=g^*(X_i)-\theta$. The proof consists of 3 steps:\\
Step 1: 
$\hat b^{2^{(k)}}=\tilde b^{2^{(k)}}+o_P(N^{-1})$, $\hat\sigma_\varepsilon^{2^{(k)}}=\tilde\sigma_\varepsilon^{2^{(k)}}+o_P(N^{-1})$.\\
Step 2:
$\tilde b^{2^{(k)}}=\check b^{2^{(k)}}+o_P(N^{-\frac{1}{2}})$, $\tilde\sigma_\varepsilon^{2^{(k)}}=\check\sigma_\varepsilon^{2^{(k)}}+o_P(N^{-\frac{1}{2}})$.\\
Step 3:
$n^{1/2}\{V(\sigma_{\varepsilon,\mathrm{gen}}^2)\}^{-1/2}(\hat\sigma_{\varepsilon,\mathrm{gen}}^2-\sigma_{\varepsilon,\mathrm{gen}}^2)\to  N (0,1)$, $n^{1/2}\{V(b_\mathrm{gen}^2)\}^{-1/2}(\hat b_\mathrm{gen}^2-b_\mathrm{gen}^2) \to  N (0,1)$.

Step 1. Let $\Delta_1=M^{-1}\sum_{i\in J_k}\hat g^{(-k)}(X_i)-\mu^{(-k)}$, $\Delta_2=\hat\theta_\mathrm{gen}-\theta$ and $\delta_i=\hat g^{(-k)}(X_i)-\mu^{(-k)}-\tilde g^*(X_i)$ . Then, $\Delta_2=O_P(n^{-1/2})$, and
\[
\Delta_1=M^{-1}\sum_{i\in J_k}\delta_i+M^{-1}\sum_{i\in J_k}\tilde g^*(X_i).
\]
By Lemma \ref{l1} and \ref{l2},
\[
M^{-1}\sum_{i\in J_k}\delta_i=o_P(M^{-1/2}),\qquad M^{-1}\sum_{i\in J_k}\tilde g^*(X_i)=O_P(M^{-1/2})
\]
and hence $\Delta_1=O_P(M^{-1/2})$. Observe that
\begin{align*}
\hat b^{2^{(k)}}&=\tilde b^{2^{(k)}}-\Delta_1^2+\Delta_1(\Delta_2-\Delta_1)-\Delta_1N^{-1}\sum_{i\in I_k}\left\{Y_i-\theta-\hat g^{(-k)}(X_i)+\mu^{(-k)}\right\}\\
&\qquad+(\Delta_1-\Delta_2)N^{-1}\sum_{i\in I_k}\left\{\hat g^{(-k)}(X_i)-\mu^{(-k)}\right\},
\end{align*}
where by Lemma \ref{l1} and Lemma \ref{l3},
\[
N^{-1}\sum_{i\in I_k}\left\{Y_i-\theta-\hat g^{(-k)}(X_i)+\mu^{(-k)}\right\}=o_P(1),\quad N^{-1}\sum_{i\in I_k}\left\{\hat g^{(-k)}(X_i)-\mu^{(-k)}\right\}=o_P(1).
\]
Therefore,
\begin{equation}
\hat b^{2^{(k)}}=\tilde b^{2^{(k)}}+o_P(N^{-1/2}).\label{hb2-tildeb2}
\end{equation}
Besides,
\[
\hat\sigma_\varepsilon^{2^{(k)}}=\tilde\sigma_\varepsilon^{2^{(k)}}-2(\Delta_1-\Delta_2)N^{-1}\sum_{i\in I_k}(\varepsilon_i-\delta_i)+(\Delta_1-\Delta_2)^2=\tilde\sigma_\varepsilon^{2^{(k)}}+o_P(N^{-1/2}).
\]

Step 2. Observe that
\begin{align*}
\tilde b^{2^{(k)}}&=\check b^{2^{(k)}}+M^{-1}\sum_{i\in J_k}\delta_i\{\delta_i+2\tilde g^*(X_i)\}+2N^{-1}\sum_{i\in I_k}\delta_i\{\varepsilon_i-\tilde g^*(X_i)-\delta_i\},\\
\tilde\sigma_\varepsilon^{2^{(k)}}&=\check\sigma_\varepsilon^{2^{(k)}}-2N^{-1}\sum_{i\in I_k}\varepsilon_i\delta_i+N^{-1}\sum_{i\in I_k}\delta_i^2.
\end{align*}
By Lemma \ref{l1},
\begin{align*}
&M^{-1}\sum_{i\in J_k}\delta_i\{\delta_i+2\tilde g^*(X_i)\}=E_{I_k^c}(\delta^2)+2 E_{I_k^c}\{\delta\tilde g^*(X)\}+o_P(M^{-\frac{1}{2}}),\\
&N^{-1}\sum_{i\in I_k}\delta_i\{\varepsilon_i-\tilde g^*(X_i)-\delta_i\}=E_{I_k^c}(\delta\varepsilon)- E_{I_k^c}\{\delta\tilde g^*(X)\}- E_{I_k^c}(\delta^2)+o_P(N^{-\frac{1}{2}}),\\
&N^{-1}\sum_{i\in I_k}\varepsilon_i\delta_i=E_{I_k^c}(\delta\varepsilon)+o_P(N^{-\frac{1}{2}}),\\
&N^{-1}\sum_{i\in I_k}\delta_i^2=E_{I_k^c}(\delta^2)+o_P(N^{-\frac{1}{2}}).
\end{align*}
Hence,
\begin{align}
\tilde b^{2^{(k)}}&=\check b^{2^{(k)}}+2 E_{I_k^c}(\delta\varepsilon)- E_{I_k^c}(\delta^2)+o_P(N^{-\frac{1}{2}}),\label{cancel1}\\
\tilde\sigma_\varepsilon^{2^{(k)}}&=\check\sigma_\varepsilon^{2^{(k)}}-2 E_{I_k^c}(\delta\varepsilon)+ E_{I_k^c}(\delta^2)+o_P(N^{-\frac{1}{2}}).\label{cancel2}
\end{align}
By assuming $ E_{I_k^c}(\delta\varepsilon)=o_P(N^{-\frac{1}{2}})$ and $ E_{I_k^c}(\delta^2)=o_P(N^{-\frac{1}{2}})$, we have
\begin{equation}
\tilde b^{2^{(k)}}=\check b^{2^{(k)}}+o_P(N^{-\frac{1}{2}}),\qquad\tilde\sigma_\varepsilon^{2^{(k)}}=\check\sigma_\varepsilon^{2^{(k)}}+o_P(N^{-\frac{1}{2}}).\label{tildeb2-checkb2}
\end{equation}
Step 3. Observe that
\begin{align*}
&K^{-1}\sum_{k=1}^K\check b^{2^{(k)}}=2n^{-\frac{1}{2}}\sum_{i=1}^n\varepsilon_i\tilde g^*(X_i)+(m+n)^{-1}\sum_{i=1}^n\{\tilde g^*(X_i)\}^2+(m+n)^{-1}\sum_{i=n+1}^{m+n}\{\tilde g^*(X_i)\}^2,\\
&K^{-1}\sum_{k=1}^K\check\sigma_\varepsilon^{2^{(k)}}=\sum_{i=1}^n\varepsilon_i^2.
\end{align*}
By Lindeberg-Feller Central Limit Theorem, as $m,n,p\to\infty$,
\[
\frac{n^{1/2}(K^{-1}\sum_{k=1}^K\check b^{2^{(k)}}-b_\mathrm{gen}^2)}{\{V(b_\mathrm{gen}^2)\}^{1/2}}\to N(0,1),\quad\frac{n^{1/2}(K^{-1}\sum_{k=1}^K\check\sigma_\varepsilon^2-\sigma_{\varepsilon,\mathrm{gen}}^2)}{\{V(\sigma_{\varepsilon,\mathrm{gen}}^2)\}^{1/2}}\to N(0,1).
\]
Hence,
\[
n^{1/2}\{V(b_\mathrm{gen}^2)\}^{-1/2}(\hat b_\mathrm{gen}^2-b_\mathrm{gen}^2) \to  N (0,1),\quad n^{1/2}\{V(\sigma_{\varepsilon,\mathrm{gen}}^2)\}^{-1/2}(\hat\sigma_{\varepsilon,\mathrm{gen}}^2-\sigma_{\varepsilon,\mathrm{gen}}^2)\to  N (0,1).
\]
 
Now, we show the asymptotic normal result for the variance estimator. Recall from \eqref{cancel1} and \eqref{cancel2},
\begin{align*}
&\hat\sigma_{Y,\mathrm{gen}}^2=\hat b_\mathrm{gen}^2+\hat\sigma_{\varepsilon,\mathrm{gen}}^2\\
&=K^{-1}\sum_{k=1}^K\left\{\check b^{2^{(k)}}+2 E_{I_k^c}(\delta\varepsilon)-E_{I_k^c}(\delta^2)+\check\sigma_\varepsilon^{2^{(k)}}-2 E_{I_k^c}(\delta\varepsilon)+ E_{I_k^c}(\delta^2)+o_P(N^{-\frac{1}{2}})\right\}\\
&=K^{-1}\sum_{k=1}^K\left\{\check b^{2^{(k)}}+\check\sigma_\varepsilon^{2^{(k)}}+o_P(N^{-\frac{1}{2}})\right\},
\end{align*}
where the bias terms $2 E_{I_k^c}(\delta\varepsilon)$ and $E_{I_k^c}(\delta^2)$ canceled out. By Lindeberg-Feller Central Limit Theorem and Slutsky's Theorem, as $m,n,p\to\infty$,
\[
\frac{n^{1/2}(\hat\sigma_{Y,\mathrm{gen}}^2-\sigma_Y^2)}{\{V(\sigma_Y^2)\}^{1/2}} \to  N (0,1).
\]
Lastly, for the PVE estimation, by Step 1 and 2, we showcase that $\hat b^{2^{(k)}}=\check b^{2^{(k)}}+o_P(N^{-1/2})$ and $\hat\sigma_Y^{2^{(k)}}=\check\sigma_Y^{2^{(k)}}+o_P(N^{-1/2})$, where $\check\sigma_Y^{2^{(k)}}=\check\sigma_\varepsilon^{2^{(k)}}+\check b^{2^{(k)}}$. Besides, we also have $\hat\sigma_Y^{2^{(k)}}=\sigma_Y^2+o_P(1)$ by Lemma \ref{l1}. Hence, for each $k\leq K$,
\begin{align*}
&n^{1/2}\left(\frac{\hat b^{2^{(k)}}}{\hat\sigma_Y^{2^{(k)}}}-\frac{b_\mathrm{gen}^2}{\sigma_Y^2}\right)=n^{1/2}\left\{\frac{\sigma_Y^2(\hat b^{2^{(k)}}-b_\mathrm{gen}^2)-b_\mathrm{gen}^2(\hat\sigma_Y^{2^{(k)}}-\sigma_Y^2)}{\sigma_Y^2\hat\sigma_Y^{2^{(k)}}}\right\}\\
&\qquad=n^{1/2}\left[\frac{\sigma_Y^2\{\check b^{2^{(k)}}-b_\mathrm{gen}^2+o_P(N^{-1/2})\}-b_\mathrm{gen}^2\{\check\sigma_Y^{2^{(k)}}-\sigma_Y^2+o_P(N^{-1/2})\}}{\sigma_Y^2\{\sigma_Y^2+o_P(1)\}}\right]\\
&\qquad=n^{1/2}\sigma_Y^{-2}(\check b^{2^{(k)}}-b_\mathrm{gen}^2)-n^{1/2}\sigma_Y^{-4}b_\mathrm{gen}^2(\check\sigma_Y^{2^{(k)}}-\sigma_Y^2)+o_P(1).
\end{align*}
It follows that
\begin{align*}
&n^{1/2}(R_\mathrm{gen}^2-PVE)\\
&\qquad=n^{1/2}\sigma_Y^{-2}K^{-1}\sum_{k=1}^K(\check b^{2^{(k)}}-b_\mathrm{gen}^2)-n^{1/2}\sigma_Y^{-4}b_\mathrm{gen}^2K^{-1}\sum_{k=1}^K(\check\sigma_Y^{2^{(k)}}-\sigma_Y^2)+o_P(1)\\
&\qquad=n^{1/2}\sum_{i=1}^{n}\left(\sigma_Y^{-4}\sigma_{\varepsilon,\mathrm{gen}}^2[\{\tilde g^*(X_i)\}^2+\varepsilon_i\tilde g^*(X_i)-b_\mathrm{gen}^2]-\sigma_Y^{-4}b_\mathrm{gen}^2(\varepsilon_i^2-\sigma_\varepsilon^2)\right)\\
&\qquad\qquad+n^{1/2}\sum_{i=n+1}^{m+n}\sigma_Y^{-4}\sigma_{\varepsilon,\mathrm{gen}}^2[\{\tilde g^*(X_i)\}^2-b_\mathrm{gen}^2].
\end{align*}
By Lindeberg-Feller Central Limit Theorem,
\[
n^{1/2}V^{-1/2}(R_\mathrm{gen}^2)(R_\mathrm{gen}^2-\mathrm{PVE})\to N(0,1).
\]
\end{proof}

\end{document}